\title{A Short Course on Error-Correcting Codes}
\author{Mario Blaum  \\
{\tt mblaum@hotmail.com}}
 \date{\copyright Copyright 2009\\
All Rights Reserved}
 \documentstyle [12pt]{book}
 \newtheorem{theo}{Theorem}[section]
 \newtheorem{lemma}{Lemma}[section]
 
 \newtheorem{defin}{Definition}[section]
 \newtheorem{ex}{Example}[section]
 \newtheorem{alg}{Algorithm}[section]
 \newtheorem{cor}{Corollary}[section]
 \newtheorem{prob}{}[section]

\newcommand{\qed}{\hfill$\Box$\vspace{.8cm}}
\newcommand{\pr}{\vspace{.8cm} {\bf Problems}}
\newcommand{\sol}{\vspace{.8cm} {\bf Solutions}}
\newcommand{\pf}{{\bf Proof: }}
\newcommand{\al}{\mbox{$\alpha$}}
\newcommand{\be}{\mbox{$\beta$}}
\newcommand{\ga}{\mbox{$\gamma$}}
\newcommand{\si}{\mbox{$\sigma$}}
\newcommand{\om}{\mbox{$\omega$}}
\newcommand{\eq}{\mbox{$\,=\,$}}
\newcommand{\ur}{\mbox{$\underline{r}$}}

\newcommand{\ua}{\mbox{$\underline{a}$}}
\newcommand{\ub}{\mbox{$\underline{b}$}}

\newcommand{\uv}{\mbox{$\underline{v}$}}
\newcommand{\ue}{\mbox{$\underline{e}$}}

\newcommand{\us}{\mbox{$\underline{s}$}}
\newcommand{\uu}{\mbox{$\underline{u}$}}
\newcommand{\uc}{\mbox{$\underline{c}$}}

\newcommand{\uw}{\mbox{$\underline{w}$}}
\newcommand{\up}{\mbox{$\underline{p}$}}
\newcommand{\uq}{\mbox{$\underline{q}$}}
\newcommand{\ug}{\mbox{$\underline{g}$}}
\newcommand{\hS}{\mbox{$\hat{S}$}}

\newcommand{\0}{\mbox{$\underline{0}$}}
\newcommand{\1}{\mbox{$\underline{1}$}}
\newcommand{\ra}{\mbox{$\rightarrow$}}

\newcommand{\la}{\mbox{$\leftarrow$}}

\newcommand{\lra}{\mbox{$\leftrightarrow$}}

\newcommand{\lf}{\mbox{$\lfloor$}}
\newcommand{\rf}{\mbox{$\rfloor$}}

\newcommand{\xor}{\mbox{$\oplus$}}
\newcommand{\C}{\mbox{${\cal C}$}}

\newcommand{\th}{\mbox{${\tilde{h}}$}}
\newcommand{\G}{\mbox{${\cal G}$}}
\newcommand{\E}{\mbox{${\cal E}$}}

\newcommand{\bS}{\mbox{${\bf S}$}}

\begin{document}
 \parindent=0pt
 \maketitle

\chapter{Basic Concepts in Error Correcting Codes}
\label{ch1}
\section{Introduction}
\label{sec1}
When digital data are transmitted over a noisy channel, it is important
to have a mechanism allowing recovery against a limited number of
errors. Normally, a user string of 0's and 1's, called bits, is encoded
by adding a number of redundant bits to it. When the receiver attempts
to reconstruct the original message sent, it starts by examining a
possibly corrupted version of the encoded message, and then makes a
decision. This process is called the decoding.

The set of all possible encoded messages is called an error-correcting
code. The field was started in the late 40's by the work of Shannon
and Hamming, and since then thousands of papers on the subject
have been published. There are also several very good books touching
different aspects of error-correcting
codes~\cite
{ad,be,bl,cf,cl,d,gl,im,hi,ho,kk,lee,lin,l,mac,mc,mi,moon,mz,pl,rao,rhee,r,sw,sw2,vans,pw,w,wigg}.
Programs implementing different codes can be found in~\cite{br}.

The purpose of this course is giving an introduction to the theory
and practice of error-correcting codes.

Unless otherwise stated, we will assume that our information
symbols are bits, i.e., 0 and 1. The set $\{0,1\}$ has a field
structure under the exclusive-OR ($\xor$)
and product operations. We denote this field $GF(2)$, which means
Galois field of order 2.

Roughly, there are two types of error-correcting codes: codes of
block type and codes of convolutional type. Codes of block type
encode a fixed number of bits, say $k$ bits, into a vector of length
$n$. So, the information string is divided into blocks of $k$ bits
each. Convolutional codes take the string of information bits
globally and slide a window over the data in order to encode.
A certain amount of memory is needed by the encoder.

In this course, we concentrate on block codes.

As said above, we encode $k$ information bits into $n$ bits.
So, we have a 1-1 function $f$,

$$f:GF(2)^k\ra GF(2)^n.$$

The function $f$ defines the encoding procedure. The set of
$2^k$ encoded vectors of length $n$ is called a code of {\em length}
$n$ and {\em dimension} $k$, and we denote it
as an $[n,k]$ code.
We call codewords the elements of the code
while we call words  the vectors of length $n$ in general.
The ratio $k/n$ is called the {\em rate} of the code.

Apart from the length and the dimension, a third parameter is needed
in order to define the error-correcting power of the code. This
parameter is the so called minimum (Hamming) distance of the code.
Formally:

\begin{defin}
\label{def1}
{\em
Given two vectors of length $n$, say $\ua$ and $\ub$, we call the
Hamming distance between $\ua$ and $\ub$ the number of coordinates
in which they differ (notation, $d_H(\ua,\ub)$).

Given a code $\C$ of length $n$ and dimension $k$, let

$$d=\min\{d_H(\ua,\ub)\;:\;\ua\neq \ub\;,\;\ua,\ub\in\C\}.$$

We call $d$ the minimum (Hamming) distance of the code $\C$ and we say
that $\C$ is an $[n,k,d]$ code.
}
\end{defin}

It is easy to verify that $d_H(\ua,\ub)$ verifies the axioms of
distance (Problem~\ref{pr1.1}), i.e.,

\begin{enumerate}
\item $d_H(\ua,\ub)=d_H(\ub,\ua)$.

\item $d_H(\ua,\ub)=0$ if and only if $\ua=\ub$.

\item $d_H(\ua,\uc)\leq d_H(\ua,\ub)+d_H(\ub,\uc)$.

\end{enumerate}

We call a sphere of radius $r$ and center $\ua$ the set of vectors
that are at distance at most $r$ from $\ua$.

The relation between $d$ and the maximum number of errors that code
$\C$ can correct is given by the following lemma:

\begin{lemma}
\label{lemma1}
{\em
The maximum number of errors that an $[n,k,d]$ code can correct is
$\lf{d-1\over 2}\rf$, where $\lf x\rf$ denotes the largest integer
smaller or equal than $x$.
}
\end{lemma}

{\bf Proof:} Assume that vector $\ua$ was transmitted but a possibly
corrupted version of $\ua$, say $\ur$, was received. Moreover, assume
that no more than
$\lf{d-1\over 2}\rf$ errors have occurred.

Consider the set of $2^k$
spheres of radius
$\lf{d-1\over 2}\rf$ whose centers are the codewords in $\C$.
By the definition of $d$, all these spheres are disjoint. Hence,
$\ur$ belongs to one and only one sphere: the one whose center is
codeword $\ua$. So, the decoder looks for the sphere in which $\ur$
belongs, and outputs the center of that sphere as the decoded vector.
As we see, whenever the number of errors is at most
$\lf{d-1\over 2}\rf$, this procedure will give the correct answer.

Moreover, $\lf (d-1)/2\rf$ is the maximum number of errors that the code
can correct. For let $\ua,\ub\in\C$ such that
$d_H(\ua,\ub)=d$. Let $\uu$ be a vector such that
$d_H(\ua,\uu)=1+\lf (d-1)/2\rf$ and
$d_H(\ub,\uu)=d-1-\lf (d-1)/2\rf$. We easily verify that
$d_H(\ub,\uu)\leq d_H(\ua,\uu)$, so, if $\ua$ is transmitted and
$\uu$ is received (i.e., $1+\lf (d-1)/2\rf$ errors have occurred),
the decoder cannot decide that the transmitted codeword was
$\ua$, since codeword
$\ub$ is at least as close to $\uu$ as $\ua$.\qed

\begin{ex}
\label{ex1}
{\em
Consider the following 1-1 relationship between $GF(2)^2$ and
$GF(2)^5$ defining the encoding:

\begin{eqnarray*}
00&\lra &00000\\
10&\lra &00111\\
01&\lra &11100\\
11&\lra &11011
\end{eqnarray*}

The 4 codewords in $GF(2)^5$ constitute
a $[5,2,3]$ code $\C$. From Lemma~\ref{lemma1}, $\C$ can correct 1 error.

For instance, assume that we receive the word $\ur=10100$.
The decoder looks into the 4 spheres of radius 1 (each sphere has
6 elements!) around each codeword. In effect, the sphere with
center 11100 consists of the center and of the 5 words at distance 1
from such center: 01100, 10100, 11000, 11110 and 11101. Notice that $\ur$
belongs in the sphere with center 11100. 

If we look at the table above, the final
output of the decoder is the information block 01.

However, let's assume that the transmitted codeword was 00000, and
two errors occur such that the received word is 00101. We can see
that this received word belongs in the sphere with center the
codeword 00111, so it will be erroneously decoded. This happens
because the number of errors has exceeded the maximum allowed by the
error-correcting capability of the code. \qed
}
\end{ex}

Example \ref{ex1}
shows that the decoder has to make at most 24
checks before arriving to the correct decision. When large codes are
involved, as is the case in applications, this decoding procedure
is not practical, since it amounts to an exhaustive search over
a huge set of vectors. A large part of this course will be devoted
to finding codes with efficient decoding procedures.

One of the goals in the theory of error-correcting codes is finding
codes with high rate and minimum distance as large as possible. The
possibility of finding codes with the right properties is often
limited by bounds that constrain the choice of parameters $n$, $k$
and $d$. We give some of these bounds in the next section.

Let us point out that error-correcting codes can be used for
detection instead of correction of errors. The simplest example
of an error-detecting code is given by a parity code: a parity
is added to a string of bits in such a way that the total number
of bits is even (a more sophisticated way of saying this, is that
the sum modulo-2 of the bits has to be 0). For example,
0100 is encoded as 01001. If an error occurs, or, more generally,
an odd number of errors, these errors will be detected since the
sum modulo 2 of the received bits will be 1. Notice that 2 errors
will be undetected. In general, if an $[n,k,d]$ code is used for
detection only, the decoder checks whether the received vector
is in the code or not. If it is not, then errors are detected.
It is easy to see that an $[n,k,d]$ code can detect up to
$d-1$ errors. Also, we can choose to correct less than
$\lf {d-1\over 2}\rf$ errors, say $s$ errors, by taking disjoint
spheres of radius $s$ around codewords, and using the remaining
capacity to detect errors. In other words, we want to correct
up to $s$ errors or detect up to $s+t$ errors when more than
$s$ errors occur. The relation between $s$, $t$ and the minimum
distance $d$ is given in Problem~\ref{pr1.2}.

Another application of error-correcting codes is in erasure correction.
An erased bit is a bit that cannot be read, so the decoder has
to decide if it was a 0 or a 1. An erasure is normally denoted
with the symbol $?$. For instance, 01?0 means that we cannot
read the third symbol. Obviously, it is easier to correct erasures
than to correct errors, since in the case of erasures we already
know the location, we simply have to find what the erased bit was.
It is not hard to prove that an $[n,k,d]$ code can correct up
to $d-1$ erasures. We may also want to simultaneously correct errors
and erasures. This situation is treated in
Problem~\ref{pr1.3}, which gives the number
of errors and erasures that a code with minimum
distance $d$ can correct. In fact, both
Problems~\ref{pr1.2} and~\ref{pr1.3} can be viewed as generalizations
of Lemma~\ref{lemma1}.

\vspace{.8cm}
 
\pr
 
\begin{prob}
\label{pr1.1}
{\em
Prove that the Hamming distance $d_H$ satisfies the axioms of
distance.
}
\end{prob}
 
\begin{prob}
\label{pr1.2}
{\em
Let $\C$ be a code with minimum distance $d$ and let $s$ and
$t$ be two numbers such that $2s+t\leq d-1$.
Prove that $\C$
can correct up to $s$ errors or detect up to $s+t$ errors when more
than $s$ errors occurred.
}
\end{prob}
 
\begin{prob}
\label{pr1.3}
{\em
Prove that a code $\C$ with minimum distance $d$
can correct $s$ errors together with $t$ erasures
whenever $2s+t\leq d-1$
}
\end{prob}
 
\sol
\vspace{.8cm}

{\bf Problem~\ref{pr1.1}}
 
Let $\ua=(a_1,a_2,\ldots,a_n)$,
$\ub=(b_1,b_2,\ldots,b_n)$ and
$\uc=(c_1,c_2,\ldots,c_n)$.
It is clear that $d_H(\ua,\ub)=0$ if and only if $\ua=\ub$ and that
$d_H(\ua,\ub)=d_H(\ub,\ua)$.
So, it remains to be proved the triangle inequality
 
$$d_H(\ua,\uc)\leq d_H(\ua,\ub)+d_H(\ub,\uc).$$
 
Let $S(\ua,\uc)$ be the set of coordinates where
$\ua$ and $\uc$ differ, i.e.,
$S(\ua,\uc)=\{i:a_i\neq c_i\}$.
Notice that $d_H(\ua,\uc)=|S(\ua,\uc)|$, where, if $S$ is a set,
$|S|$ denotes the cardinality of the set $S$.
 
Similarly, we define $S(\ua,\ub)$ and $S(\ub,\uc)$.
Claim:
$S(\ua,\uc)\subseteq S(\ua,\ub)\cup S(\ub,\uc)$.
 
In effect, if $i\in S(\ua,\uc)$ and
$i\not\in S(\ua,\ub)$, then $a_i\neq c_i$ and $a_i=b_i$; hence
$b_i\neq c_i$ and $i\in S(\ub,\uc)$, so the claim follows.
Hence,
 
$$d_H(\ua,\uc)=|S(\ua,\uc)|\leq
|S(\ua,\ub)|+|S(\ub,\uc)|=d_H(\ua,\ub)+d_H(\ub,\uc).$$
 
\vspace{.8cm}
 
{\bf Problem~\ref{pr1.2}}
 
Consider the spheres with radius $s$ and the codewords of $\C$
as centers. These spheres are disjoint, hence, when $s$ or fewer
errors occur, they will be corrected (see Lemma~1.1.1).
Assume that $\uu\in\C$ is transmitted and
$l$ errors occurred, where $s+1\leq l\leq s+t$.
Let $\ur$ be the received vector.
Since $d_H(\uu,\ur)>s$, $\ur$ is not in the sphere with center
$\uu$ and radius $s$. Assume that $\ur$ is in the sphere with
center $\uv$ and radius $s$, for some $\uv\in\C$, $\uv\neq\uu$.
In this case,
we would have incorrect decoding and the $l$ errors would be
undetected. But
 
$$d_H(\uu,\uv)\leq d_H(\uu,\ur)+d_H(\ur,\uv)\leq (s+t)+s=2s+t\leq d-1.$$
 
This is a contradiction, since any two codewords are at distance
at least $d$ apart.
 
\vspace{.8cm}
 
{\bf Problem~\ref{pr1.3}}
 
Let $\uu$ be the transmitted codeword and $\ur$ the received word.
Let $T$ be the set of erased locations and $S$ be the set
of locations in error; hence $|T|=t$ and $|S|=s$. Assume that
$\ur$ is decoded as a codeword $\uv\neq\uu$, where $\uv$ has suffered
at most $s'=\lf(d-1-t)/2\rf$ errors in a set of locations $S'$.
Hence, $\uu$ and $\uv$ may differ only in the set of erasures $T$
and in the error sets $S$ and $S'$. Hence,
 
$$d_H(\uu,\uv)\leq |T|+|S|+|S'|\leq t+2s'\leq d-1.$$
 
This is a contradiction.

\section{Linear Codes}
\label{sec2}

We have seen in the previous section that a binary code of length
$n$ is  a subset of $GF(2)^n$. Notice that, being $GF(2)$ a field,
$GF(2)^n$ has a structure
of vector space over $GF(2)$.
We say that a code $\C$
is linear if it is a subspace of $GF(2)^n$, i.e.:

\begin{enumerate}

\item $\underline{0}\in\C$.

\item $\forall\;\ua,\ub\in\C$, $\ua\xor\ub\in\C$.

\end{enumerate}

The symbol $\underline{0}$ denotes the all-zero vector.
In general, we denote vectors with underlined letters, otherwise
letters denote scalars.

In Section~\ref{sec1}, we assumed that
a code had $2^k$ elements, $k$ being the dimension. However, we
can define a code of length $n$ as any subset of $GF(2)^n$.
To a large extent, this course is about picking out the right subset
of $GF(2)^n$ to form codes with a rich structure.

There are many interesting combinatorial
questions regarding non-linear codes.
Probably, the most important question is the following: given the
length $n$ and the minimum distance $d$, what is the maximum number
of codewords that a code can have? For more about non-linear codes,
the reader is referred to~\cite{mac}.
From now on, when we say code, we assume that the code is linear
(unless otherwise stated).
Linear codes are in general easier to encode and decode than
their non-linear counterparts, hence they are more suitable for
implementation in applications.

In order to find
the minimum distance of a linear code, it is enough to find its minimum
{\em weight}. We say that the
(Hamming) weight of a vector $\uu$ is the distance
between $\uu$ and the zero vector. In other words, the weight of
$\uu$, denoted $w_H(\uu)$, is the number of non-zero coordinates of the
vector $\uu$. The minimum weight of a code is the minimum between
all the weights of the non-zero codewords. The proof of the
following lemma is left as a problem.

\begin{lemma}
\label{lemma2}
{\em
Let $\C$ be a linear
$[n,k,d]$ code. Then, the minimum distance and the minimum weight of
$\C$ are the same.
}
\end{lemma}

Next, we introduce two important matrices that define a linear
error-correcting code. Since a code $\C$
is now a subspace, the dimension $k$ of $\C$ is the cardinality of
a basis of $\C$. We denote by
$[n,k,d]$, as in the previous section, a code of length $n$, dimension
$k$ and minimum distance $d$.
We say that a $k\times n$ matrix $G$ is a {\em generator}
matrix of a code
$\C$ if the rows of $G$ are a basis of $\C$.
Given a generator matrix, the encoding process is simple.
Explicitly, let $\uu$ be an information vector of length $k$ and
$G$ a $k\times n$ generator matrix, then $\uu$ is encoded
into the $n$-vector $\uv$ given by

\begin{equation}
\label{eq1}
\uv=\uu\,G.
\end{equation}

\begin{ex}
\label{ex2}
{\em
Let $G$ be the $2\times 5$ matrix

$$G=\left(
\begin{array}{ccccc}
0&0&1&1&1\\
1&1&1&0&0
\end{array}\right)$$

It is easy to see that $G$ is a generator matrix of the $[5,2,3]$
code described in Example~\ref{ex1}.
\phantom{x}\qed
}
\end{ex}

Notice that, although a code may have many generator matrices, the
encoding depends on the particular matrix chosen, according to
Equation~(\ref{eq1}). We say that $G$ is a {\em systematic} generator
matrix if $G$ can be written as

\begin{equation}
\label{eq2}
G=(I_k|V),
\end{equation}

where $I_k$ is the $k\times k$ identity matrix and $V$ is a
$k\times (n-k)$ matrix. A systematic generator matrix has the
following advantage: given an information vector $\uu$ of length $k$,
the encoding given by~(\ref{eq1}) outputs a codeword
$(\uu,\uw)$, where $\uw$ has length $n-k$. In other words, a systematic
encoder adds $n-k$ redundant bits to the $k$ information bits, so
information and redundancy are clearly separated. This also simplifies
the decoding process, since, after decoding, the redundant bits are
simply discarded. For that reason, most encoders used in applications
are systematic.

A permutation of the columns of a generator matrix gives a new generator
matrix defining a new code. The codewords of the new code
are permutations of the coordinates of the codewords
of the original code. We then
say that the two codes are {\em equivalent}.
Notice that equivalent codes have the same distance properties, so
their error correcting capabilities are exactly the same.

By permuting the columns of the generator matrix in Example~\ref{ex2},
we obtain the following generator matrix $G'$:

\begin{equation}
\label{eq3}
G'=\left(
\begin{array}{cccccc}
1&0&\phantom{0}&0&1&1\\
0&1&&1&1&0
\end{array}\right)
\end{equation}

The matrix $G'$ defines a systematic encoder for a code that is
equivalent to the one given in Example~\ref{ex1}. For instance,
the information vector $11$ is encoded into $11\;101$.

In fact, by row operations and column permutations, any generator
matrix can be transformed into a systematic generator matrix, so
it is always possible to find a systematic encoder
for a linear code. However, when we permute columns, we obtain an
equivalent code to the original one, not the original code itself.
If we want to obtain exactly the same code, only row operations are
allowed in order to obtain a systematic generator matrix.

The second important matrix related to a code is the so called
{\em parity check} matrix.
We say that an $(n-k)\times n$ matrix $H$ is a parity check matrix
of an $[n,k]$ code $\C$ if and only if, for any $\uc\in\C$,

\begin{equation}
\label{eq4}
\uc\,H^T=\0,
\end{equation}

where $H^T$ denotes the transpose of matrix $H$ and $\0$ is
a zero vector
of length $n-k$. We say that the parity check matrix
$H$ is in systematic form if

\begin{equation}
\label{eq5}
H=(W|I_{n-k}),
\end{equation}

where $I_{n-k}$ is the $(n-k)\times (n-k)$ identity matrix and
$W$ is an $(n-k)\times k$ matrix.

Given a systematic generator matrix $G$ of a code $\C$, it is easy
to find the systematic parity check matrix $H$ (and conversely).
Explicitly, if $G$ is given by~(\ref{eq2}), $H$ is given by

\begin{equation}
\label{eq6}
H=(V^T|I_{n-k})
\end{equation}

We leave the proof of this fact to the reader.

For example, the systematic parity check matrix of the code whose
systematic generator matrix is given by~(\ref{eq3}), is

\begin{equation}
\label{eq7}
H=\left(
\begin{array}{cccccc}
0&1&\phantom{0} &1&0&0\\
1&1&&0&1&0\\
1&0&&0&0&1
\end{array}\right)
\end{equation}

We state now an important property of parity check matrices.

\begin{lemma}
\label{lemma3}
{\em
Let $\C$ be a linear
$[n,k,d]$ code and $H$ a parity-check matrix.
Then, any $d-1$ columns of $H$ are linearly independent.
}
\end{lemma}

{\bf Proof:} Numerate the columns of $H$ from 0 to $n-1$.
Assume that columns $0\leq i_1<i_2<\ldots <i_m\leq n-1$ are linearly
dependent, where $m\leq d-1$. Without loss of generality,
we may assume that the
sum of these columns is equal to the column vector zero.
Let $\uv$ be a vector of length $n$ whose non-zero coordinates
are in locations $i_1,i_2,\ldots,i_m$. Then, we have

$$\uv\,H^T\;=\;\0,$$

hence $\uv$ is in $\C$. But $\uv$ has weight $m\leq d-1$, contradicting
the fact that $\C$ has minimum distance $d$.\hfill $\Box$

\vspace{.8cm}

\begin{cor}
\label{cor2}
{\em
For any linear $[n,k,d]$ code, the minimum distance $d$ is the
smallest number $m$ such that there is a subset of
$m$ linearly dependent columns.
}
\end{cor}

{\bf Proof:} It follows immediately from Lemma~\ref{lemma3}.
\hfill $\Box$

\vspace{.8cm}

\begin{cor}[Singleton Bound]
\label{cor1}
{\em
For any linear $[n,k,d]$ code,

$$d\leq n-k+1.$$
}
\end{cor}

{\bf Proof:} Notice that, since
$H$ is an $(n-k)\times n$ matrix,
any $n-k+1$ columns
are going to be linearly dependent, so if $d>n-k+1$ we would contradict
Corollary~\ref{cor2}.\hfill $\Box$

\vspace{.8cm}

Codes meeting the Singleton bound are called Maximum Distance
Separable (MDS).
In fact, except for
trivial cases, binary codes are not MDS (Problem~\ref{pr1.7}).
In order to obtain MDS codes, we will define codes over larger fields,
like the so called Reed Solomon codes, to be described
later in the course.

We also give a second bound relating the redundancy and
the minimum distance of an $[n,k,d]$ code: the
so called Hamming or volume bound. Let us denote by
$V(r)$ the number of elements in a sphere of radius $r$ whose
center is an element in $GF(2)^n$.
It is easy to verify that

\begin{equation}
\label{eq8}
V(r)\,=\,\sum_{i=0}^r\,{n\choose i}.
\end{equation}

We then have:

\begin{lemma}[Hamming bound]
\label{lemma4}
{\em
Let $\C$ be a linear
$[n,k,d]$ code, then

\begin{equation}
\label{eq9}
n-k\,\geq\,\log_2 V\left(\lf (d-1)/ 2\rf\right).
\end{equation}
}
\end{lemma}

{\bf Proof:} Notice that the $2^k$ spheres
with the $2^k$ codewords as centers and radius $\lf (d-1)/2\rf$
are disjoint. The total number of vectors contained in these spheres
is $2^k\,V\left(\lf (d-1)/ 2\rf\right)$. This number has to
be smaller than
or equal to the total number of vectors in the space,
i.e.,

\begin{equation}
\label{eq9'}
2^n\geq 2^k\,V\left(\lf (d-1)/ 2\rf\right).
\end{equation}

Inequality~(\ref{eq9}) follows immediately from~(\ref{eq9'}).
\hfill $\Box$

\vspace{.8cm}

A {\em perfect} code is a code
for which Inequality~(\ref{eq9}) is in effect equality.
Geometrically, a perfect code is a code for which
the $2^k$ spheres of radius $\lf (d-1)/2\rf$ and the codewords
as centers cover the whole space.

There are not many perfect codes. In the binary case, the only
non-trivial linear
perfect codes are the Hamming codes and the $[23,12,7]$
Golay code, to be presented later in this chapter. However, the proof
of this fact is beyond the scope
of this course. We refer the interested reader to~\cite{l}.

\vspace{.8cm}

\pr

\begin{prob}
\label{pr1.4}
{\em
Prove Lemma~\ref{lemma2}.
}\end{prob}
 
\begin{prob}
\label{pr1.5}
{\em
Prove that if $G$ is a systematic generator matrix of a code
given by~(\ref{eq2}), then a systematic parity check matrix
of the code is given by~(\ref{eq6}).
}\end{prob}
 
\begin{prob}
\label{pr1.6}
{\em
Let $\C_1$ be the code formed by all the vectors of length
$n$ and even weight and $\C_2$ be the code whose only codewords
are the all-zero and the all-1 vectors (also of length $n$).
Find the minimum distance and systematic generator and parity
check matrices for both $\C_1$ and $\C_2$.
}\end{prob}
 
\begin{prob}
\label{pr1.7}
{\em
Find all binary linear MDS codes. Prove your statement.
}\end{prob}
 
\begin{prob}
\label{pr1.8}
{\em
Let $\C$ be an
$[n,k]$ code with parity check matrix $H$. Let $\C'$ be a code
obtained by adding a parity check bit to every codeword of $\C$.
$\C'$ is called an extended $\C$ code. In particular, notice that
if $\C$ is an
$[n,k,2t+1]$ code, then
$\C'$ is an $[n+1,k,2t+2]$ code.
 
Find a parity check matrix $H'$ for $\C'$ as a function of $H$.
}\end{prob}

\sol
\vspace{.8cm}

{\bf Problem~\ref{pr1.4}}
 
Let $w$ be the minimum weight of $\C$. In particular, $d\leq w$.
 
Assume that $\uu,\uv\in GF(2)^n$.
Claim: $d_H(\uu,\uv)=w_H(\uu\oplus\uv)$.
In effect, let $u_i$ and $v_i$ be the $i$-th coordinates
in $\uu$ and $\uv$ respectively. If $u_i=v_i$, then $u_i\oplus v_i=0$,
otherwise $u_i\oplus v_i=1$. So, the number of coordinates in which
$\uu\oplus \uv$ is 1 coincides with the number of coordinates in which
$\uu$ and $\uv$ differ, hence, the claim follows.
 
Now, assume that $\uu,\uv\in\C$ and $d_H(\uu,\uv)=d$.
Since $\C$ is linear, $\uu\oplus\uv\in\C$. By the claim above,
$w_H(\uu\oplus\uv)=d$, hence, $w\geq d$. This completes the proof.
 
\vspace{.8cm}
 
{\bf Problem~\ref{pr1.5}}
 
Since the rows of $G$ form a basis of the code, it is enough to prove
that the rows of $G$ and the rows of $H$ are orthogonal. In other words,
we have to prove that
 
$$GH^T=\0_{k\times (n-k)}$$
 
where we denote by $\0_{k\times (n-k)}$ a $k\times (n-k)$ 0-matrix.
Performing this matrix product, we obtain
 
$$GH^T=(I_k|V)(V^T|I_{n-k})^T=(I_k|V)\left(
\begin{array}{c}V\\ \hline I_{n-k}\end{array}\right)=V\oplus V=
\0_{k\times (n-k)},$$
 
completing the proof.
 
\vspace{.8cm}
 
{\bf Problem~\ref{pr1.6}}
 
Clearly, $\C_1$ and $\C_2$ are linear codes, so it is enough to find
the minimum weight in both. Since all codewords have even weight,
the minimum weight of $\C_1$ is 2, while $\C_2$ has only one non-zero
codeword, hence its minimum weight is $n$.
 
Since exactly half of the vectors in $GF(2)^n$ have even weight,
$\C_1$ has dimension $n-1$, i.e., $\C_1$ is an $[n,n-1,2]$ code.
A systematic generator matrix for $\C_1$ is given by
 
$$G_1=\left(
\begin{array}{cccccc}
1&0&0&\ldots&0&1\\
0&1&0&\ldots&0&1\\
0&0&1&\ldots&0&1\\
\vdots&\vdots&\vdots&\ddots&\vdots&\vdots\\
0&0&0&\ldots&1&1\end{array}\right)=
(I_{n-1}|(\1_{n-1})^T),$$
 
where $\1_{n-1}$ denotes an all-1 vector of length $n-1$.
A systematic parity check matrix is given by $H_1=\1_n$,
$\1_n$ the all-1 vector of length $n$.
 
For $\C_2$, the roles are reversed. We verify immediately
that $\C_2$ is an $[n,1,n]$ code.
A systematic generator matrix for $\C_2$ is given by
$G_2=H_1$ and a systematic parity check matrix by $H_2=G_1$.
 
$\C_1$ and $\C_2$ are duals of each other, i.e., $C_1=C_2^{\perp}$.
$\C_1$ is called the parity-check code of length $n$ and $\C_2$ the
repetition code of length $n$.
 
\vspace{.8cm}
 
{\bf Problem~\ref{pr1.7}}
 
Let us find all the binary MDS codes of length $n$. From the previous
problem, we see that both the $[n,n-1,2]$ even weight code and the
$[n,1,n]$ repetition code are MDS. Also, the whole space $GF(2)^n$
is an $[n,n,1]$ code, hence it is MDS.
 
We claim that those are the only binary MDS codes. In effect, assume
that $\C$ is an $[n,k,n-k+1]$ binary code, $k<n$.
Let $G$ be a systematic generator matrix, i.e.,
$G=(I_k|V)$, $V$ a $k\times (n-k)$ matrix.
Since $d=n-k+1$, in particular, each row in $G$ has weight $\geq n-k+1$,
hence, $V$ is an all-1 matrix. If $n-k=1$, we obtain the generator matrix
corresponding to the even weight code, so assume that $n-k>1$.
In particular, $d=n-k+1>2$.
 
If $k=1$,
we obtain the generator matrix corresponding to the repetition code, so
assume also that $k>1$. Let
$\ug_1$ and $\ug_2$ be the first and second rows
in $G$ respectively, then,
 
$$\ug_1\oplus\ug_2=(11,\0_{n-2})\in\C.$$
 
But this codeword has weight 2, contradicting the fact that
the minimum distance is greater than 2.
 
\vspace{.8cm}
 
{\bf Problem~\ref{pr1.8}}
 
Let $H$ be an $(n-k)\times k$ parity-check matrix for $\C$,
then a parity check matrix for $\C'$ is given by
the $(n+1-k)\times (n+1)$ matrix
 
$$H'=\left(
\begin{array}{c}
\begin{array}{c|c}
H&
\begin{array}{c}
0\\
0\\
\vdots\\
0
\end{array}
\end{array}\\
\hline
1\,1\,\ldots\,1
\end{array}
\right)=
\left(\begin{array}{c}
H|(\0_{n-k})^T\\
\hline \1_{n+1}\end{array}
\right).$$
 
In effect, if $\uu\in\C'$, notice that, in particular, the first
$n$ bits of $\uu$ are in $\C$, so its inner product with any of the
first $n-k$ rows of $H'$ will be zero. Finally, since the exclusive-OR
of all the bits in $\uu$ is zero, this is equivalent to say that
its inner product with the all-1 vector is zero.

\section{Syndromes, Cosets and Standard Array Decoding}
\label{sec3}

Let $\C$ be an $[n,k,d]$ code with parity check matrix $H$.
Let $\uu$ be a transmitted vector and $\ur$ a possibly corrupted
received version of $\uu$. We say that the syndrome of $\ur$
is the vector $\us$ of length $n-k$ given by

\begin{equation}
\label{eq10}
\us\,=\,\ur H^T.
\end{equation}

Notice that, if no errors occurred, the syndrome of $\ur$ is the
zero vector. The syndrome, however,
tells us more than a vector being in the
code or not. Say, as before, that $\uu$ was transmitted and $\ur$
was received, where $\ur=\uu\xor\ue$, $\ue$ an error vector.
Notice that,

$$\us=\ur H^T=(\uu\xor\ue)H^T=\uu H^T\xor \ue H^T=\ue H^T,$$

since $\uu$ is in $\C$. Hence, the syndrome does not depend on
the received vector but on the error vector.
In the next lemma, we show that to every error vector
of weight $\leq (d-1)/2$ corresponds a unique syndrome.

\begin{lemma}
\label{lemma5}
{\em
Let $\C$ be a linear
$[n,k,d]$ code with parity check matrix $H$. Then, there is a 1-1
correspondence between errors of weight $\leq (d-1)/2$
and syndromes.
}
\end{lemma}

\pf Let $\ue_1$ and $\ue_2$ be two distinct
error vectors of
weight $\leq (d-1)/2$ with syndromes $\us_1=\ue_1 H^T$ and
$\us_2=\ue_2 H^T$. If $\us_1=\us_2$, then
$\us=(\ue_1\xor\ue_2)H^T=\us_1\xor\us_2=\0$, hence
$\ue_1\xor\ue_2\in\C$. But
$\ue_1\xor\ue_2$ has weight $\leq d-1$, a contradiction.
\qed

Lemma~\ref{lemma5} gives the key for a decoding method that is more
efficient than exhaustive search. We can construct a table
with the 1-1 correspondence between syndromes and error patterns
of weight $\leq (d-1)/2$ and decode by look-up table. In other words,
given a received vector, we first find its syndrome and then we
look in the table to which error pattern it corresponds. Once we
obtain the error pattern, we add it to the received vector, retrieving
the original information. This procedure may be efficient for small
codes, but it is still too complex for large codes.

\begin{ex}
\label{ex3}
{\em
Consider the code whose parity matrix $H$ is given by~(\ref{eq7}).
We have seen that this is a $[5,2,3]$ code. We have 6 error patterns
of weight $\leq 1$.
The 1-1 correspondence between these error patterns and the syndromes,
can be immediately verified to be

\begin{eqnarray*}
00000&\lra &000\\
10000&\lra &011\\
01000&\lra &110\\
00100&\lra &100\\
00010&\lra &010\\
00001&\lra &001
\end{eqnarray*}

For instance, assume that we receive the vector $\ur=10111$.
We obtain the syndrome $\us=\ur H^T=100$. Looking at the table above,
we see that this syndrome corresponds to the error pattern $\ue=00100$.
Adding this error pattern to the received vector,
we conclude that the transmitted vector was $\ur\xor\ue=10011$.\qed
}
\end{ex}

We say that a coset of a code $\C$ is a set of elements
$\uv\oplus\C$, where $\uv$ is any vector.
Notice that if $\uv$ and $\uw$ are in the same coset, then
$\uv\oplus\uw$ is in the code. Also, if $\uv$ and $\uw$ are
in the same coset, then $\uv\oplus\C=\uw\oplus\C$.
Cosets are disjoint and the union of all of them gives a partition of
the space $GF(2)^n$. We prove these facts in the Problems.

\begin{lemma}
\label{lemma6}
{\em
Let $\C$ be a linear $[n,k,d]$ code, then, there is a 1-1
onto correspondence between cosets and syndromes.
}
\end{lemma}

\pf Observe that all elements in the same coset have the same
syndrome. Assume that the elements $\uu$ and $\uv$
have the same syndrome $\us$; then $\uu\oplus\uv$
is in $\C$, hence, $\uu$ and $\uv$ are in the same coset, showing that
to every coset correponds a unique syndrome.

Conversely, let $H$ be a systematic
parity check matrix of $\C$ as in~(\ref{eq5}). Given a syndrome
$\us$, the vector $(\0,\us)$ has syndrome $\us$, where
$\0$ is a zero vector of length $k$. Hence, to $\us$ corresponds
the coset defined by $(\0,\us)$, which is unique.

Let us give another proof using linear algebra. Let
$f:GF(2)^n\ra GF(2)^{n-k}$, $f(\uv)=\uv H^T$. By the definition of $H$,
$\ker (f)=\C$. Hence, $n=\dim (\ker(f))+\dim(f(GF(2)^n))=
k+\dim(f(GF(2)^n))$, i.e.,
$\dim(f(GF(2)^n))=n-k$ and $f$ is onto.\qed

In each coset, an element of minimum weight is called a
{\em coset leader}. If there is an element of weight $\leq (d-1)/2$,
then, by Lemma~\ref{lemma5}, this element is the coset leader and is
unique.

\begin{defin}
\label{def2}
{\em
A standard array of an $[n,k,d]$ code $\C$
is a $2^{n-k}\times 2^k$ matrix such that:

\begin{enumerate}

\item Its entries are the $2^n$ vectors in the space.

\item The entries in each row are the elements of the different
cosets of $\C$.

\item The first element in each row corresponds to a coset leader
in the coset.

\item The first row corresponds to $\C$.

\end{enumerate}
}
\end{defin}

The next example illustrates a decoding method using the standard
array of a code.

\begin{ex}
\label{ex4}
{\em
Consider the code $\C$ with parity check matrix
$H$ given by~(\ref{eq7}). Below we give the standard array of $\C$.

$$
\begin{array}{|l||c|c|c|c||c|}
\hline
{\rm message}&00&01&10&11&{\rm syndrome}\\
\hline
\hline
{\rm code} & 00000 & 01110 & 10011 & 11101 & 000\\
\hline
{\rm coset}&10000 & 11110 & 00011 & 01101 & 011    \\
\hline
{\rm coset}&01000 & 00110 & 11011 & 10101 & 110    \\
\hline
{\rm coset}&00100 & 01010 & 10111 & 11001 & 100      \\
\hline
{\rm coset}&00010 & 01100 & 10001 & 11111 & 010        \\
\hline
{\rm coset}&00001 & 01111 & 10010 & 11100 & 001          \\
\hline
{\rm coset}&11000 & 10110 & 01011 & 00101 & 101            \\
\hline
{\rm coset}&10100 & 11010 & 00111 & 01001 & 111              \\
\hline
\end{array}
$$

The second row
contains the code itself, while the remaining rows contain the cosets.
The first column contains
the coset leaders. For convenience, we have included a
first row with the information string and a fifth column
with the syndromes. As in Example~\ref{ex3}, assume that we want
to decode the vector $\ur=10111$.
We obtain the syndrome $\us=\ur H^T=100$.
We then proceed to locate vector
$\ur$ in the row corresponding to this syndrome in the standard array.
We can see that $\ur$ is in the third entry of the row.
The decoded vector is then the one corresponding to the third entry
in the code row, i.e., codeword 10011, since this codeword is obtained
by adding the
received vector to the coset leader 00100, which is
{\em the} error pattern.
In general, since we are only interested in the information bits,
the final output of the decoder is 10.\qed
}
\end{ex}

Decoding by standard array has more conceptual than practical
application. In this course we will study some codes with more
efficient decoding algorithms.

Observe that standard array decoding can be used to decode
beyond the minimum distance of the code. In general, given a
$\uv\in GF(2)^n$ and $\C$ a code of length $n$,
we say that {\em maximum likelihood decoding} of $\uv$
with respect to
$\C$ is finding the closest codeword in $\C$ (in Hamming
distance) to $\uv$. This closest codeword might be
at a distance that exceeds the minimum distance of the code.
Also, the closest codeword might not necessarily be unique.
For instance, consider the standard array in
Example~\ref{ex4}. If the syndrome is 101, the decoder decides that
the error is the coset leader 11000. But it could as well have
decided that the error was 00101: both possibilities are equally
likely.

In general, maximum likelihood decoding is a difficult problem.
Most decoding methods decode up to the minimum distance of the code.

\pr

\begin{prob}
\label{pr1.9}
{\em
Let $H$ be a systematic parity check matrix of a code $\C$ as
given by~(\ref{eq5}). Assume that $\C$ can correct up to $t$ errors.
Let $\ur$ be a received vector whose syndrome $\us=\ur H^T$ has weight
$\leq t$. Prove that the only error pattern of weight $\leq t$
is $\ue=(\0_k|\us)$, where
$\0_k$ is an all-0 vector of length $k$.
}\end{prob}
 
\begin{prob}
\label{pr1.10}
{\em
Let $\C$ be a code of length $n$,
$\uv$ any vector in $GF(2)^n$ and
$\uv\oplus\C$ the coset of $\C$ corresponding to $\uv$.
Prove that:
 
\begin{enumerate}
 
\item
If $\uw\in\uv\oplus\C$, then,
$\uv\oplus\uw\in\C$ and $\uv\oplus\C=\uw\oplus\C$.
 
\item
If $\uw\not\in\uv\oplus\C$, then,
$\uv\oplus\C\cap\uw\oplus\C=\emptyset$.
 
\end{enumerate}
 
}\end{prob}
 
\begin{prob}
\label{pr1.11}
{\em
Consider the code whose parity check matrix is given
by~(\ref{eq7}). Do maximum likelihood
decoding of the vector 00111 with respect to this code. Is the
answer unique? If not, find all possible answers.
}\end{prob}
 
\sol
\vspace{.8cm}

{\bf Problem~\ref{pr1.9}}
 
We can easily verify that $\ue H^T=(\0_k|\us)H^T=\us$. Since the
code can correct up to $t$ errors, by Lemma~\ref{lemma5}, the
syndrome is unique, so, if $t$ or less errors have occurred,
the error pattern is given by $\ue$.
 
This problem is important because of the following:
if we assume that the first $k$ information bits carry information,
an error pattern given by $\ue$ means that the errors occurred in
the redundant part. So, the decoder may choose to ignore the
redundant bits and output the first $k$ bits whenever the syndrome
has weight $\leq t$. We use this fact in Section~\ref{sec6} when
decoding the Golay code.
 
\vspace{.8cm}
 
{\bf Problem~\ref{pr1.10}}
 
(a) If $\uw\in\uv\oplus\C$, there is a $\uc\in\C$ such that
$\uw=\uv\oplus\uc$. Hence, $\uw\oplus\uv=\uc\in\C$.
 
Now, let $\uw\oplus\uc'\in\uw\oplus\C$.
Hence, $\uw\oplus\uc'=\uv\oplus (\uc\oplus\uc')\in\uv\oplus\C$, since
$\uc\oplus\uc'\in\C$.
So, $\uw\oplus\C\subseteq\uv\oplus\C$. Similarly, we prove
$\uv\oplus\C\subseteq\uw\oplus\C$, completing the proof.
 
(b) Assume $\uu\in\uv\oplus\C\cap\uw\oplus\C$. Hence,
$\uu=\uv\oplus\uc=\uw\oplus\uc'$, where $\uc,\uc'\in\C$.
In particular, $\uw=\uv\oplus (\uc\oplus\uc')\in\uv\oplus\C$,
since $\uc\oplus\uc'\in\C$.
This is a contradiction.
 
\vspace{.8cm}
 
{\bf Problem~\ref{pr1.11}}
 
Computing the syndrome of 00111, this syndrome is 111.
Looking at the standard array in Example~\ref{ex4}, we see that
00111 belongs in the last row. If we consider the error to be
the coset leader 10100, 00111 is decoded as 10011. However, there
is another error pattern of weight 2 in the coset, 01001. If we choose
this pattern as the error vector, 00111 is decoded as 01110. Those are
the two possible solutions of maximum likelihood decoding, i.e., there
are no vectors in $\C$ at distance 1 or less from
00111, and there are exactly two vectors at distance 2,
10011 and 01110.
 
\section{Hamming Codes}
\label{sec4}

In this section, we study the first important family of codes,
the so called Hamming codes. As we will see, Hamming codes
can correct up to one error.

Given a number $r$ of redundant bits, we say that a
$[2^r-1,2^r-r-1,3]$ Hamming code is a code having an $r\times (2^r-1)$
parity check matrix $H$
such that its columns are all the different
non-zero vectors of length $r$.

A Hamming code has
minimum distance 3. This follows
from its definition and Corollary~\ref{cor2}:
notice that any 2 columns in $H$, being different, are linearly
independent. Also, if we take any two different columns and their
sum, these 3 columns are linearly dependent, proving our assertion.

A natural way of writing the columns of $H$ in a Hamming code,
is by considering them as binary numbers on base 2 in increasing
order. This means, the first column is 1 on base 2, the second columns
is 2, and so on. The last column is $2^r-1$ on base 2, i.e.,
$(1,1,\ldots,1)^T$. This parity check matrix, although non-systematic,
makes the decoding very simple.

In effect, let $\ur$ be a received vector such that $\ur=\uv\oplus\ue$,
where $\uv$ was the transmitted codeword and $\ue$ is an error vector
of weight 1. Then, the syndrome is $\us=\ue H^T$, which gives the
column corresponding to the location in error. This column, as a number
on base 2, tells us exactly where the error has occurred, so the
received vector can be corrected.

\begin{ex}
\label{ex5}
{\em
Consider the $[7,4,3]$ Hamming code
$\C$ with parity check matrix

\begin{equation}
\label{eq11}
H=\left(
\begin{array}{ccccccc}
0&0&0&1&1&1&1\\
0&1&1&0&0&1&1\\
1&0&1&0&1&0&1
\end{array}\right)
\end{equation}

Assume that vector $\ur=1100101$ is received. The syndrome is
$\us=\ur H^T=001$, which is the binary representation of the number~1.
Hence, the first location is in error, so the decoder estimates that
the transmitted vector was $\uv=0100101$.\qed
}
\end{ex}

We can obtain 1-error correcting codes of any length simply by
shortening a Hamming code. This procedure works as follows:
assume that we want to encode $k$ information bits into a
1-error correcting code. Let $r$ be the smallest
number such that $k\leq 2^r-r-1$.
Let $H$ be the parity-check matrix of a $[2^r-1,2^r-r-1,3]$ Hamming
code. Then construct a matrix $H'$ by eliminating some
$2^r-r-1-k$ columns from $H$. The code whose parity-check matrix is
$H'$ is a $[k+r,k,d]$ code with $d\geq 3$, hence it can correct one
error.
We call it a shortened Hamming code. For instance, the $[5,2,3]$ code
whose parity-check matrix is given by~(\ref{eq7}), is a shortened
Hamming code.

In general, if $H$ is the parity-check matrix of a code $\C$,
$H'$ is a matrix obtained by eliminating a certain number of columns
from $H$ and $\C'$ is the code with parity-check matrix $H'$,
we say that $\C'$ is obtained by shortening $\C$.

A $[2^r-1,2^r-r-1,3]$ Hamming code can be extended to a
$[2^r,2^r-r-1,4]$ Hamming code by adding to each codeword a parity
bit that is the exclusive-OR of the first $2^r-1$ bits. The new
code is called an extended Hamming code.

\pr

\begin{prob}
\label{pr1.12}
{\em
Prove that $[2^r-1,2^r-r-1,3]$ Hamming codes are perfect.
}\end{prob}
 
\begin{prob}
\label{pr1.13}
{\em
Let
 
$$
H=\left(
\begin{array}{ccccccc}
0&1&1&1&0&0\\
1&0&1&0&1&0\\
1&1&0&0&0&1
\end{array}\right)
$$
 
be a systematic parity check matrix
for a (shortened) $[6,3,3]$ Hamming code.
Construct the standard array for the code.
Add a row for the information symbols and
a column for the syndromes.
}\end{prob}
 
\begin{prob}
\label{pr1.14}
{\em
Find systematic generator and parity-check matrices for the
extended $[8,4,4]$ Hamming code.
}\end{prob}
 
\begin{prob}
\label{pr1.15}
{\em
Given two vectors $\uu=u_0,u_1,\ldots,u_{n-1}$ and
$\uv=v_0,v_1,\ldots,v_{n-1}$, we say that the inner product
between $\uu$ and $\uv$, denoted $\uu\cdot\uv$, is the bit
 
$$\uu\cdot\uv=\bigoplus_{i=0}^{n-1}\,u_iv_i.$$
 
Given a code $\C$, we say that the dual of $\C$, denoted
$\C^{\perp}$, is the set of all vectors $\uv$ such that
$\uv\cdot\uu=0$ for all $\uu\in\C$.
If $\uv\cdot\uu=0$, we say that $\uu$ and $\uv$ are {\em orthogonal}.
 
Let $\C$ be an $[n,k]$ code with generator matrix $G$ and parity check
matrix $H$. Prove:
 
\begin{enumerate}
 
\item $G$ is a parity check matrix and $H$ is a generator matrix
of $\C^{\perp}$.
 
\item $\dim(\C^{\perp})=n-\dim(\C)$.
 
\item $\C=(\C^{\perp})^{\perp}$.
 
\end{enumerate}
}\end{prob}
 
\begin{prob}
\label{pr1.16}
{\em
Let $\C$ be the $[7,4,3]$ Hamming code with $H$ in systematic form.
Find $\C^{\perp}$ together with its parity
check and generator matrices.
What is the minimum distance of $\C^{\perp}$?
}\end{prob}
 
\begin{prob}
\label{pr1.17}
{\em
We say that an $[n,k]$ code $\C$ is {\em self-dual}
if $\C=\C^{\perp}$. Let $G$ be a generator matrix of $\C$. Prove that
$\C$ is self-dual if and only if any two (not necessarily distinct)
rows of $G$ are orthogonal and $k=n/2$.
Is the $[8,4,4]$ extended Hamming code self-dual
(see Problem~\ref{pr1.14})?
}\end{prob}
 
\sol
\vspace{.8cm}
 
{\bf Problem~\ref{pr1.12}}
 
Notice that, according to~(\ref{eq4}),
$V(\lf (d-1)/2\rf)=V(1)=1+(2^r-1)=2^r$, so,
$r=\log_2 V(1)$, proving that the Hamming bound~(\ref{eq9}) is met
with equality.
 
\vspace{.8cm}
 
{\bf Problem~\ref{pr1.13}}
 
Using the matrix $H$, the standard array of the code is
 
$$
\begin{array}{|l||c|c|c|c|c|c|c|c||c|}
\hline
{\rm message}&000&001&010&100&011&101&110&111&{\rm synd}\\
\hline
\hline
{\rm code} &000000&001110&010101&100011&011011&101101&110110&111000&000\\
\hline
{\rm coset}&000001&001111&010100&100010&011010&101100&110111&111001&001\\
\hline
{\rm coset}&000010&001100&010111&100001&011001&101111&110100&111010&010\\
\hline
{\rm coset}&000100&001010&010001&100111&011111&101001&110010&111100&100\\
\hline
{\rm coset}&001000&000110&011101&101011&010011&100101&111110&110000&110\\
\hline
{\rm coset}&010000&011110&000101&110011&001011&111101&100110&101000&101\\
\hline
{\rm coset}&100000&101110&110101&000011&111011&001101&010110&011000&011\\
\hline
{\rm coset}&100100&101010&110001&000111&111111&001001&010010&011100&111\\
\hline
\end{array}
$$
 
The first row carries the uncoded messages, the second row the code
itself and the other rows the cosets. We write the coset leaders in
the first column and the syndromes in the last one.
 
\vspace{.8cm}
 
{\bf Problem~\ref{pr1.14}}
 
A (systematic) parity check matrix for the $[7,4,3]$ Hamming code
is given by
 
\begin{equation}
\label{eqpr1}
H=\left(
\begin{array}{ccccccc}
0&1&1&1&1&0&0\\
1&0&1&1&0&1&0\\
1&1&0&1&0&0&1
\end{array}\right)
\end{equation}
 
By Problem~\ref{pr1.8}, a parity check matrix for the extended
$[8,4,4]$ Hamming code is given by
 
$$H'=\left(
\begin{array}{cccccccc}
0&1&1&1&1&0&0&0\\
1&0&1&1&0&1&0&0\\
1&1&0&1&0&0&1&0\\
1&1&1&1&1&1&1&1
\end{array}\right)
$$
 
Replacing the last row by the exclusive-OR of the 4 rows, we obtain
the following systematic parity check matrix for the $[8,4,4]$ extended
Hamming code:
 
$$H''=\left(
\begin{array}{cccccccc}
0&1&1&1&1&0&0&0\\
1&0&1&1&0&1&0&0\\
1&1&0&1&0&0&1&0\\
1&1&1&0&0&0&0&1
\end{array}\right)
$$
 
By~(\ref{eq2}) and~(\ref{eq6}), a systematic generator matrix for the
code is given by
 
$$G''=\left(
\begin{array}{cccccccc}
1&0&0&0&0&1&1&1\\
0&1&0&0&1&0&1&1\\
0&0&1&0&1&1&0&1\\
0&0&0&1&1&1&1&0
\end{array}\right)
$$
 
\vspace{.8cm}
 
{\bf Problem~\ref{pr1.15}}
 
1. Every row in $H$ is orthogonal to every element in $\C$, by the
definition of parity check matrix, hence, every row is in $\C^{\perp}$.
Also, the $n-k$ rows are linearly independent. If there would be another
codeword in $\C^{\perp}$ that is independent from the $n-k$
rows of $H$, the space $\C$ would satisfy $n-k+1$ independent
linear homogeneous equations. This contradicts the fact that $\C$ has
dimension $k$ (this argument can be seen also by taking the generator
and parity check matrices to systematic form).
 
So, each element in $\C^{\perp}$ is generated by
the rows of $H$, i.e., $H$ is a generator matrix for $\C^{\perp}$.
 
An analogous argument may be used to show that $G$ is a parity check
matrix for $\C^{\perp}$.
 
2. Since $H$ is a generator matrix for
$\C^{\perp}$, then $\dim(\C^{\perp})=n-k$.
 
3. Let $\uu\in\C$. Let $\uv$ be any element in $\C^{\perp}$.
Hence, $\uv\cdot\uu=0$, i.e., $\uu\in (\C^{\perp})^{\perp}$.
Thus,
$\C\subseteq (\C^{\perp})^{\perp}$.
 
On the other hand,
$\dim ((\C^{\perp})^{\perp})=n-\dim (\C^{\perp})=n-(n-k)=k=
\dim (\C)$.
 
An even easier argument, using part 1 of the problem: notice that
$\C$ and
$(\C^{\perp})^{\perp}$ have the same generator matrix $G$, so they
must be equal.
 
\vspace{.8cm}
 
{\bf Problem~\ref{pr1.16}}
 
A systematic parity check matrix for the $[7,4,3]$ Hamming code is given
by~(\ref{eqpr1}).
By~(\ref{eq2}) and~(\ref{eq6}), a systematic generator matrix for the
code is given by
 
$$G=\left(
\begin{array}{ccccccc}
1&0&0&0&0&1&1\\
0&1&0&0&1&0&1\\
0&0&1&0&1&1&0\\
0&0&0&1&1&1&1
\end{array}\right).
$$
 
By Problem~\ref{pr1.15}, $H$ is a generator matrix for $\C^{\perp}$
and $G$ is a parity check matrix. Using $H$, we can find the 8
codewords in $\C^{\perp}$:
 
$$\C^{\perp}=\{0000000,0111100,1011010,1101001,1100110,1010101,0110011,
0001111\}$$
 
We can see that the minimum distance of
$\C^{\perp}$ is 4, hence,
$\C^{\perp}$ is a $[7,3,4]$ code. Moreover, every non-zero codeword
has constant weight 4.
 
In general, it can be proven that the dual of a
$[2^r-1,2^r-r-1,3]$ Hamming
code is a $[2^r-1,r,2^{r-1}]$ code called
a {\em simplex} code. Each non-zero codeword in
a $[2^r-1,r,2^{r-1}]$ simplex code has constant weight $2^{r-1}$.
 
\vspace{.8cm}
 
{\bf Problem~\ref{pr1.17}}
 
Assume that any two rows in $G$ are orthogonal and
$\dim(\C)=n/2$. Then, any two codewords in $\C$ are orthogonal, since
they are linear combinations of the rows of $G$. Hence,
$\C\subseteq \C^{\perp}$. On the other hand,
$\dim(\C)=\dim (\C^{\perp})=n/2$, so,
$\C=\C^{\perp}$.
 
Conversely, assume that $\C=\C^{\perp}$.
By Problem~\ref{pr1.15},
$\dim(\C^{\perp})=n-\dim(\C)=\dim(\C)$, so,
$\dim(\C)=n/2$. In particular, since any two rows in $G$ are in
$\C^{\perp}$ they are orthogonal.
 
The $[8,4,4]$ extended Hamming code is self-dual. In effect,
if we consider the generator matrix $G''$ of the code given in
Problem~\ref{pr1.14}, we see that any two rows are orthogonal.
Since the dimension of the code is 4=8/2, the result follows.

\section{Probabilities of Errors}
\label{sec5}

In the discussion of the previous sections, we have
omitted so far considering
an important parameter: the probability that a bit is in
error. In this section, we
assume that the
channel is a {\em binary symmetric channel} (BSC) with probability $p$:
it is equally likely
that a transmitted
0 is received as a
1 or a transmitted
1 is received as a 0
with probability $p$.
The probability that a transmitted bit remains unchanged is
$1-p$. The BSC is illustrated in Figure~1.1.

A first question is, what is the probability of decoding error?
Assume that the information string is encoded into
an $[n,k,2t+1]$ code $\C$, and that
every occurrence of at least
$t+1$ errors produces an incorrect
decoding. This is a conservative assumption: for many codes, when the
error correcting capability of the code is exceeded, the errors are
detected (although not corrected). Denote the probability of incorrect
decoding by
$P_{\rm err}$.
So, $P_{\rm err}$ is upper-bounded by the probability
that the number of errors exceeds $t$, i.e.,

\begin{equation}
\label{eq12}
P_{\rm err}\leq\sum_{i=t+1}^n{n\choose i}p^i(1-p)^{n-i}=
1-\sum_{i=0}^t{n\choose i}p^i(1-p)^{n-i}.
\end{equation}

If $p$ is a small number, the first term might
dominate the sum, so, usually the following is a good
approximation:

\begin{equation}
\label{eq13}
P_{\rm err}\leq {n\choose t+1}p^{t+1}(1-p)^{n-t-1}.
\end{equation}

For instance, if $\C$ is the $[5,2,3]$ code whose standard array is
given in Example~\ref{ex4} and $p=.01$, we have, using~(\ref{eq12}),

\begin{equation}
\label{eq13'}
P_{\rm err}\leq 0.00098.
\end{equation}

If we use the approximation given by~(\ref{eq13}), we obtain

$$
P_{\rm err}\leq 0.00097.
$$

\begin{figure}
\label{fig1}
$$
\begin{picture}(130,70)
\put(10,0){\circle*{3}}
\put(10,50){\circle*{3}}
\put(110,50){\circle*{3}}
\put(110,0){\circle*{3}}
\put(10,0){\line(2,1){100}}
\put(10,50){\line(2,-1){100}}
\put(10,0){\line(1,0){100}}
\put(10,50){\line(1,0){100}}
\put(50,-10){$1-p$}
\put(50,55){$1-p$}
\put(20,10){$p$}
\put(20,34){$p$}
\put(0,-3){1}
\put(0,47){0}
\put(115,-3){1}
\put(115,47){0}
\end{picture}
$$
\caption{BSC}
\end{figure}

As we can see, the two values are very close to each other.

After decoding, some symbols may be in error and some may not.
A more important parameter than
$P_{\rm err}$ is the average
probability of bit error after decoding,
that we denote
$p_{\rm err}$.

After decoding, the output of the decoder are the $k$ information
bits. Let $p_i$ denote the probability
that bit $i$ is in error after decoding, $0\leq i\leq k-1$, then we
have

\begin{equation}
\label{eq12'}
p_{\rm err}={1\over k}\sum_{i=0}^{k-1}p_i
\end{equation}

Finding an exact expression for
$p_{\rm err}$ is a difficult problem in general. An analysis of the
$[5,2,3]$ code
with
standard array given
in Example~\ref{ex4} will illustrate this point.
Once the error is corrected,
the decoder outputs the first 2 information bits,
so~(\ref{eq12'}) becomes

\begin{equation}
\label{eq12''}
p_{\rm err}={1\over 2}(p_0+p_1)
\end{equation}

Let us start by finding $p_0$.
Since all codewords are equally likely to be transmitted, without
loss of generality, assume that the 0-codeword was the transmitted one.
Therefore, the error pattern will be equal to the received vector.
Looking at the
standard array given
in Example~\ref{ex4} we see that the first bit will be in error
only when an error pattern in the third or in the fourth columns
of the array has occurred. In these two columns,
there are 5 patterns of weight 2,
7 patterns of weight 3,
3 patterns of weight 4 and
1 pattern of weight 5. Hence,

$$p_0=
5p^2(1-p)^3+
7p^3(1-p)^2+
3p^4(1-p)+
p^5
$$

Similarly, the second bit will be in error only when an error
pattern in columns 2 or 4 has occurred, hence, an analysis similar
to the one above shows that $p_1=p_0$, and by~(\ref{eq12''}),
$p_{\rm err}=p_0$. This gives

\begin{equation}
\label{eq14}
p_{\rm err}=
5p^2(1-p)^3+
7p^3(1-p)^2+
3p^4(1-p)+
p^5
\end{equation}

The example above illustrates the difficulty of finding an exact
expression for $p_{\rm err}$ in general.
As in the case of $P_{\rm err}$, when $p$ is small,
the first term usually gives a good approximation.

If we take $p=.01$, (\ref{eq14}) gives
$p_{\rm err}=.00049$. If we took only the first term, we would obtain
$p_{\rm err}=.00048$.
As we can see, this simple coding scheme
considerably lowers the average probability of bit error.

A fundamental question is the following: given a BSC with bit
error probability $p$, does it exist a code of high rate that
can arbitrarily lower $p_{\rm err}$?
The answer, due to Shannon, is yes, provided that the code has
rate below a parameter called the capacity of the channel.

\begin{defin}
\label{def3}
{\em
Given a BSC with probability of bit error $p$, we say that the
capacity of the channel is

\begin{equation}
\label{eq15}
C(p)=1+p\log_2p+(1-p)\log_2(1-p)
\end{equation}
}
\end{defin}

\begin{theo}[Shannon]
\label{theo1}
{\em
For any $\epsilon >0$ and $R<C(p)$, there is an $[n,k]$ binary code
of rate $k/n\geq R$ with $P_{\rm err}<\epsilon$.
}
\end{theo}

For a proof of Theorem~\ref{theo1} and its generalizations,
the reader is referred
to~\cite{ga}\cite{mc}, or even to Shannon's original paper~\cite{sha}.

Theorem~\ref{theo1} has enormous theoretical importance: it shows
that reliable communication is not limited
in the presence of noise, only the rate of communication is.
For instance, if $p=.01$ as in the examples above, the capacity of
the channel is $C(.01)=.9192$. Hence, there are codes of rate $\geq .9$
with $P_{\rm err}$ arbitrarily small.
It also tells us not to look for codes with rate .92 making
$P_{\rm err}$ arbitrarily small.

The proof of Theorem~\ref{theo1}, though,
is based on probabilistic methods and the assumption
of arbitrarily
large values of $n$. In practical applications, $n$ cannot
be too large. The theorem does not tell us how to construct efficient
codes, it just asserts their existence. Moreover, when we construct
codes, we want them to have efficient encoding and decoding
algorithms. One of the goals
of this course is exhibiting some of the
most widely used codes in applications together with their encoding
and decoding procedures.

\pr

\begin{prob}
\label{pr1.18}
{\em
Let $\C$ be a perfect code. Prove that Inequality~(\ref{eq12})
becomes equality for $\C$.
}\end{prob}
 
\begin{prob}
\label{pr1.19}
{\em
Find an exact expression for $P_{\rm err}$ when the
$[5,2,3]$ code with standard array given in Example~\ref{ex4}
is used.
Calculate the value of $P_{\rm err}$
for $p=.01$.
}\end{prob}
 
\begin{prob}
\label{pr1.20}
{\em
Prove that $p_{\rm err}<P_{\rm err}$ when standard array
decoding is used.
}\end{prob}
 
\begin{prob}
\label{pr1.21}
{\em
Assume that only the rows with coset leaders of weight
$\leq 1$ in the standard array of Example~\ref{ex4}
are used for decoding,
while the last 2 rows are used for error detection. In other words,
if the syndrome is either 101 or 111 the decoder declares an
uncorrectable error since it knows that more than one error has occurred.
We denote by
$P_{\rm det}$ the probability that the decoder detects errors but
does not correct them. With this decoding scheme, find
$P_{\rm err}$, $p_{\rm err}$ and $P_{\rm det}$.
Calculate the value of each of these expressions for $p=.01$.
}\end{prob}
 
\begin{prob}
\label{pr1.22}
{\em
Consider the standard array of
the $[6,3,3]$ shortened Hamming code of
Problem~\ref{pr1.13}.
Assume that the last row is used for error detection only.
Find exact expressions for
$P_{\rm err}$, $p_{\rm err}$ and $P_{\rm det}$.
Calculate the values  of these expressions for $p=.01$.
}\end{prob}
 
\sol
\vspace{.8cm}

{\bf Problem~\ref{pr1.18}}
 
Assume that $\C$ is a perfect $[n,k,2t+1]$ code, then, the spheres
of radius $t$ around each codeword cover the whole space.
If $t+1$ or more errors occur, then the received word will fall into
a sphere that is different to the one corresponding to the transmitted
codeword. Since the decoder outputs the center of the sphere where
the received word belongs, whenever $\geq t+1$ errors occur we have
incorrect decoding. Hence, Inequality~(\ref{eq12}) becomes equality
in this case.
 
\vspace{.8cm}
 
{\bf Problem~\ref{pr1.19}}
 
Without loss of generality, assume that 00000 has been transmitted.
Looking at the standard array in Example~\ref{ex4}, we see that
a decoding error will occur only when the received vector is not in
the first column (which corresponds to the coset leaders).
The second, third and fourth column contain every vector of
weight $\geq2$, except two. Hence, we obtain
 
$$P_{\rm err}=
\left(\sum_{i=2}^5 {5\choose i}p^i(1-p)^{5-i}\right)-2p^2(1-p)^3=
8p^2(1-p)^3+10p^3(1-p)^2+5p^4(1-p)+p^5.$$
 
For $p=.01$, the expression above gives
$P_{\rm err}=.00079$. The reader should compare this value with the
upper bound given in~(\ref{eq13'}).
 
\vspace{.8cm}
 
{\bf Problem~\ref{pr1.20}}
 
Let $p_i$ be the probability that bit $i$ is in error after decoding
when a standard array for an $[n,k,]$ code is used, $0\leq i\leq k-1$.
Let $p_j$ be the maximum of these values, so, since
$p_{\rm err}$ is the average of the $p_i$'s,
$p_{\rm err}\leq p_j$. If we take a first row in the standard array
for the information symbols as in
Example~\ref{ex4}, we see that bit $j$ will be incorrectly decoded
only when the error pattern belongs in one of the columns corresponding
to an information vector for which bit $j$ is 1. Notice that there
are exactly $2^{k-1}$ such columns, while we have incorrect
decoding occurs when the error pattern is in any of the $2^k-1$
columns excluding the first
one. In particular, we have incorrect decoding
when bit $j$ is
incorrectly decoded; hence
$p_j\leq P_{\rm err}$.
Notice that we have equality only when $k=1$.
 
 
\vspace{.8cm}
 
{\bf Problem~\ref{pr1.21}}
 
With the decoding system of this problem, incorrect decoding
occurs only when the error pattern belongs in the rows corresponding
to coset leaders of weight $\leq 1$ and in any column except the
first one. We see that there are 6 error patterns of weight 2,
6 error patterns of weight 3,
5 error patterns of weight 4 and
1 error pattern of weight 5. Hence, we have
 
$$P_{\rm err}=6p^2(1-p)^3+6p^3(1-p)^2+5p^4(1-p)+p^5.$$
 
Similarly, the first bit will be decoded in error if the error
pattern is either in the third or in the fourth column, but not
in the last 2 rows. We have
3 error patterns of weight 2,
5 error patterns of weight 3,
3 error patterns of weight 4 and
1 error pattern of weight 5, hence,
 
$$p_0=3p^2(1-p)^3+5p^3(1-p)^2+3p^4(1-p)+p^5.$$
 
The second bit will be decoded in error if the error
pattern is either in the second or in the fourth column, but not
in the last 2 rows. We can see that $p_0=p_1$, hence,
$p_{\rm err}=(p_0+p_1)/2=p_0$. Thus,
 
$$p_{\rm err}=3p^2(1-p)^3+5p^3(1-p)^2+3p^4(1-p)+p^5.$$
 
Finally, an error will be detected (but not corrected) only if
the error pattern belongs in one of the last two rows of the
standard array. We see that there are
4 error patterns of weight 2 and
4 error patterns of weight 3, so,
 
$$P_{\rm det}=4p^2(1-p)^3+4p^3(1-p)^2.$$
 
The expresions above for $p=.01$ give
$P_{\rm err}=.00059$,
$p_{\rm err}=.00030$ and
$P_{\rm det}=.00039$.
 
\vspace{.8cm}
 
{\bf Problem~\ref{pr1.22}}
 
Using the standard array of Problem~\ref{pr1.13}, a decoding error
occurs when the error pattern belongs in a column different from
the first one and in a row different from the last one.
We see that there are
12 error patterns of weight 2,
16 error patterns of weight 3,
15 error patterns of weight 4 and
6 error patterns of weight 5. So,
 
\begin{eqnarray*}
P_{\rm err}&=&12p^2(1-p)^4+16p^3(1-p)^3+15p^4(1-p)^2+6p^5(1-p).
\end{eqnarray*}
 
The first information bit will be decoded in error only when the
error pattern belongs in the 4th, 6th, 7th and 8th column, but
not in the last row.
We see that there are
6 error patterns of weight 2,
10 error patterns of weight 3,
8 error patterns of weight 4 and
4 error patterns of weight 5. So,
 
$$p_0=6p^2(1-p)^4+10p^3(1-p)^3+8p^4(1-p)^2+4p^5(1-p).$$
 
Similarly, the second
information bit will be decoded in error only when the
error pattern belongs in the 3rd, 5th, 7th and 8th column, but
not in the last row.
In this case, we see that
 
$$p_1=6p^2(1-p)^4+10p^3(1-p)^3+8p^4(1-p)^2+4p^5(1-p)+p^6.$$
 
Finally, the third
information bit will be decoded in error only when the
error pattern belongs in the 2nd, 5th, 6th and 8th column, but not
in the last row; hence,
 
$$p_2=6p^2(1-p)^4+10p^3(1-p)^3+8p^4(1-p)^2+4p^5(1-p).$$
 
This gives
 
$$p_{\rm err}={p_0+p_1+p_2\over 3}=
6p^2(1-p)^4+10p^3(1-p)^3+8p^4(1-p)^2+4p^5(1-p)+{1\over 3}p^6.$$
 
We detect an error when the error pattern belongs in the last row, i.e.,
 
\begin{eqnarray*}
P_{\rm det}&=&3p^2(1-p)^4+4p^3(1-p)^3+p^6.
\end{eqnarray*}
 
If $p=.01$, we obtain
$P_{\rm err}=.00117$, $p_{\rm err}=.00052$ and
$P_{\rm det}=.00027$.
 
\section{The Golay Code}
\label{sec6}
The Golay code, denoted $\G_{23}$,
is the $[23,12]$ code whose parity check matrix
is given by

\begin{equation}
\label{eq18}
H=(P\mid I_{11})
\end{equation}

where $I_{11}$ is the $11\times 11$ identity matrix and
$P$ is the $11 \times 12$ matrix

\begin{equation}
\label{eq19}
P=\left( \begin{array}{cccccccccccc}
1 & 0 & 1 & 0 & 0 & 0 & 1 & 1 & 1 & 0 & 1 & 1\\
1 & 1 & 0 & 1 & 0 & 0 & 0 & 1 & 1 & 1 & 0 & 1\\
0 & 1 & 1 & 0 & 1 & 0 & 0 & 0 & 1 & 1 & 1 & 1\\
1 & 0 & 1 & 1 & 0 & 1 & 0 & 0 & 0 & 1 & 1 & 1\\
1 & 1 & 0 & 1 & 1 & 0 & 1 & 0 & 0 & 0 & 1 & 1\\
1 & 1 & 1 & 0 & 1 & 1 & 0 & 1 & 0 & 0 & 0 & 1\\
0 & 1 & 1 & 1 & 0 & 1 & 1 & 0 & 1 & 0 & 0 & 1\\
0 & 0 & 1 & 1 & 1 & 0 & 1 & 1 & 0 & 1 & 0 & 1\\
0 & 0 & 0 & 1 & 1 & 1 & 0 & 1 & 1 & 0 & 1 & 1\\
1 & 0 & 0 & 0 & 1 & 1 & 1 & 0 & 1 & 1 & 0 & 1\\
0 & 1 & 0 & 0 & 0 & 1 & 1 & 1 & 0 & 1 & 1 & 1
\end{array} \right)
\end{equation}

The matrix $P$ has a very particular structure.
Let $\up_0,\up_1,\ldots,\up_{10}$ be the first 11 bits of each
row of $P$.
Denote by $\rho^i(\uv)$ $i$ cyclic
rotations to the right of a vector $\uv$.
We observe that each $\up_i$ is a rotation to the right of the
previous $\up_i$, i.e., $\up_i=\rho^i(\up_0)$, $0\leq i\leq 10$.

The extended Golay code is the
$[24,12]$ code obtained by adding a parity bit to each codeword
of the Golay code (see Problem~\ref{pr1.8}).
We denote by $\G_{24}$ the extended Golay code.
A systematic parity check matrix for $\G_{24}$ is given by

\begin{equation}
\label{eq20}
H_1=(Q\mid I_{12}),
\end{equation}

where $Q$ is the $12\times 12$ matrix given by

\begin{equation}
\label{eq21}
Q=\left(
\begin{array}{c}
P\\
\hline
\1_{11},0
\end{array}
\right),
\end{equation}

$P$ is given by~(\ref{eq19}) and
$\1_{11}$ is the all-1 vector of length 11.

\begin{lemma}
\label{dual}
{\em
The code $\G_{24}$ is self-dual, i.e.,
$\G_{24}=\G_{24}^{\perp}$.
}
\end{lemma}

\pf $H_1$ is a generator matrix for
$\G_{24}^{\perp}$. According to Problem~\ref{pr1.17},
it is enough to prove that any two rows in $H_1$ are orthogonal.
This follows immediately from Problem~\ref{pr1.24}.\qed

As a corollary of Lemma~\ref{dual}, $H_1$ is also a generator matrix
for $\G_{24}$. Moreover:

\begin{cor}
\label{cdual}
{\em
Let $H_1$ be the parity check matrix of $\G_{24}$ given
by~(\ref{eq20}). Then, $H_1$ is also a generator matrix of $\G_{24}$
and so is

\begin{equation}
\label{eq22}
H_2=(I_{12}\mid Q^T).
\end{equation}

Also, each codeword in $\G_{24}$ has weight divisible by 4.
}
\end{cor}

\pf The claims about the parity check and generator matrices
are immediate following the fact that $\G_{24}$ is
self dual. The fact that
every codeword has weight divisible by 4 follows
from Problem~\ref{pr1.25}.\qed

The next lemma is the main result concerning the Golay code.

\begin{lemma}
\label{weight}
{\em
The minimum distance of
the code $\G_{24}$ is 8, i.e.,
$\G_{24}$ can correct three errors and detect four.
}
\end{lemma}

\pf
According to Corollary~\ref{cdual}, it is enough to prove that
there are no codewords of weight 4. Assume that there is
a codeword of weight 4, say
$(\uu\mid\uv)$,
where $\uu$ and $\uv$
have length 12. If $w_H(\uu)=0$,
using the generator matrix $H_2$
given by~(\ref{eq22}), $\uu$ is encoded uniquely into the zero vector,
a contradiction. If $w_H(\uu)=1$, then
$(\uu\mid\uv)$ is a row in $H_2$, a contradiction since $\uv$ has
weight 3 and cannot be in $Q^T$.
If $w_H(\uu)=2$, then $\uv$ is the sum of two
rows of $Q^T$. But $\uv$ cannot have weight 2 by Problem~\ref{pr1.24}.

If $\uv$ has weight 1 or 0, a similar proof follows with respect
to the generator matrix $H_1$.

This shows that there are no codewords of weight 4, so,
by Corollary~\ref{cdual}, the next
possibility is codewords of weight 8. Notice that
there are codewords of
weight 8. For instance any of the first eleven
rows in the generator matrix $H_1$ is
a codeword of weight 8.
Hence, the minimum distance in $\G_{24}$ is 8.\qed

\begin{cor}
\label{cweight}
{\em
The minimum distance of
the code $\G_{23}$ is 7, i.e.,
$\G_{23}$ can correct three errors.
}
\end{cor}

Having determined that $\G_{24}$ has minimum distance 8, the next
step is providing a decoding algorithm that will correct 3 errors
and detect 4. There are many methods to decode the Golay code.
We give one of them.

Let $\uu=(\uu_1\mid\uu_2)$ be a transmitted codeword, where each
part $\uu_1$ and $\uu_2$ has length 12.
We may assume that the first 12 bits (i.e., the vector $\uu_1$),
carry the information, while the last 12 bits (i.e., $\uu_2$),
represent the redundancy.
Let $\ur=(\ur_1\mid\ur_2)$ be a possibly corrupted version of $\uu$
and $\ue=(\ue_1\mid\ue_2)$ be the error
vector, $w_H(\ue)\leq 4$.
Hence, $\ur=\uu\oplus\ue$. The decoder is interested in estimating
the information bits only.

Assume first that
$w_H(\ue)\leq 3$. If
$w_H(\ue_1)=0$, then, if we calculate the syndrome
$\us_1=\ur H_1^T$, we see that $w_H(\us_1)\leq 3$. Moreover,
the error pattern is exactly $\ue=(\0\mid\us_1)$ (see
Problem~\ref{pr1.9}). This means, there were no errors in
the information part and the decoder outputs $\ur_1$ as an estimate
of $\uu_1$.

Similarly, if
$w_H(\ue_2)=0$, then
the error vector is exactly $\ue=(\us_2\mid\0)$, where
$\us_2=\ur H_2^T$. Hence, the decoder outputs $\ur_1\oplus\us_2$ as
estimate of the information bits.

So, if
$w_H(\us_1)>3$ and
$w_H(\us_2)>3$, then
$\ue_1\neq 0$ and
$\ue_2\neq 0$. Since
$w_H(\ue)\leq 3$, then either
$w_H(\ue_1)=1$ or
$w_H(\ue_2)=1$.

Let
$\ur^{(i)}$, $0\leq i\leq 23$, be the
the received vector
$\ur$ with location $i$ complemented.

If
$w_H(\ue_1)=1$ and location
$i$, $0\leq i\leq 11$, is in error,
then the syndrome
$\us_1^{(i)}=\ur^{(i)} H_1^T$ has weight $\leq 2$.
The error vector is then $(\delta_i\mid\us_1^{(i)})$, where
$\delta_i$ denotes a vector of length 12 with a 1 in location $i$,
0 elsewhere. The decoder outputs
$\ur_1\oplus\delta_i$ as an estimate of the information bits.
This operation is repeated at most
12 times in order to check if exactly one
of the first 12 bits is in error.

If none of the syndromes
$\us_1^{(i)}=\ur^{(i)} H_1^T$, $0\leq i\leq 11$, has weight $\leq 2$,
a similar procedure is implemented for
$\ur^{(i)}$, $12\leq i\leq 23$.
We now check the 12 syndromes
$\us_2^{(i)}=\ur^{(i)} H_2^T$,
$12\leq i\leq 23$. If one of them, say $i$,
has weight $\leq 2$, then  the error vector is
$(\us_2^{(i)}\mid \delta_{i-12})$
and the estimate of the information part
is $\ur_1\oplus\us_2^{(i)}$.

If after the 24 checks described above neither
$\us_1^{(i)}$ nor $\us_2^{(i)}$
have weight $\leq 2$, then the decoder decides that 4 errors have
occurred and declares an uncorrectable error.

As a result of the discussion above, we obtain
the following algorithm:

\begin{alg}[Decoding Algorithm for the Extended Golay Code]
\label{algolay}
{\em Let \\
$\ur=(\ur_1\mid\ur_2)$ be a received word, and let
$\us_1=\ur H_1^T$ and
$\us_2=\ur H_2^T$.
Denote by $\uq_0,\uq_1,\ldots,\uq_{11}$ the rows of $Q$, where $Q$ is
given by~(\ref{eq21}),
by $\uq'_0,\uq'_1,\ldots,\uq'_{11}$ the rows of $Q^T$, and
by $\delta_i$ a vector of length 12 with a 1 in location $i$,
$0\leq i\leq 11$, 0 elsewhere. Then:

\begin{tabbing}
If $w_H(\us_1)\leq 3$, output $\ur_1$ and stop.\\
Else, \=
if $w_H(\us_2)\leq 3$, output $\ur_1\oplus\us_2$ and stop.\\
\>Else, \= while $0\leq i\leq 11$, do:\\
\> \>
$\us_1^{(i)}\la\us_1\oplus\uq'_i$. If $w_H(\us_1^{(i)})\leq 2$
for some $i$, then
output $\ur_1\oplus \delta_i$ and stop.\\
\> Else, \= while $0\leq i\leq 11$, do:\\
\> \>
$\us_2^{(i)}\la\us_2\oplus\uq_i$. If $w_H(\us_2^{(i)})\leq 2$
for some $i$, then
output $\ur_1\oplus
\us_2^{(i)}$ and stop.\\
\> Else, declare an uncorrectable error.
\end{tabbing}
}
\end{alg}

\begin{ex}
\label{exgolay}
{\em
Let $\ur=0 1 1 1 1 0 1 1 1 0 1 0\;0 0 1 1 0 0 0 0 0 0 10$.
According to Algorithm~\ref{algolay},\\
$\us_1\la\ur H_1^T=1 0 0 0 1 0 1 1 1 1 1 0$.
Since $w_H(\us_1)>3$, we go on with the recursion of the Algorithm.
Eventually, for $i=6$, notice that adding
row 6 of $Q^T$ to $\us_1$, we obtain
$\us_1^{(6)}=0 0 0 0 0 0 0 0 1 0 0 1$, which has weight 2. Hence, there
was an error in bit 6 of the information bits and $\ur$ is decoded
as $\uu=0 1 1 1 1 0 0 1 1 0 1 0$.\qed
}
\end{ex}

The Golay code was introduced for the first time in~\cite{golay}.

\pr
\begin{prob}
\label{pr1.23}
{\em
Prove that matrix $H_1$ given by~(\ref{eq20}) is a systematic
parity check matrix for $\G_{24}$.
}\end{prob}
 
\begin{prob}
\label{pr1.24}
{\em
Prove that the distance between
any two rows (resp. columns)
of $Q$ in~(\ref{eq21}) is
6 and the inner product of any two distinct
rows of $Q$ is 0.
}\end{prob}
 
\begin{prob}
\label{pr1.25}
{\em
Prove that if $\C$ is a self dual
code with generator matrix $G$,
and each row of $G$ has weight divisible by 4, then every codeword
in $\C$ has weight divisible by 4.
}\end{prob}
 
\begin{prob}
\label{pr1.26}
{\em
Prove that $\G_{23}$ is a perfect 3-error correcting code.
}\end{prob}
 
\begin{prob}
\label{pr1.27}
{\em
Decode the following vectors in
$\left(GF(2)\right)^{23}$ with respect to
$\G_{23}$ (give as output only the 12 information bits):
 
$\ur=1 1 0 0 1 1 1 0 1 1 0 0 0 1 1 1 1 0 1 1 1 0 1$ and
$\ur=0 1 0 0 1 1 1 1 1 0 1 1 0 1 1 1 1 0 0 0 0 0 0$.
 
}\end{prob}
 
\begin{prob}
\label{pr1.28}
{\em
Write a computer program implementing Algorithm~\ref{algolay}.
}\end{prob}
 

\sol
\vspace{.8cm}
 
{\bf Problem~\ref{pr1.23}}
 
By Problem~\ref{pr1.8}, a parity-check matrix
is given by
 
$$H'=
\left(\begin{array}{c|l}
\hspace{1cm} H \hspace{1cm}&
\left.
\begin{array}{l}
0\\ 0\\ \vdots \\ 0
\end{array}
\right\}{\scriptstyle 11}\\
\hline
\underbrace{11 \ldots 1}_{23}&
\begin{array}{l}
1
\end{array}
\end{array}
\right)=
\left(\begin{array}{c|c|l}
P & I_{11}&
\left.
\begin{array}{l}
0\\ 0\\ \vdots \\ 0
\end{array}
\right\}{\scriptstyle 11}\\
\hline
\underbrace{11 \ldots 1}_{12}&
\underbrace{11 \ldots 1}_{11}&
\begin{array}{l}
1
\end{array}
\end{array}
\right).$$
 
Replacing the last row by the sum of all the rows in $H'$, we obtain
the systematic parity-check matrix $H_1$.
 
\vspace{.8cm}
 
{\bf Problem~\ref{pr1.24}}
 
Calling $\up_i$ the first 11 bits of each row in $P$, $0\leq i\leq 10$,
we had, $\up_i=\rho^i(\up_0)$. Similarly, if we denote by
$\up'_i$ the transpose of
each of the first 11 columns in $P$, $0\leq i\leq 10$,
we verify that $\up'_i=\rho^i(\up'_0)$.
 
Hence, the distance between
row $i$, $0\leq i\leq 10$, of $Q$ and row 11,
is equal to
$d_H(\up_0\oplus\1_{11})+1=(11-w_H(\up_0))+1=6$. We similarly prove
that the distance between
column $i$, $0\leq i\leq 10$, of $Q$ and
column 11, is 6.
 
Consider now the distance between
rows (resp. columns) $i$ and $j$, $0\leq i<j\leq 10$.
It is enough to consider
the distance between $\up_i$ and $\up_j$
(resp. $\up'_i$ and $\up'_j$), since
the last bit in these rows (resp. columns) is 1.
 
Notice that $d_H(\up_i,\up_j)=
d_H(\rho^i(\up_0),\rho^j(\up_0))=
d_H(\up_0,\rho^{j-i}(\up_0))=
d_H(\up_0,\up_{j-i})$.
Hence, it is enough to verify that the distance between the first
row of $P$ and any other row has weight 6, which is easily done.
A similar proof holds for columns (or,
observe that column $q_i$ is equal to row $p_i$ plus
$(0,1,1,\ldots,1$)).
 
To prove that the inner product between any two rows of $Q$
is 0, it is
enough to show that the set where any two rows is 1 is an even
number. Following a procedure similar to the one described above
(essentially, by comparing any row to the first row), we see
that, from rows 0 to 10, the set where both rows are 1 has cardinality 4.
The set where one of the first 11 rows and row 12 are 1 has cardinality
6. Hence, the result follows.
 
A similar proof is valid for columns.
 
\vspace{.8cm}
 
{\bf Problem~\ref{pr1.25}}
 
Let $\uu$ and $\uv$ be two orthogonal vectors whose weight is divisible
by 4. Since the vectors are orthogonal, the number of coordinates where
the two vectors are 1 is an even number. Let us call this number $2l$.
Hence, $w_H(\uu\oplus\uv)=w_H(\uu)+w_H(\uv)-4l$.
This number is divisible by 4.
 
Now, since any two rows of $G$ are orthogonal and their weight is
divisible by 4, their sum is also divisible by 4.
In particular, the same is true for the sum of any finite number
of rows of $G$, i.e., for any codeword of $\C$.
 
\vspace{.8cm}
 
{\bf Problem~\ref{pr1.26}}
 
$\G_{23}$ is a $[23,12,7]$ code. According to~(\ref{eq8}),
$V(3)={23\choose 0}+{23\choose 1}+{23\choose 2}+{23\choose 3}=2048=
2^{11}$, hence,
$n-k=11=\log_2V(3)=\log_2V(\lf (d-1)/2\rf)$ and
Inequality~(\ref{eq9}) is met with equality.
 
\vspace{.8cm}
 
{\bf Problem~\ref{pr1.27}}
 
Let $\ur=1 1 0 0 1 1 1 0 1 1 0 0\;0 1 1 1 1 0 1 1 1 0 1$. Consider
the vector $\ur,0\in GF(2)^{24}$.
If we apply the decoding algorithm to $\ur,0$, we see that
its syndrome is
$s_1=(\ur,0)H_1^T= 1 1 1 0 1 0 1 0 0 0 1 1$.
For $i=1$, we see that
$s_1^{(1)}=s_1\oplus \uq'_2=1 0 0 0 0 1 0 0 0 0 0 0$.
Hence, $w_H(s_1^{(1)})\leq 2$, so the error in the first 12 bits has
occured in the second bit (we count from 0).

So, the output of the decoder is
$\ur_1\oplus \delta_1= 1 0 0 0 1 1 1 0 1 1 0 0$.
 
Consider now
$\ur=0 1 0 0 1 1 1 1 1 0 1 1\;0 1 1 1 1 0 0 0 0 0 0$.
If we take the vector $\ur,0$ as before, we can verify that the
algorithm declares an uncorrectable error (i.e., 4 errors have occurred).
So, we consider $\ur,1$. Let $\us_1$ and $\us_2$ be the syndromes
of $\ur,1$ as defined by the algorithm.
We can see that
$\us_1=\ur H_1^T=1 1 0 0 0 1 1 0 0 1 0 0$ and
$\us_2=\ur H_2^T=0 1 1 0 0 1 1 1 0 0 1 1$.
Neither of them has weight $\leq 3$. We also verify that
$w_H(s_1^{(i)})>2$, for all $0\leq i\leq 11$.
On the other hand, we can see that
$\us_2^{(1)}=\us_2\oplus \uq_{1}=0 0 1 0 0 0 0 0 0 1 0 0$, hence,
$w_H(\us_2^{(1)})=2$. The output of the decoder is
$\ur_1\oplus \us_2^{(1)}=0 1 1 0 1 1 1 1 1 1 1 1$.

\chapter{Finite Fields and RS Codes}
\label{ch2}
\section{Introduction}
\label{introd2}
In this chapter, we want to introduce the family of multiple
error-correcting Reed Solomon (RS) codes. RS codes operate not over
bits, as was the case of the codes studied in the previous chapter,
but over bytes. Each byte is a vector composed by several bits.
Typical cases in magnetic and optical recording involve 8-bit
bytes. In order to operate with bytes, we need a method to multiply
them. To this end,
we develop the theory of finite fields.
In the previous chapter, we considered codes whose coordinates
were elements of the binary
field $GF(2)$. In this chapter the codes will have coordinates
over any finite field.

\section{Finite Fields}
\label{sec2.2}

This section contains an introduction to the theory of
finite fields. For a more complete treatment, the reader is referred
to~\cite{mac}, chapter~4, and to~\cite{ln,mce,m}.

Essentially, the elements of a finite field are vectors of a certain
length $\nu$, that we call bytes. In most applications,
the bytes are
binary vectors, although we will not be bound by
this restriction in our study.

We know how to add
two binary vectors: we simply exclusive-OR them componentwise.
What we need now is a rule that allows us to multiply bytes while
preserving associative, distributive, and multiplicative inverse
properties, i.e., a product
that gives to the set of bytes of length $\nu$ the structure of a
field. To this end, we will define a multiplication
between vectors that satisfies the associative and
commutative properties,
it has a 1 element, each non-zero element is invertible and
it is distributive with respect to the sum operation.

Recall the definition of
the ring $Z_m$ of integers modulo $m$:
$Z_m$ is the set $\{0,1,2,\ldots,m-1\}$, with a sum and product
of any two elements defined as the residue of dividing by $m$ the
usual sum or product. $Z_m$ is a field if and only if $m$ is a prime
number (see Problem~\ref{pr2.1}).
From now on $p$ denotes a prime number and $Z_p$ will be
denoted as $GF(p)$.

Consider the vector space $(GF(p))^{\nu}$ over the field $GF(p)$.
We can view each vector as a polynomial of degree
$\leq \nu-1$ as follows:
the vector $\ua=(a_0,a_1,\ldots,a_{\nu-1})$ corresponds to the polynomial
$a(\al)=a_0+a_1\al+\ldots +a_{\nu-1}\al^{\nu-1}$.

The goal now is to give to
$\left(GF(p)\right)^{\nu}$
the structure of a field. We will denote such a field
by $GF(p^{\nu})$. The sum in
$GF(p^{\nu})$ is the usual sum of vectors in
$\left(GF(p)\right)^{\nu}$. We need now to define a product.

Let $f(x)$ be an irreducible polynomial of degree $\nu$
whose coefficients are in $GF(p)$.
Let $a(\al)$ and $b(\al)$ be two elements of $GF(p^{\nu})$.
We define the product between $a(\al)$ and $b(\al)$ in
$GF(p^{\nu})$ as the unique polynomial
$c(\al)$ of degree $\leq\nu-1$ such
that $c(\al)$ is congruent to the product
$a(\al)b(\al)$ modulo $f(\al)$. In other words,
$c(\al)$ is the residue of dividing $a(\al)b(\al)$ by $f(\al)$.

The sum and product operations defined above will give to
$GF(p^{\nu})$ a field structure.
From now on, we denote the elements in $GF(p^{\nu})$ as polynomials
in $\al$ of degree $\leq \nu -1$ with coefficients in $GF(p)$.
Given two polynomials $a$ and $b$ with coefficients in $GF(p)$,
$a(\al)b(\al)$ denotes the product in
$GF(p^{\nu})$, while $a(x)b(x)$
denotes the regular product of polynomials.
Notice that, in particular $f(\al)=0$ over
$GF(p^{\nu})$, since $f(x)\equiv\;0\;(\bmod\; f(x))$.

So, the set $GF(p^{\nu})$ given by the irreducible polynomial $f(x)$
of degree $\nu$, is the set of polynomials of degree $\leq \nu -1$,
where the sum operation is the regular sum of polynomials, and the
product operation is the residue of dividing by $f(x)$ the regular
product of two polynomials. The next lemma proves that
$GF(p^{\nu})$ is indeed a field.

\begin{lemma}
\label{lemma2.1}
{\em
The set
$GF(p^{\nu})$ defined by an irreducible polynomial $f$
of degree $\nu$ is a field.
}
\end{lemma}

\pf It is clear that the usual associative, commutative, additive
inverse, existence of 0 and 1, hold for both sum and product.
The only difficulty is showing the existence of multiplicative
inverse.

We have to prove that for
every $a(\al)\in GF(p^{\nu})$,
$a(\al)\neq 0$, there is a $b(\al)$ such that
$a(\al)b(\al)=1$. Since $f(x)$ is irreducible and
$\deg(a(x))<\deg(f(x))$, $a(x)$ and $f(x)$ are relatively prime, i.e.,
$\gcd (a(x),f(x))=1$.
By Euclid's algorithm for polynomials,
there are polynomials $b(x)$ and $c(x)$ such that

$$b(x)a(x)+c(x)f(x)=1.$$

The equation above means

\begin{equation}
\label{modulo}
b(x)a(x)\equiv 1\;(\bmod f(x)).
\end{equation}

We can also assume that
$\deg(b(x))\leq \nu-1$ (if not, we take the residue
of dividing $b(x)$ by $f(x)$); hence, translating~(\ref{modulo})
to an equality in
$GF(p^{\nu})$, we obtain
$a(\al)b(\al)=1$.\qed

We have shown how to construct a finite field of cardinality
$p^{\nu}$: we simply take the polynomials
of degree $\leq \nu-1$ with
coefficients in $GF(p)$ and consider them modulo an irreducible
polynomial $f(x)$ of degree $\nu$.

In fact, every finite field has cardinality a power of a prime (see
Problem~\ref{pr2.3}).
Moreover, every finite field is {\em isomorphic}
to a field as described above.
Given two fields $F$ and $F'$ with zero elements 0 and 0' and
one elements 1 and 1' respectively, we say that $F$ and $F'$
are isomorphic if there is a 1-1 onto function
$g:F\ra F'$ preserving sums and products.

If $q$ is a prime power, we denote by $GF(q)$ the finite field
with $q$ elements (up to isomorphism).

Another important property of a finite field is that its non-zero elements
are a cyclic group, i.e., there is an element in the field whose powers
generate all the non-zero elements. In order to prove this, we need
an auxiliary lemma.

Let $G$ be a finite multiplicative abelian group.
Consider the powers of an element $a\in G$, say,
$1=a^0,a=a^1,a^2,\ldots,a^{l-1}$, and assume that $l$ is the first value
such that $a^l=1$. We say that $l$ is the {\em order} of $a$.

\begin{lemma}
\label{lemma2.11}
{\em
Let $G$ be a finite multiplicative abelian group. Then,

\begin{enumerate}

\item Let $a\in G$ and the order of $a$ is $l$.
Assume that $a^{l'}=1$ for some $l'$. Prove that $l$ divides $l'$.

\item Assume that $l$ is the order of $a\in G$ and $j$ divides $l$.
Prove that $a^j$ has order $l/j$.

\item Assume that $a$ has order $l$ and $n$ is relatively prime to
$l$. Prove that $a^n$ has order $l$.

\item If $a$ and $b$ are elements in $G$ having orders $m$ and $n$
respectively, $m$ and $n$ relatively prime, prove that $ab$ has
order $mn$.

\item Let $m$ be the
highest possible order of an element in $G$; $m$ is called the
{\em exponent} of $G$. Prove that the order of any element in
$G$ divides the exponent.

\end{enumerate}
}
\end{lemma}

We give the proof of Lemma~\ref{lemma2.11} as a problem (Problem~\ref{pr2.4}).
We are ready now to prove that $F-\{0\}$ is cyclic for any finite field $F$.

\begin{lemma}
\label{pr2.5}
{\em
Let $F$ be a finite field. Then, $F-\{0\}$ is a cyclic
group with respect to the product operation.
}
\end{lemma}

\pf
Let $m$ be the exponent of the multiplicative group
$F-\{0\}$. We have to prove that
$m=|F|-1$. Consider the polynomial $x^m-1$.
By Lemma~\ref{lemma2.11}, the order of
every element in $F-\{0\}$ divides $m$. In particular, if
$a\in F-\{0\}$, $a^m=1$. In other words, $a$
is a root
of $x^m-1$. Since $x^m-1$ has at most $m$ different roots,
and every element in
$F-\{0\}$ is a root, $m=|F|-1$.\qed

\begin{ex}
\label{finite}
{\em
Let us construct the field $GF(8)$.
Consider the polynomials of degree $\leq 2$ over $GF(2)$.
Let $f(x)=1+x+x^3$. Since $f(x)$ has no roots over $GF(2)$, it is
irreducible (notice that such an assessment can be made only
for polynomials of degree 2 or 3). Let us consider the powers of
$\al$ modulo $f(\al)$. Notice that $\al^3=\al^3+f(\al)=1+\al$.
Also, $\al^4=\al\al^3=\al (1+\al)=\al+\al^2$.
Similarly, we obtain
$\al^5=\al\al^4=\al (\al+\al^2)=\al^2+\al^3=1+\al+\al^2$,
and $\al^6=\al \al^5=\al+\al^2+\al^3=1+\al^2$. Finally,
$\al^7=\al \al^6=\al+\al^3=1$.

As we can see, every element in GF(8) can be obtained as a power of
the element $\al$. In this case, $\al$ is called a {\em primitive}
element and the irreducible polynomial $f(x)$ that defines the field
is called a {\em primitive} polynomial.
Since the multiplicative group of a finite
field is cyclic, (Lemma~\ref{pr2.5}),
there is always a primitive element.

A convenient description of $GF(8)$ is given in Table~\ref{table1}.

\begin{table}
$$
\begin{array}{|c|c|c|c|}
\hline
{\rm Vector}&{\rm Polynomial}&{\rm Power\;of}\;\al&{\rm Logarithm}\\
\hline
000&0&0&-\infty\\
100&1&1&0\\
010&\al &\al &1\\
001&\al^2&\al^2&2\\
110&1+\al&\al^3&3\\
011&\al+\al^2&\al^4&4\\
111&1+\al+\al^2&\al^5&5\\
101&1+\al^2&\al^6&6\\
\hline
\end{array}
$$
\caption{The finite field $GF(8)$ generated by $1+x+x^3$}
\label{table1}
\end{table}

The first column in Table~\ref{table1}
describes the element of the field in vector
form, the second one as a polynomial in $\al$ of degree $\leq 2$,
the third one as a power of $\al$,
and the last one gives the logarithm (also called Zech logarithm):
it simply indicates the corresponding power of $\al$.
As a convention, we denote by $-\infty$ the logarithm corresponding
to the element 0.\qed
}
\end{ex}

It is often convenient to express the
elements in a finite field as powers
of $\al$: when we multiply two of
them, we obtain a new power of $\al$ whose
exponent is the sum of the two exponents modulo $q-1$.
Explicitly, if $i$ and $j$ are the logarithms of two elements in
$GF(q)$, then their product has logarithm
$i+j\;\left(\bmod (q-1)\right)$.
In the example above, if we want to multiply the vectors 101 and 111,
we first look at their logarithms. They are 6 and 5 respectively, so
the logarithm of the product is $6+5\;(\bmod\; 7)=4$, corresponding to
the vector 011.

In order to add vectors, the best way is to express them in vector
form and add coordinate to coordinate in the usual way.

\pr

\begin{prob}
\label{pr2.1}
{\em
Prove that $Z_m$ is a field if and only if $m$ is a prime number.
}\end{prob}
 
\begin{prob}
\label{pr2.2}
{\em
Let $F$ be a finite field. Prove that there is a prime number $p$
such that, for any $a\in F$, $a\neq 0$,
$\overbrace{a+a+\ldots +a}^p=0$. The prime number $p$ is called the
{\em characteristic} of the finite field $F$.
}\end{prob}
 
\begin{prob}
\label{pr2.3}
{\em
Let $F$ be a finite field with characteristic $p$. Prove that
the cardinality of $F$ is a power of $p$.
}\end{prob}
 
\begin{prob}
\label{pr2.4}
{\em
Prove Lemma~\ref{lemma2.11}. 
 
}\end{prob}
 
 
 
\begin{prob}
\label{pr2.7}
{\em
Find all the irreducible polynomials of degree~4 over $GF(2)$.
Determine which ones of those polynomials are primitive.
}\end{prob}
 
\begin{prob}
\label{pr2.8}
{\em
Using a primitive polynomial in the previous problem, construct
the field $GF(16)$ by providing a table similar to Table~\ref{table1}.
}\end{prob}
 
\begin{prob}
\label{pr2.9}
{\em
Using the irreducible polynomial $2+x+x^2$ over $GF(3)$, construct
the table of the finite field $GF(9)$. Is the polynomial primitive?
}\end{prob}

\sol
\vspace{.8cm}

{\bf Problem~\ref{pr2.1}}
 
Assume that $Z_m$ is a field. If $m$ is not prime, say,
$m=ab$ with $a>1$ and $b>1$, then $ab\equiv 0\;(\bmod\; m)$, i.e.,
the product of $a$ and $b$ is 0 in $Z_m$, contradicting the
fact that $Z_m$ is a field.
 
Conversely, if $m$ is prime, let $a\in Z_m$, $a\neq 0$. We have to
prove that $a$ is invertible. Since $a$ and $m$ are relatively prime
in $Z$, there is a $c$ and a $d$ such that
$ca+dm=1$, i.e., $ca\equiv 1\;(\bmod m)$. Without loss, we can take
$c$ in $Z_m$; hence the product of $c$ and $a$ is 1 in $Z_m$, meaning
that $c$ is the multiplicative inverse of $a$.
 
\vspace{.8cm}
 
{\bf Problem~\ref{pr2.2}}
 
Denote by $\odot$ the product in $F$.
Without loss of generality, since $a=a\odot 1$, we can take $a=1$.
We call
$m=\overbrace{1+1+\cdots +1}^m$.
Since the field is finite, there is an $m$ such that
$m=\overbrace{1+1+\cdots +1}^m=0$.
Let $m$ be the smallest such number. We claim, $m$ is prime.
If not, $m=ab$ for $a>1$ and $b>1$. In this case,
$m=\overbrace{1+1+\cdots +1}^m=
\overbrace{1+1+\cdots +1}^a\odot
\overbrace{1+1+\cdots +1}^b=a\odot b=0$.
This contradicts the fact that $F$ is a field.
 
\vspace{.8cm}
 
{\bf Problem~\ref{pr2.3}}
 
Let $F_p$ be the subset of $F$ formed by sums of 1, i.e.,
 
$$F_p=\{m\;:\; m=\overbrace{1+1+\cdots +1}^m\}.$$
 
$F_p$ is a subfield of $F$, it is isomorphic to $Z_p$.
Consider $F$ as a vector space over $F_p$. Then, $F$ has a certain
dimension, say $k$, as a vector space over $F_p$. Hence, any
element $\uv$ in $F$ can be written uniquely as a linear combination
of a basis $\uv_1,\uv_2,\ldots,\uv_k$, i.e., there are unique
$t_1,t_2,\ldots,t_k$ in $F_p$ such that
$\uv=t_1\uv_1+t_2\uv_2+\cdots +t_k\uv_k$. There is a total of
$p^k$ different linear combinations of $\uv_1,\uv_2,\ldots ,\uv_k$,
so the cardinality of $F$ is $p^k$.
 
\vspace{.8cm}
 
{\bf Problem~\ref{pr2.4}}
 
 
 

\begin{enumerate}

\item If we divide $l'$ by $l$, we obtain
$l'=ql+r$, where $0\leq r<l$. Now, $a^{l'}=a^{ql+r}=
(a^l)^qa^r=a^r=1$. Since $l$ is the order of $a$, then $r=0$.
 
\item Notice that
$(a^j)^{l/j}=a^l=1$. Now, assume that
$(a^j)^s=1$ for some $s$, then,
$a^{js}=1$, and since $l$ is the order of $a$, $l$ divides $js$.
Since $j$ divides $l$, $l/j$ divides $s$, proving that $l/j$ is
the order of $a^j$.
 
\item Notice that $(a^n)^l\eq (a^l)^n\eq 1$. Now, assume that 
$(a^n)^m\eq a^{nm}\eq 1$ for some $m$. Then, $l$ divides $nm$, and
since $l$ and $n$ are relatively prime, $l$ divides $m$.

\item  
It is clear that $(ab)^{mn}=a^{mn}b^{mn}=(a^m)^n(b^n)^m=1$.
We have to show now that, if
there is an $l>0$ such that $(ab)^l=1$, then $mn\leq l$.
Notice that, if $(ab)^l=a^lb^l=1$, then, 
$a^l\eq b^{-l}$. Thus, $(a^l)^n\eq (b^{-l})^n\eq (b^{n})^{-l}\eq 1$, so
$a^{ln}\eq 1$ and then, $m$ divides $ln$. Since $m$ and $n$ are relatively
prime, $m$ divides $l$. Similarly,
$n$ divides $l$, and since $m$ and $n$ are relatively prime,
$mn$ divides $l$.
 
\item Let $m$ be the exponent of $G$ corresponding to the order of
an element $a$. Let $b$ be an element in
$G$ of order $n$, we will show that $n$ divides $m$.
Let $p$ be a prime dividing $n$, then we can write
$n=p^in'$, where $p$ and $n'$ are relatively prime (in other words,
$p^i$ is the largest power of $p$ dividing $n$). Similarly, we
can write $m=p^jm'$, where $p$ and $m'$ are relatively prime.
We will show that $i\leq j$. Notice that the element
$a^{p^j}$ has order $m'$ and the element $b^{n'}$ has order $p^i$.
Since $m'$ and $p^i$ are relatively prime, the element
$a^{p^j}b^{n'}$ has order $p^im'$.
Since the exponent of the group is $m\eq p^jm'$, we have
$p^im'\leq p^jm'$, i.e., $i\leq j$.
 
Since any prime power dividing $n$ also divides $m$, then
$n$ divides $m$.
 
\end{enumerate}

\vspace{.8cm}
 
{\bf Problem~\ref{pr2.7}}
 
The binary polynomials of degree 4 have the form
$a_0+a_1x+a_2x^2+a_3x^3+x^4$, where $a_i\in GF(2)$.
If the polynomial is irreducible, $a_0=1$, if not 0 would be a root.
The polynomial
$1+x+x^2+x^3+x^4$ is irreducible, since it has no roots and is not
the product of two irreducible polynomials of degree 2. The only
binary irreducible polynomial of degree 2 is in fact $1+x+x^2$.
The square of this polynomial is $1+x^2+x^4$. The polynomials of
weight 4 cannot be irreducible, since 1 is a root of them.
So, the two remaining irreducible polynomials of degree 4 are
$1+x+x^4$ and $1+x^3+x^4$.
 
The polynomial
$1+x+x^2+x^3+x^4$ is not primitive. In effect, replace $x$ by $\al$
and describe $GF(16)$ as polynomials in $\al$ modulo
$1+\al+\al^2+\al^3+\al^4$. Notice that $\al^5=1$, hence, the polynomial
is not primitive. The other two polynomials are primitive. One of
them is shown in the next problem, the other one behaves similarly.
 
\vspace{.8cm}
 
{\bf Problem~\ref{pr2.8}}
 
If we consider the primitive polynomial $1+x+x^4$, the finite field
is represented by 
 
$$
\begin{array}{|c|c|c|c|}
\hline
{\rm Vector}&{\rm Polynomial}&{\rm Power\;of}\;\al&{\rm Logarithm}\\
\hline
0000&0&0&-\infty\\
1000&1&1&0\\
0100&\al &\al &1\\
0010&\al^2&\al^2&2\\
0001&\al^3&\al^3&3\\
1100&1+\al&\al^4&4\\
0110&\al+\al^2&\al^5&5\\
0011&\al^2+\al^3&\al^6&6\\
1101&1+\al +\al^3&\al^7&7\\
1010&1+\al^2&\al^8&8\\
0101&\al+\al^3&\al^9&9\\
1110&1+\al+\al^2&\al^{10}&10\\
0111&\al+\al^2+\al^3&\al^{11}&11\\
1111&1+\al+\al^2+\al^3&\al^{12}&12\\
1011&1+\al^2+\al^3&\al^{13}&13\\
1001&1+\al^3&\al^{14}&14\\
\hline
\end{array}
$$
 
\vspace{.8cm}
 
{\bf Problem~\ref{pr2.9}}
 
The field is represented by

$$
\begin{array}{|c|c|c|c|}
\hline
{\rm Vector}&{\rm Polynomial}&{\rm Power\;of}\;\al&{\rm Logarithm}\\
\hline
00&0&0&-\infty\\
10&1&1&0\\
01&\al &\al &1\\
12&1+2\al&\al^2&2\\
22&2+2\al&\al^3&3\\
20&2&\al^4&4\\
02&2\al&\al^5&5\\
21&2+\al&\al^6&6\\
11&1+\al&\al^7&7\\
\hline
\end{array}
$$

Since $\al$ generates all the non-zero elements, $2+x+x^2$ is indeed
a primitive polynomial over $GF(3)$. 
 
\section{Cyclic Codes}
\label{sec2.3}

In the same way we defined codes over the binary field $GF(2)$, we
can define codes over any finite field $GF(q)$. Now, a code
of length $n$ is a subset of
$\left(GF(q)\right)^n$, but since we study only
linear codes, we require that such a subset is a vector space.
Similarly, we define the minimum (Hamming) distance
and the generator and parity-check matrices of a code.
Some properties of
binary linear codes, like the Singleton bound, remain the same
in the general case. Others, like the Hamming bound, require some
modifications (Problem~\ref{pr2.10'}).

Consider a linear code $\C$ over $GF(q)$ of length $n$.
We say that $\C$ is cyclic if, for any codeword
$(c_0,c_1,\ldots,c_{n-1})\in\C$, then
$(c_{n-1},c_0,c_1,\ldots,c_{n-2})\in\C$.
In other words, the code is invariant under cyclic
shifts to the right.

If we write the codewords as polynomials of degree $<n$ with
coefficients in $GF(q)$, this
is equivalent to say that if $c(x)\in\C$, then
$xc(x)\bmod(x^n-1)\in\C$. Hence, if $c(x)\in\C$, then, given any
polynomial $f(x)$, the residue of dividing $f(x)c(x)$ by $x^n-1$ is
in $\C$. In particular, if the degree of $f(x)c(x)$ is smaller than
$n$, then $f(x)c(x)\in\C$.
A more fancy way of describing the above property, is by saying that
a cyclic code of length $n$ is an ideal in the ring of polynomials
over $GF(q)$ modulo $x^n-1$~\cite{h}.

From now on, we write
the elements of a cyclic code $\C$ as polynomials modulo $x^n-1$.

\begin{theo}
\label{cyclic}
{\em
$\C$ is an $[n,k]$ cyclic code over $GF(q)$ if and only if
there is a (monic) polynomial
$g(x)$ of degree $n-k$ such that $g(x)$ divides $x^n-1$ and
each $c(x)\in\C$ is a multiple
of $g(x)$, i.e., $c(x)\in\C$ if and only if
$c(x)=f(x)g(x)$, $\deg (f)<k$. We call $g(x)$
a generator polynomial of $\C$.
}
\end{theo}

\pf
Let $g(x)$ be a monic polynomial in $\C$ such that $g(x)$ has minimal
degree. If $\deg (g)=0$ (i.e., $g=1$), then $\C$ is the whole space
$\left(GF(q)\right)^n$,
so assume $\deg (g)\geq 1$. Let $c(x)$ be any element in
$\C$. We can write $c(x)=f(x)g(x)+r(x)$, where $\deg (r)<\deg (g)$.
Since $\deg (fg)<n$, $g\in\C$ and $\C$ is cyclic, in particular,
$f(x)g(x)\in\C$. Hence, $r(x)=c(x)-f(x)g(x)\in\C$. If $r\neq 0$,
we would contradict the fact that $g(x)$ has minimal degree, hence,
$r=0$ and $c(x)$ is a multiple of $g(x)$.

Similarly, we can prove that $g$ divides $x^n-1$. Let
$x^n-1=h(x)g(x)+r(x)$, where $\deg (r)<\deg (g)$. In particular,
$h(x)g(x)\equiv -r(x)\bmod (x^n-1)$, hence, $r(x)\in\C$. Since
$g(x)$ has minimal degree, $r=0$, so $g(x)$ divides $x^n-1$.

Conversely, assume that every element in $\C$ is
a multiple of $g(x)$ and $g$ divides $x^n-1$.
It is immediate that the code is linear and that it has dimension $k$.
Let $c(x)\in\C$, hence, $c(x)=f(x)g(x)$ with
$\deg (f)<k$. Also, since $g(x)$ divides $x^n-1$,
$x^n-1\eq h(x)g(x)$. Assume that
$c(x)\eq c_0+c_1x+c_2x^2+\cdots +c_{n-1}x^{n-1}$, then,
$xc(x)\equiv c_{n-1}+c_0x+\cdots +c_{n-2}x^{n-1}\;(\bmod\;x^n-1)$.
We have to prove that
$c_{n-1}+c_0x+\cdots +c_{n-2}x^{n-1}\eq q(x)g(x)$, where $q(x)$ has degree
$\leq k-1$. Notice that

\begin{eqnarray*}
c_{n-1}+c_0x+\cdots +c_{n-2}x^{n-1}&=&
c_{n-1}+c_0x+\cdots +c_{n-2}x^{n-1}+c_{n-1}x^n-c_{n-1}x^n\\
&=&c_0x+\cdots +c_{n-2}x^{n-1}+c_{n-1}x^n-c_{n-1}(x^n-1)\\
&=&xc(x)-c_{n-1}(x^n-1)\\
&=&xf(x)g(x)-c_{n-1}h(x)g(x)\\
&=&(xf(x)-c_{n-1}h(x))g(x),
\end{eqnarray*}

proving that the element is in the code. \qed

Theorem~\ref{cyclic} gives a method to find all cyclic codes
of length $n$: simply take all the (monic) factors of $x^n-1$. Each one
of them is the generator polynomial of a cyclic code.

\begin{ex}
\label{ex2.0}
{\em
Consider the $[8,3]$ cyclic code over $GF(3)$ generated by
$g(x)=2+x^2+x^3+2x^4+x^5$.
We can verify that $x^8-1=g(x)(1+x^2+x^3)$, hence, $g(x)$ indeed
generates a cyclic code.

In order to encode an information polynomial
over $GF(3)$ of degree $\leq 2$ into a codeword, we multiply it
by $g(x)$.

Say that we want to encode $\uu=(2,0,1)$, which in polynomial form
is $u(x)=2+x^2$. Hence, the encoding gives
$c(x)=u(x)g(x)=1+x^2+2x^3+2x^4+2x^6+x^7$.
In vector form, this gives $\uc=(1\;0\;1\;2\;2\;0\;2\;1)$.\qed
}
\end{ex}

The encoding method of a cyclic code with generator polynomial
$g$ is then very simple: we multiply the information
polynomial by $g$. However, this encoder is not systematic.
A systematic encoder of a cyclic code is given by the following
algorithm:

\begin{alg}[Systematic Encoding Algorithm for Cyclic Codes]
\label{alg2.1}
{\em Let $\C$ be a \\
cyclic $[n,k]$ code over $GF(q)$
with generator polynomial $g(x)$.
Let $u(x)$ be an information polynomial, $\deg(u)<k$.
Let $r(x)$ be the residue of dividing $x^{n-k}u(x)$ by
$g(x)$. Then, $u(x)$ is encoded into the polynomial
$c(x)=u(x)-x^kr(x)$.
}\end{alg}

In order to prove that Algorithm~\ref{alg2.1} really provides a
systematic encoder for cyclic codes, we have to show two facts:
one is that the encoder is really systematic. This is easily
seen, since $0\leq \deg(r)<n-k$, hence, $k\leq\deg(x^kr(x))<n$.
If $u(x)$ and $r(x)$ in vector notation are given by
$\uu=(u_0,u_1,\ldots,u_{k-1})$ and
$\ur=(r_0,r_1,\ldots,r_{n-k-1})$, then $c(x)$ in vector notation
is given by $\uc=(u_0,u_1,\ldots,u_{k-1},-r_0,-r_1,\ldots,-r_{n-k-1})$.

The second fact we need to verify is that $c(x)$ in effect belongs
in $\C$, i.e., $g(x)$ divides $c(x)$.
Notice that, by the definition of $r(x)$,
$x^{n-k}u(x)=q(x)g(x)+r(x)$, for a certain $q(x)$.
Multiplying both sides of this equality by $x^k\bmod (x^n-1)$,
we obtain $u(x)\equiv x^kq(x)g(x)+x^kr(x)\bmod (x^n-1)$. Hence,
$c(x)=u(x)-x^kr(x)\equiv x^kq(x)g(x)\bmod (x^n-1)$, i.e.,
$c(x)\in\C$.

\begin{ex}
\label{ex2.0'}
{\em
Consider the $[8,3]$ cyclic code over $GF(3)$ of Example~\ref{ex2.0}.
If we want to encode systematically the information vector
$\uu=(2,0,1)$ (or $u(x)=2+x^2$),
we have to obtain first the residue of dividing
$x^5u(x)=2x^5+x^7$ by $g(x)$. This residue is
$r(x)=2+x+2x^2$. Hence, the output of the encoder is
$c(x)=u(x)-x^kr(x)=2+x^2+x^3+2x^4+x^5$. In vector form,
this gives $\uc=(2\;0\;1\;1\;2\;1\;0\;0)$.\qed
}
\end{ex}

In the next section, we define the very important
family of Reed Solomon codes.

\pr

\begin{prob}
\label{pr2.10'}
{\em
Let $\C$ be an $[n,k,d]$
linear code over $GF(q)$. Prove that Lemma~\ref{lemma3},
Corollary~\ref{cor2} and
the Singleton bound (Corollary~\ref{cor1}) for
binary codes also hold in this case.
Is the same true for the Hamming bound given by Lemma~\ref{lemma4}?
If not, give an appropriate version of the Hamming bound for $\C$.
}\end{prob}
 
\begin{prob}
\label{pr2.10}
{\em
Let $\C$ be a cyclic code over $GF(q)$. Prove that there is
a (monic) polynomial $h(x)$ of degree $k$ such that, for every
$c(x)\in\C$, $c(x)h(x)\equiv 0\bmod (x^n-1)$.
The polynomial $h(x)$ is called the parity check polynomial
of the code $\C$.
}\end{prob}
 
\begin{prob}
\label{pr2.11}
{\em
Given a cyclic code $\C$ with generator polynomial $g(x)$
and parity check polynomial $h(x)$, find a generator matrix and
a parity check matrix for the code.
}\end{prob}
 
\begin{prob}
\label{pr2.12}
{\em
Find all the cyclic codes of length 4 over $GF(3)$.
}\end{prob}
 
\begin{prob}
\label{pr2.13}
{\em
Consider the cyclic code $\C$
over $GF(2)$ with generator polynomial
$g(x)=1+x+x^3$ and length 7.
 
\begin{enumerate}
 
\item Verify that $\C$ is cyclic (i.e., $g$ divides $x^7-1$).
 
\item Find a generator matrix and a parity check matrix for $\C$.
 
\item Find the minimum distance of $\C$.
 
\item Encode systematically the information vector 1011 using
Algorithm~\ref{alg2.1}.
 
\end{enumerate}
 
}\end{prob}
 
\begin{prob}
\label{pr2.13'}
{\em
Consider the $[8,5]$ cyclic code over $GF(3)$ generated by
$g(x)=1+x+x^3$.
 
\begin{enumerate}
 
\item Prove that the code is in effect cyclic.
 
\item Prove that the code is the dual of the one given in
Example~\ref{ex2.0}.
 
\item Encode systematically the information vector 21011.
 
\end{enumerate}
 
}\end{prob}

\sol
\vspace{.8cm}
 
{\bf Problem~\ref{pr2.10'}}
 
Let us prove Lemma~\ref{lemma3} in this general case.
Denote the columns of $H$ $\uc_0,\uc_1,\ldots,\uc_{n-1}$.
Assume that columns $0\leq i_1<i_2<\ldots <i_m\leq n-1$ are linearly
dependent, where $m\leq d-1$. Hence, there exist
$a_1,a_2,\ldots,a_m$ in $GF(q)$ such that
$\sum_{l=1}^m\,a_l\uc_{i_l}=\0$, where $\0$ denotes the all-zero column.
 
Without loss of generality, assume that the $a_l$'s are non-zero.
Let $\uv$ be a vector of length $n$ and weight $m$
whose non-zero coordinates
are $a_1,a_2,\ldots,a_m$
in locations $i_1,i_2,\ldots,i_m$. Thus, we have
 
$$\uv\,H_T\;=\sum_{l=1}^m\,a_l\uc_{i_l}\;=\0;$$
 
hence $\uv$ is in $\C$. But $\uv$ has weight $m\leq d-1$, contradicting
the fact that $\C$ has minimum distance $d$.
 
Corollaries~\ref{cor2} and~\ref{cor1} (Singleton Bound) are
analogous to the binary case.
 
The Hamming bound, though, does not look exactly the same.
Let us denote by
$V_q(r)$ the number of elements in a sphere of radius $r$ whose
center is an element in $GF(q)^n$.
An easy counting argument gives
 
\begin{equation}
\label{eq8'}
V_q(r)\,=\,\sum_{i=0}^r\,{n\choose i}(q-1)^i.
\end{equation}
 
Notice that~(\ref{eq8'}) generalizes the case $q=2$ given
by~(\ref{eq8}).
The Hamming bound in the general case then becomes:
 
\begin{equation}
\label{eq9''}
n-k\,\geq\,\log_q V_q\left(\lf (d-1)/ 2\rf\right).
\end{equation}
 
The proof of~(\ref{eq9''}) is similar to the proof of
Lemma~\ref{lemma4}.
 
\vspace{.8cm}
 
{\bf Problem~\ref{pr2.10}}
 
Let $g(x)$ be the generator polynomial of $\C$. Since $g$ divides
$x^n-1$, there is an $h(x)$ such that $g(x)h(x)=x^n-1$.
Let $c(x)\in\C$. By Theorem~\ref{cyclic},
$c(x)=u(x)g(x)$ for a certain $u(x)$. Hence,
 
$$c(x)h(x)=u(x)g(x)h(x)=u(x)(x^n-1)\equiv 0\;(\bmod\; x^n-1).$$
 
\vspace{.8cm}
 
{\bf Problem~\ref{pr2.11}}
 
Let $g(x)=g_0+g_1x+\cdots +g_{n-k-1}x^{n-k-1}+g_{n-k}x^{n-k}$ and
$h(x)=h_0+h_1x+\cdots +h_{k-1}x^{k-1}+h_kx^k$.
Notice that the $k$ codewords
$g(x),xg(x),x^2g(x),\ldots,x^{k-1}g(x)$ are linearly independent.
Since $\C$ has dimension $k$, they form a basis for the code and
can be taken as the rows of a generator matrix.
 
If we write the matrix explicitly, we obtain
 
$$G\;=\;\left.
\left(
\begin{array}{l}
\begin{array}{llllllllll}
g_0 &g_1&g_2&\ldots &g_{n-k}
&0&0&0&\ldots &0\\
0&g_0 &g_1&\ldots &g_{n-k-1}&g_{n-k}
&0&0&\ldots &0 \\
\vdots&
\vdots&
\vdots&
\ddots&
\vdots&
\vdots&
\vdots&
\vdots&
\ddots&
\vdots
\end{array}\\
\begin{array}{lllllllllll}
0&0&0&\ldots &\ldots &0&g_0&g_1&\ldots &\ldots &g_{n-k}
\end{array}
\end{array}
\right)\right\}k
.$$
 
If the parity check polynomial $h(x)=(x^n-1)/g(x)$ is given by
$h(x)=h_0+h_1x+h_2x^2+\cdots +h_kx^k$,
a parity check matrix for $\C$ is given by the following matrix:
 
$$H\;=\;\left.
\left(
\begin{array}{l}
\begin{array}{llllllllll}
h_k&h_{k-1}&h_{k-2}&\ldots &h_0&0&0&0&\ldots &0\\
0&h_k&h_{k-1}&\ldots &h_1&h_0&0&0&\ldots &0\\
\vdots&
\vdots&
\vdots&
\ddots&
\vdots&
\vdots&
\vdots&
\vdots&
\ddots&
\vdots
\end{array}\\
\begin{array}{lllllllllll}
0&0&0&\ldots &\ldots &0&h_k&h_{k-1}&\ldots &\ldots &h_0
\end{array}
\end{array}
\right)\right\}n-k
.$$
 
If we denote by $\th(x)$ the polynomial
$\th(x)=h_k+h_{k-1}x+h_{k-2}x^2+\cdots +h_1x^{k-1}+h_0x^k$, the rows of
$H$ are given by the polynomials $x^j\th(x)$, $0\leq j\leq n-k-1$.
 
It remains to be proved that $H$ is in effect a parity check matrix
for $\C$. To this end, we have to show that any row in $H$
is orthogonal to any row in $G$. We denote the inner product
of polynomials with the symbol ``$\cdot$", to differentiate it from
polynomial product. We have to show that
$x^ig(x)\cdot x^j\th(x)=0$ for $0\leq i\leq k-1$, $0\leq j\leq n-k-1$.
If $i\leq j$,
$x^ig(x)\cdot x^j\th(x)=
g(x)\cdot x^{j-i}\th(x)$, while if
$j\leq i$,
$x^ig(x)\cdot x^j\th(x)=
x^{i-j}g(x)\cdot \th(x)$.
Hence, it is enough to show that the first row in $G$ is orthogonal
to every row in $H$, and that the first row in $H$ is orthogonal
to every row in $G$.
 
Notice that for $0\leq j\leq n-k-1$,
 
\begin{equation}
\label{inner}
g(x)\cdot x^j\th(x)=g_jh_k+g_{j+1}h_{k-1}+\cdots g_{n-k}h_{j+2k-n}.
\end{equation}
 
Since $x^n-1=
g(x)h(x)=\sum_{l=0}^n(\sum_{\nu+\mu=l}g_{\nu}h_{\mu})x^l$,
in particular,
$\sum_{\nu+\mu=k+j}g_{\nu}h_{\mu}=0$ for $0\leq j\leq n-k-1$.
But this is the sum appearing in~(\ref{inner}), so
$g(x)$ and $x^j\th(x)$ are orthogonal for $0\leq j\leq n-k-1$.
 
In a completely analogous way, we prove that $x^ig(x)$ and
$\th(x)$ are orthogonal for $0\leq i\leq k-1$.
 
\vspace{.8cm}
 
{\bf Problem~\ref{pr2.12}}
 
Notice that $x^4-1=(x-1)(x+1)(x^2+1)$, so, excluding trivial cases
(i.e., $(GF(3)^4$ and the 0-code), the codes are generated
by the factors of $x^4-1$. They are:
 
\begin{enumerate}
 
\item The $[4,3]$ code generated by $x-1=x+2$.
 
\item The $[4,3]$ code generated by $x+1$.
 
\item The $[4,2]$ code generated by $x^2+1$.
 
\item The $[4,2]$ code generated by $x^2-1=x^2+2$.
 
\item The $[4,1]$ code generated by $(x-1)(x^2+1)=x^3+2x^2+x+2$.
 
\item The $[4,1]$ code generated by $(x+1)(x^2+1)=x^3+x^2+x+1$.
 
\end{enumerate}
 
\vspace{.8cm}
 
{\bf Problem~\ref{pr2.13}}
 
\begin{enumerate}
 
\item We easily verify that
$1+x^7=(1+x+x^3)(1+x+x^2+x^4)$ over $GF(2)$, so
$g(x)$ divides $x^7+1$ (or $x^7-1$) and the code $\C$
is a cyclic $[7,4]$ code.
 
\item By Problem \ref{pr2.11}, a generator matrix for $\C$ is given
by
 
$$G=\left( \begin{array}{ccccccc}
1&1&0&1&0&0&0\\
0&1&1&0&1&0&0\\
0&0&1&1&0&1&0\\
0&0&0&1&1&0&1
\end{array}\right),$$
 
while a parity check matrix for $\C$ is given by
 
$$H=\left( \begin{array}{ccccccc}
1&0&1&1&1&0&0\\
0&1&0&1&1&1&0\\
0&0&1&0&1&1&1
\end{array}\right).$$
 
\item By observing the parity check matrix $H$ above, we see that
the columns of $H$ are all the possible vectors of length 3 over
$GF(2)$, hence, $\C$ is equivalent to a Hamming code and has
minimum distance 3.
 
\item In polynomial form, 1011 corresponds to
$u(x)=1+x^2+x^3$. The residue of dividing $x^3(1+x^2+x^3)$ by
$g(x)=1+x+x^3$ is $r(x)=1$, so, the encoded polynomial is
$u(x)-x^4r(x)=1+x^2+x^3+x^4$. In vector form, this corresponds
to codeword 1011100.
 
\end{enumerate}
 
\vspace{.8cm}
 
{\bf Problem~\ref{pr2.13'}}
 
\begin{enumerate}
 
\item We verify that
$x^8-1=2+x^8=(1+x+x^3)(2+x+2x^2+2x^3+x^5)$, so $g(x)$ divides
$x^8-1$ and the code is cyclic.
 
\item By Problem \ref{pr2.11}, it is enough to observe that
$g(x)=\th(x)$, where $h(x)$ is the parity check
polynomial of the $[8,3]$ code of Example~\ref{ex2.0}, since the
parity check matrix of one is the generator matrix of the other.
 
\item In polynomial form, 21011 corresponds to
$u(x)=2+x+x^3+x^4$. The residue of dividing $x^3(2+x+x^3+x^4)$ by
$g(x)=1+x+x^3$ is $r(x)=1+2x+2x^2$,
so, the encoded polynomial is
$u(x)-x^5r(x)=2+x+x^3+x^4+2x^5+x^6+x^7$.
In vector form, this corresponds
to codeword 21011211.
 
\end{enumerate}
 
\section{Reed Solomon Codes}
\label{sec2.4}

Throughout this section, the codes considered are over the field $GF(q)$,
where $q>2$. Let $\al$ be a primitive element in $GF(q)$, i.e.,
$\al^{q-1}=1$, $\al^i\neq 1$ for $i\not\equiv 0\bmod q-1$.
A Reed Solomon (RS) code of length $n=q-1$ and dimension
$k$ is the cyclic code generated by

$$g(x)=(x-\al )(x-\al^2)\ldots (x-\al^{n-k-1})(x-\al^{n-k}).$$

Since each $\al^i$ is a root of unity, $x-\al^i$ divides $x^n-1$,
hence $g$ divides $x^n-1$ and the code is cyclic.

An equivalent way of describing a RS code, is as the set of polynomials
over $GF(q)$
of degree $\leq n-1$ with roots $\al ,\al^2,\ldots,\al^{n-k}$, i.e.,
$F$ is in the code if and only if $\deg (F)\leq n-1$ and
$F(\al)=F(\al^2)=\ldots =F(\al^{n-k})=0$.

This property allows us immediately to find a parity check matrix
for a RS code. Say that $F(x)=F_0+F_1x+\ldots+F_{n-1}x^{n-1}$ is in
the code. Let $1\leq i\leq n-k$, then

\begin{equation}
\label{parity}
F(\al^i)=F_0+F_1\al^i+\ldots+F_{n-1}\al^{i(n-1)}=0.
\end{equation}

In other words, (\ref{parity}) tells us that
codeword $(F_0,F_1,\ldots,F_{n-1})$ is orthogonal to the vectors
$(1,\al^i,\al^{2i},\ldots,\al^{i(n-1)})$, $1\leq i\leq n-k$.
Hence these vectors are the rows of a parity check matrix for the
RS code. The parity check matrix of an $[n,k]$ RS code over $GF(q)$
is then

\begin{equation}
\label{matrix}
H=\left(
\begin{array}{ccccc}
1&\al  &\al^2&\ldots &\al^{n-1}\\
1&\al^2&\al^4&\ldots &\al^{2(n-1)}\\
\vdots&\vdots&\vdots&\ddots&\vdots\\
1&\al^{n-k}&\al^{(n-k)2}&\ldots &\al^{(n-k)(n-1)}
\end{array}\right)
\end{equation}

In order to show that $H$ is in fact a parity check matrix, we need
to prove that the rows of $H$ are linearly independent.
The next lemma provides an even stronger result.

\begin{lemma}
\label{MDS}
{\em
Any set of $n-k$ columns in matrix $H$ defined by~(\ref{matrix}) is
linearly independent.
}
\end{lemma}

\pf Take a set $0\leq i_1< i_2<\ldots <i_{n-k}\leq n-1$ of columns
of $H$. Denote $\al^{i_j}$ by $\al_j$, $1\leq j\leq n-k$.
Columns $i_1,i_2,\ldots,i_{n-k}$ are linearly independent if and only
if their determinant is non-zero, i.e., if and only if

\begin{equation}
\label{vander}
\det\left(
\begin{array}{cccc}
\al_1&\al_2&\ldots &\al_{n-k}\\
(\al_1)^2&(\al_2)^2&\ldots &(\al_{n-k})^2\\
\vdots&\vdots&\ddots&\vdots\\
(\al_1)^{n-k}&(\al_2)^{n-k}&\ldots &(\al_{n-k})^{n-k}
\end{array}\right)\neq 0.
\end{equation}

Let

\begin{equation}
\label{vander2}
V(\al_1,\al_2,\ldots,\al_{n-k})=
\det\left(
\begin{array}{cccc}
1&1&\ldots&1\\
\al_1&\al_2&\ldots &\al_{n-k}\\
\vdots&\vdots&\ddots&\vdots\\
(\al_1)^{n-k-1}&(\al_2)^{n-k-1}&\ldots &(\al_{n-k})^{n-k-1}
\end{array}\right)
\end{equation}

We call the determinant
$V(\al_1,\al_2,\ldots,\al_{n-k})$ a
{\em Vandermonde determinant}:
it is the determinant of
an $(n-k)\times (n-k)$ matrix whose rows are the powers of
vector $\al_1,\al_2,\ldots,\al_{n-k}$, the powers running
from 0 to $n-k-1$.
By properties of determinants, if we consider the determinant
in~(\ref{vander}), we have

\begin{equation}
\label{vander3}
\det\left(
\begin{array}{cccc}
\al_1&\al_2&\ldots &\al_{n-k}\\
(\al_1)^2&(\al_2)^2&\ldots &(\al_{n-k})^2\\
\vdots&\vdots&\ddots&\vdots\\
(\al_1)^{n-k}&(\al_2)^{n-k}&\ldots &(\al_{n-k})^{n-k}
\end{array}\right)=
\al_1\al_2\ldots \al_{n-k}
V(\al_1,\al_2,\ldots,\al_{n-k}).
\end{equation}

Hence, by~(\ref{vander}) and ~(\ref{vander3}), since the
$\al_j$'s are non-zero, it is enough to prove that\\
$V(\al_1,\al_2,\ldots,\al_{n-k})\neq 0$.
By Problem~\ref{pr2.14'}, we have that

\begin{equation}
\label{monde}
V(\al_1,\al_2,\ldots,\al_{n-k})=\prod_{1\leq i<j\leq n-k}
(\al_j-\al_i).
\end{equation}

Since $\al$ is a primitive element in $GF(q)$, its powers $\al^l$,
$0\leq l\leq n-1$ are distinct. In particular, the $\al_i$'s,
$1\leq i\leq n-k$ are distinct, hence, the product at the right
hand side of~(\ref{monde}) is non-zero.\qed

\begin{cor}
\label{MDSs}
{\em
An $[n,k]$ RS code has minimum distance $n-k+1$.
}
\end{cor}

\pf Let $H$ be the parity check matrix of the RS code defined
by~(\ref{matrix}). Notice that, since {\em any} $n-k$ columns in $H$ are
linearly independent, $d\geq n-k+1$ by Lemma~\ref{lemma3} (see
Problem~\ref{pr2.10'}).

On the other hand, $d\leq n-k+1$ by the Singleton bound
(Corollary~\ref{cor1} and Problem~\ref{pr2.10'}), so we have
equality.\qed

Since RS codes meet the Singleton bound with equality, they are MDS.
We have seen that in the binary case, the only MDS codes were
trivial ones (see Problem~\ref{pr1.7}).

\begin{ex}
\label{ex4.1}
{\em
Consider the $[7,3,5]$ RS code over $GF(8)$, where $GF(8)$ is given
by Table~\ref{table1}. The generator polynomial is

$$g(x)=(x-\al)(x-\al^2)(x-\al^3)(x-\al^4)=
\al^3+\al x+x^2+\al^3x^3+x^4.$$

Assume that we want to encode the 3 byte vector $\uu=101\;001\;111$.
Writing the bytes as powers of $\al$ in polynomial form,
we have $u(x)=\al^6+\al^2x+\al^5x^2$.

In order to encode $u(x)$, we perform

$$u(x)g(x)=
\al^2+\al^4x+\al^2x^2+\al^6x^3+\al^6x^4+\al^4x^5+\al^5x^6.$$

In vector form the output of the encoder is given by the 7 bytes
$0 0 1\; 0 1 1\; 0 0 1\; 1 0 1\; 1 0 1\; 0 1 1\; 1 1 1$.

If we encode $u(x)$ using a systematic encoder (Algorithm~\ref{alg2.1}),
then the output of the encoder is

$$\al^6+\al^2x+\al^5x^2+\al^6x^3+\al^5x^4+\al^4x^5+\al^4x^6,$$

which in vector form is
$ 1 0 1\; 0 0 1\; 1 1 1 \;1 0 1\;1 1 1\;0 1 1\;0 1 1$.\qed
}
\end{ex}

Next we make some observations:

\begin{enumerate}

\item
The definition given above for an $[n,k]$
Reed Solomon code states that
$F(x)$ is in the code if and only if
it has as roots the powers
$\al,\al^2,\ldots,\al^{n-k}$ of a primitive element $\al$.
However, it is enough to state that $F$ has as roots a set of
{\em consecutive} powers of $\al$, say,
$\al^m,\al^{m+1},\ldots,\al^{m+n-k-1}$, where $0\leq m\leq n-1$.
Although our definition (i.e., $m=1$) gives the most usual setting
for RS codes, often engineering reasons may determine different
choices of $m$.
It is easy to verify that with the more general definition of RS codes,
the minimum distance remains $n-k+1$ (Problem~\ref{pr2.16'}).

\item Our definition also assumed that $\al$ is a primitive element
in $GF(q)$ and $n=q-1$. But we really don't need
this assumption either.
If $\al$ is not primitive, $\al\neq 1$, then $\al$ has order $n$,
where $n$ divides $q-1$ and $1<n<q-1$. In this case, we can define
an $[n,k]$ RS code in a completely analogous way to
the case in which $\al$ is primitive. These codes will be shorter.
Again, there may be good engineering reasons to choose a non-primitive
$\al$. If $\al$ is a primitive element in $GF(q)$, we call the
RS code defined using consecutive powers of $\al$ a {\em primitive}
RS code. If $\al$ is not primitive, the RS code is called non-primitive.

\item Given an $[n,k]$ RS code,
there is an easy way to
shorten it and obtain an $[n-l,k-l]$ code for $l<k$.
In effect, if we have only
$k-l$ bytes of information, we add $l$ zeroes in order to
obtain an information string of length $k$. We then find the $n-k$
redundant bytes using a systematic encoder. When writing, of course,
the $l$ zeroes are not written, so we have an $[n-l,k-l]$ code, called
a shortened RS code. It is easy to verify that shortened RS codes are
also MDS. Again, engineering reasons may determine that the length of
a block may be too long at $n=q-1$, so a shortened version of a RS
code may be preferable.

\end{enumerate}

We have defined RS codes, proven that they are MDS and showed how to
encode them systematically. The next step, to be developed in the
next sections, is decoding them.

\pr

\begin{prob}
\label{pr2.14'}
{\em
Let $\al_1,\al_2,\ldots,\al_m$ be elements in a field and
$V(\al_1,\al_2,\ldots,\al_m)$ their Vandermonde determinant.
Prove that
 
$$V(\al_1,\al_2,\ldots,\al_m)=\prod_{1\leq i<j\leq m}(\al_j-\al_i).$$
 
}\end{prob}
 
\begin{prob}
\label{pr2.14}
{\em
Let $\al$ be a primitive element in $GF(q)$ and $n=q-1$.
Prove that, for $s\not\equiv 0\bmod n$,
$\sum_{i=0}^{n-1}\al^{is}=0$ and for
$s\equiv 0\bmod n$,
$\sum_{i=0}^{n-1}\al^{is}=n$.
}\end{prob}
 
\begin{prob}
\label{pr2.15}
{\em
Consider a $[15,9]$ RS code over $GF(16)$, where
$GF(16)$ was constructed in Problem~\ref{pr2.8}.
Encode systematically the polynomial
$u(x)=\al^3+\al^9x^2+\al^7x^3+\al^5x^4+\al^{10}x^6+\al^2x^7+\al^{12}x^8$.
}\end{prob}
 
\begin{prob}
\label{pr2.16}
{\em
Consider an $[8,4]$ RS code over $GF(9)$, where
$GF(9)$ was constructed in Problem~\ref{pr2.9}.
Encode systematically the polynomial
$u(x)=\al^2+\al^2x+\al^7x^2+\al^3x^3$.
}\end{prob}
 
\begin{prob}
\label{pr2.16'}
{\em
Verify that, if we define, more generally, an $[n,k]$
RS code as the set
of polynomials of degree $\leq n-1$ having as roots the
consecutive powers $\al^m,\al^{m+1},\ldots,\al^{m+n-k-1}$, the
minimum distance of the code is $n-k+1$.
}\end{prob}
 
\begin{prob}
\label{pr2.17}
{\em
Write a computer program that encodes systematically an information
polynomial of degree $\leq k-1$ into an $[n,k]$ RS code.
}\end{prob}

\sol
\vspace{.8cm}

{\bf Problem~\ref{pr2.14'}}
 
We prove the result by induction on $m$.
If $m=2$, it is clear that $V(\al_1,\al_2)=\al_2-\al_1$.
So, assume that the result is true for $m\geq 2$, let's prove that
it is true for $m+1$. Replacing $\al_1$ by $x$ in
$V(\al_1,\al_2,\ldots,\al_{m+1})$, we obtain a polynomial of degree $m$
on $x$, i.e.,
 
\begin{eqnarray}
\label{van}
f(x)&=&\det\left(
\begin{array}{cccc}
1&1&\ldots &1\\
x&\al_2&\ldots &\al_{m+1}\\
\vdots&\vdots &\ddots &\vdots\\
x^m&(\al_2)^m&\ldots &(\al_{m+1})^m
\end{array}\right)
\end{eqnarray}
 
Notice that, if we replace $x$ by $\al_i$ in~(\ref{van}),
$2\leq i\leq m+1$, we have
a repeated column in the matrix;
hence, its determinant is 0. In other
words, the elements $\al_2,\al_3,\ldots,\al_{m+1}$ are the $m$
(different) roots of $f$. So, we can write
 
\begin{eqnarray}
\label{van2}
f(x)&=&C(x-\al_2)(x-\al_3)\ldots (x-\al_{m+1})=
(-1)^mC\prod_{j=2}^{m+1}(\al_j-x),
\end{eqnarray}
 
where $C$ is the lead coefficient of $f$.
 
We also notice that
$f(\al_1)=V(\al_1,\al_2,\ldots,\al_{m+1})$.
 
By properties of
determinants, the
lead coefficient $C$ is equal to
 
\begin{eqnarray}
\label{van3}
C&=&(-1)^m
\det\left(
\begin{array}{cccc}
1&1&\ldots &1\\
\al_2&\al_3&\ldots &\al_{m+1}\\
\vdots&\vdots &\ddots &\vdots\\
(\al_2)^{m-1}&(\al_e)^{m-1}&\ldots &(\al_{m+1})^{m-1}
\end{array}\right)
\end{eqnarray}
 
But the determinant in the right is $V(\al_2,\al_3,\ldots,\al_{m+1})$,
so, by induction, (\ref{van3}) becomes
 
\begin{eqnarray}
\label{van4}
C&=&(-1)^m\prod_{2\leq i<j\leq m+1}(\al_j-\al_i).
\end{eqnarray}
 
Replacing in~(\ref{van2}) $x$ by $\al_1$ and $C$ by the value obtained
in~(\ref{van4}), we obtain the result.
 
\vspace{.8cm}
 
{\bf Problem~\ref{pr2.14}}
 
If $s\equiv 0\;(\bmod\; n)$, $\al^{is}\eq(\al^s)^i\eq 1$, so
$\sum_{i=0}^{n-1}\al^{is}=n$. So, assume that
$s\not\equiv 0\;(\bmod\; n)$; hence, since $\al$ is primitive,
$\al^s\neq 1$. Now,
 
$$\sum_{i=0}^{n-1}\al^{is}=
\sum_{i=0}^{n-1}(\al^s)^i={(\al^s)^n-1\over \al^s-1}=0,$$
 
since $(\al^s)^n=1$.
 
\vspace{.8cm}
 
{\bf Problem~\ref{pr2.15}}
 
The generator polynomial is 

\begin{eqnarray*}
\prod _{i=1}^6(x-\al^i)&=&\al^6+\al^9x+\al^6x^2+\al^4x^3+
\al^{14}x^4+\al^{10}x^5+x^6.
\end{eqnarray*}

Using Algorithm~\ref{alg2.1}, $u(x)$ is encoded as
 
$$c(x)=\al^3+\al^9x^2+\al^7x^3+
\al^5x^4+\al^{10}x^6+
\al^2x^7+\al^{12}x^8+
\al^{14}x^9+\al^{8}x^{10}+
\al^{11}x^{11}+\al^{4}x^{12}+
\al^{2}x^{13}.$$
 
\vspace{.8cm}
 
{\bf Problem~\ref{pr2.16}}
 
The generator polynomial is 

\begin{eqnarray*}
\prod _{i=1}^4(x-\al^i)&=&\al^2+\al^4x+\al^2x^2+\al^7x^3+x^4.
\end{eqnarray*}

Using
Algorithm~\ref{alg2.1}, $u(x)$ is encoded as
 
$$c(x)=\al^2+\al^2x+\al^7x^2+
\al^3x^3+\al^{3}x^4+
\al^4x^5+x^6+
\al x^7.$$
 
In vector form, this corresponds to vector
 
$$\uc\;=\;(1 2\;1 2\;1 1\;2 2\;2 2\;2 0\;1 0\;0 1).$$
 
\vspace{.8cm}
 
{\bf Problem~\ref{pr2.16'}}
 
Let $\C$ be the code formed by
the set
of polynomials of degree $\leq n-1$ having as roots the
consecutive powers $\al^m,\al^{m+1},\ldots,\al^{m+n-k-1}$,
$\al$ a primitive element.
Then, a parity check matrix for the code is given by
 
\begin{equation}
\label{matrix1}
H=\left(
\begin{array}{ccccc}
1&\al^m&(\al^m)^2&\ldots &(\al^m)^{n-1}\\
1&\al^{m+1}&(\al^{m+1})^2&\ldots &(\al^{m+1})^{n-1}\\
\vdots&\vdots&\vdots&\ddots&\vdots\\
1&\al^{m+n-k-1}&(\al^{m+n-k-1})^2&\ldots &(\al^{m+n-k-1})^{n-1}
\end{array}\right)
\end{equation}
 
We show now that
any set of $n-k$ columns in matrix $H$ defined by~(\ref{matrix1}) is
linearly independent.
 
In effect,
take a set $0\leq i_1< i_2<\ldots <i_{n-k}\leq n-1$ of columns
of $H$. Denote $\al^{i_j}$ by $\al_j$, $1\leq j\leq n-k$.
Columns $i_1,i_2,\ldots,i_{n-k}$ are linearly independent if and only
if their determinant is non-zero, i.e., if and only if
 
\begin{equation}
\label{vander1}
\det\left(
\begin{array}{cccc}
\al_1^m&\al_2^m&\ldots &\al_{n-k}^m\\
\al_1^{m+1}&\al_2^{m+1}&\ldots &\al_{n-k}^{m+1}\\
\vdots&\vdots&\ddots&\vdots\\
\al_1^{m+n-k-1}&\al_2^{m+n-k-1}&\ldots &\al_{n-k}^{m+n-k-1}
\end{array}\right)\neq 0.
\end{equation}
 
But this determinant is equal to
 
$$\al_1^m\al_2^m\ldots \al_{n-k}^mV(\al_1,\al_2,\ldots,\al_m),$$
 
which is different from 0 since the Vandermonde determinant
$V(\al_1,\al_2,\ldots,\al_m)$ is different from 0.
 
\section{Decoding of RS codes: the key equation}
\label{sec2.5}

Through this section $\C$ denotes an
$[n,k]$ RS code (unless otherwise stated).
Assume that a codeword $F(x)=\sum_{i=0}^{n-1}F_ix^i$ in
$\C$ is transmitted and
a word $R(x)=\sum_{i=0}^{n-1}R_ix^i$ is received; hence,
$F$ and $R$ are related by an error vector
$E(x)=\sum_{i=0}^{n-1}E_ix^i$, where $R(x)=F(x)+E(x)$.
The decoder will attempt to find $E(x)$.

Let us start by computing the syndromes.
For $1\leq j\leq n-k$, we have

\begin{equation}
\label{eq2.2}
S_j=R(\al^j)=\sum_{i=0}^{n-1}R_i\al^{ij}=
\sum_{i=0}^{n-1}E_i\al^{ij}
\end{equation}

Before proceeding further,
consider~(\ref{eq2.2}) in a particular case.

Take the $[n,n-2]$ 1-byte correcting RS code.
In this case, we have two syndromes $S_1$ and $S_2$, so, if exactly
one error has occurred, say in location $i$,
by~(\ref{eq2.2}), we have

\begin{equation}
\label{eq2.2''}
S_1=E_i\al^i\;{\rm and}\;S_2=E_i\al^{2i}.
\end{equation}

Hence, $\al^i=S_2/S_1$, so we can determine the location $i$ in error.
The error value is $E_i=(S_1)^2/S_2$.

\begin{ex}
\label{ex5.1}
{\em
Consider the $[7,5,3]$ RS code over $GF(8)$, where $GF(8)$ is given
by Table~\ref{table1}.

Assume that we want to decode the received vector

$$\ur=(1 0 1\; 0 0 1\; 1 1 0 \;0 0 1\;0 1 1\;0 1 0\;1 0 0),$$

which in polynomial form is

$$R(x)=\al^6+\al^2x+\al^3x^2+\al^2x^3+\al^4x^4+\al x^5+x^6.$$

Evaluating the syndromes, we obtain $S_1=R(\al)=\al^2$ and
$S_2=R(\al^2)=\al^4$.
Thus, $S_2/S_1=\al^2$, meaning that location 2 is in error.
The error value is $E_2=(S_1)^2/S_2=(\al^2)^2/\al^4=1$, which
in vector form is $100$. The output of the decoder is then

$$\uc=(1 0 1\; 0 0 1\; 0 1 0 \;0 0 1\;0 1 1\;0 1 0\;1 0 0),$$

which in polynomial form is

$$C(x)=\al^6+\al^2x+\al x^2+\al^2x^3+\al^4x^4+\al
x^5+x^6.$$
\qed
}
\end{ex}

Let $\E$ be the subset of $\{0,1,\ldots,n-1\}$ of locations in error,
i.e., $\E=\{l:E_l\neq 0\}$. With this notation, (\ref{eq2.2}) becomes

\begin{equation}
\label{eq2.2'}
S_j=\sum_{i\in\E}E_i\al^{ij}\;,\; 1\leq j\leq n-k.
\end{equation}

The decoder will
find the error set $\E$ and the
error values $E_i$ when the error correcting capability of the code
is not exceeded.
Thus, if $s$ is the number of errors and $2s\leq n-k$,
the system of equations given by~(\ref{eq2.2'}) has a unique solution.
However, this is a non-linear system, and it is very difficult to
solve it directly. We will study methods of transforming parts of the
decoding process into a linear problem.

In order to find the set of locations in error $\E$
and the corresponding error values $\{E_i:i\in\E\}$,
we define two polynomials. The first one is called
the {\em error locator polynomial}, which is the polynomial that has
as roots the values $\al^{-i}$, where $i\in\E$. We denote this
polynomial by $\si (x)$. Explicitly,

\begin{equation}
\label{eq2.7}
\si(x)=\prod_{i\in\E}(x-\al^{-i}).
\end{equation}

If somehow we can determine the polynomial $\si(x)$, by
finding its roots, we can obtain the set $\E$ of
locations in error.
Once we have the set of
locations in error, we need to find the errors
themselves. We define a second polynomial, called the {\em error
evaluator polynomial} and denoted by $\om(x)$, as follows:

\begin{equation}
\label{eq2.8}
\om(x)=\sum_{i\in\E}E_i\prod_{\scriptstyle
l\in\E\atop l\neq i}(x-\al^{-l}).
\end{equation}

Since an $[n,k]$ RS code corrects at most $(n-k)/2$ errors,
we assume that $|\E|=\deg(\si)\leq (n-k)/2$. Notice also that
$\deg(\om)\leq |\E|-1$, since $\om$ is a sum of polynomials of degree
$|\E|-1$. Moreover,

\begin{equation}
\label{eq2.9}
E_i={\om(\al^{-i})\over \si'(\al^{-i})},
\end{equation}

where $\si'$ denotes the (formal) derivative of $\si$
(see Problem~\ref{pr2.19'}).

Let us prove some of these facts in the following lemma:

\begin{lemma}
\label{lemma2.2}
{\em
The polynomials $\si(x)$ and $\om(x)$ are relatively prime, and
the error values $E_i$ are given by~(\ref{eq2.9}).
}
\end{lemma}

\pf In order to show that $\si(x)$ and $\om(x)$ are relatively prime,
it is enough to observe that they have no roots
in common. In effect, if $\al^{-j}$ is a root of $\si(x)$, then
$j\in\E$. By~(\ref{eq2.8}),

\begin{equation}
\label{eq2.10}
\om(\al^{-j})=\sum_{i\in\E}E_i
\prod_{\scriptstyle l\in\E\atop l\neq i}(\al^{-j}-\al^{-l})
=E_j\prod_{\scriptstyle l\in\E\atop l\neq j}(\al^{-j}-\al^{-l})\neq 0.
\end{equation}

Hence, $\si(x)$ and $\om(x)$ are relatively prime.

In order to prove~(\ref{eq2.9}), notice that

$$\si'(x)=\sum_{i\in\E}
\prod_{\scriptstyle l\in\E\atop l\neq i}(x-\al^{-l}),$$

hence,

\begin{equation}
\label{eq2.11}
\si'(\al^{-j})=\prod_{\scriptstyle l\in\E\atop l\neq j}
(\al^{-j}-\al^{-l}).
\end{equation}

By (\ref{eq2.10}) and (\ref{eq2.11}), (\ref{eq2.9}) follows.

\qed

The decoding methods of RS codes are based on finding the error
locator and the error evaluator polynomials. By finding the roots of
the error locator polynomial, we determine the locations in error,
while the errors themselves can be found using~(\ref{eq2.9}). We will
establish a relationship between $\si(x)$ and $\om(x)$, but first we
need to define a third polynomial, the syndrome polynomial.
We define the syndrome polynomial as the polynomial of degree
$\leq n-k-1$ whose coefficients are the $n-k$ syndromes.
Explicitly,

\begin{equation}
\label{eq2.3}
S(x)=S_1+S_2x+S_3x^2+\ldots +S_{n-k}x^{n-k-1}=
\sum_{j=0}^{n-k-1}S_{j+1}x^j.
\end{equation}

Notice that $R(x)$ is in $\C$ if and only if $S(x)=0$.

The next theorem gives the so called {\em key equation} for decoding
RS codes, and it establishes a fundamental relationship between
$\si(x)$, $\om(x)$ and $S(x)$.

\begin{theo}
\label{theo2.1}
{\em
There is a polynomial $\mu(x)$ such that
the error locator, the error evaluator and the syndrome
polynomials verify the following equation:

\begin{equation}
\label{eq2.13}
\si(x)S(x)=-\om(x)+ \mu(x) x^{n-k}.
\end{equation}

Alternatively, Equation~(\ref{eq2.13}) can be written as a congruence
as follows:

\begin{equation}
\label{eq2.12}
\si(x)S(x)\equiv -\om(x)\;(\bmod\; x^{n-k}),
\end{equation}
}
\end{theo}

\pf By~(\ref{eq2.3}) and (\ref{eq2.2'}), we have

\begin{eqnarray}
S(x)&=&\sum_{j=0}^{n-k-1}S_{j+1}x^j\nonumber\\
    &=&\sum_{j=0}^{n-k-1}\left(\sum_{i\in \E}E_i\al^{i(j+1)}\right)
x^j\nonumber\\
    &=&\sum_{i\in \E}E_i\al^i\sum_{j=0}^{n-k-1}(\al^ix)^j\nonumber\\
    &=&\sum_{i\in \E}E_i\al^i{(\al^ix)^{n-k}-1\over \al^ix-1}\nonumber\\
\label{eq2.6}
&=&\sum_{i\in \E}E_i{(\al^ix)^{n-k}-1\over x-\al^{-i}},
\end{eqnarray}

since $\sum_{l=0}^ma^l=(a^{m+1}-1)/(a-1)$ for $a\neq 1$
(Problem~\ref{pr2.18}).



Multiplying both sides of~(\ref{eq2.6}) by $\si(x)$, where $\si(x)$
is given by~(\ref{eq2.7}), we obtain

\begin{eqnarray*}
\si(x)S(x)&= & \sum_{i\in\E}E_i((\al^ix)^{n-k}-1)
\prod_{\scriptstyle l\in\E\atop l\neq i}(x-\al^{-l})\\
&= & -\sum_{i\in\E}E_i\prod_{\scriptstyle l\in\E\atop l\neq i}(x-\al^{-l})+
\left(\sum_{i\in\E}E_i\al^{i(n-k)}
\prod_{\scriptstyle l\in\E\atop l\neq i}(x-\al^{-l})\right)x^{n-k}\\
&= & -\om(x)+\mu(x)x^{n-k},
\end{eqnarray*}

since $\om(x)$ is given by~(\ref{eq2.8}). This completes the proof.\qed

The decoding methods for RS codes concentrate on solving the key
equation. In the next section we study the simplest (conceptually)
of these methods,
the Peterson-Gorenstein-Zierler decoder.

\pr

\begin{prob}
\label{pr2.18}
{\em
Prove that, for $a\neq 1$, $\sum_{i=0}^{n-1}a^i=(a^n-1)/(a-1)$.
}\end{prob}
 
\begin{prob}
\label{pr2.19}
{\em
Consider the $[15,13]$ RS code over $GF(16)$. Decode the received
word
 
\begin{eqnarray*}
R(x)&=&
\al^3+\al x+\al^6x^2+\al^5x^3+\al^8x^4+x^5
+\al^3x^6+\al^8 x^7+\al^6x^8+\al^6x^9+\al^3x^{10} \\
&&
+\al^4x^{11}
+\al^{12}x^{12}
+\al^{12}x^{13}
+\al^{13}x^{14}.
\end{eqnarray*}
}\end{prob}
 
\begin{prob}
\label{pr2.19'}
{\em
Given a polynomial $f(x)=a_0+a_1x+\cdots +a_mx^m$ with coefficients
over a field $F$, we define the (formal) derivative of $f$, denoted $f'$,
as the polynomial
 
$$f'(x)=a_1+2a_2x+\cdots +ma_mx^{m-1}.$$
 
\begin{enumerate}
 
\item If $f$ and $g$ are polynomials, prove that
$(f+g)'=f'+g'$ and $(fg)'=f'g+fg'$.
 
\item If the field $F$ has characteristic 2, find $f'$ and $f''$ for
the polynomial $f$ above.
 
\end{enumerate}
 
}\end{prob}
 
\begin{prob}
\label{pr2.20}
{\em
Let $\C$ be an $[n,k]$ RS code. Assume that $t$ erasures have occurred,
and a number of errors $s\leq (n-k-t)/2$. Let $\E_1$ be the set
of locations in error and $\E_2$ the set of erased locations
(notice, $\E_2$ is known). Let $\E=\E_1\cup \E_2$, and define
the error locator polynomial
 
\begin{equation}
\label{eq2.7'}
\si_1(x)=\prod_{i\in\E_1}(x-\al^{-i}),
\end{equation}
 
the erasure locator polynomial
 
\begin{equation}
\label{eq2.8'}
\si_2(x)=\prod_{i\in\E_2}(x-\al^{-i}),
\end{equation}
 
and the error-erasure evaluator polynomial
 
\begin{equation}
\label{eq2.8''}
\om(x)=\sum_{i\in\E}E_i\prod_{\scriptstyle
l\in\E\atop l\neq i}(x-\al^{-l}).
\end{equation}
 
Give an equivalent form of the key equation~(\ref{eq2.12}) for this case.
}\end{prob}
 
\begin{prob}
\label{pr2.81}
{\em
As in Problem~\ref{pr2.16'}, consider
an $[n,k]$ RS code as the set
of polynomials of degree $\leq n-1$ having as roots the
consecutive powers $\al^m,\al^{m+1},\ldots,\al^{m+n-k-1}$.
Give an equivalent form of the key equation~(\ref{eq2.12}) for this case.
Give also an equivalent form for errors and erasures, as in
Problem~\ref{pr2.20}.
}\end{prob}

\sol
\vspace{.8cm}
 
{\bf Problem~\ref{pr2.18}}
 
Notice that
 
$$(a-1)(1+a+\cdots a^{n-1})=
(a+a^2+\cdots a^n)-(1+a+a^2+\cdots a^{n-1})=a^n-1,$$
 
so the result follows.
 
\vspace{.8cm}
 
{\bf Problem~\ref{pr2.19}}
 
Evaluating the syndromes, we obtain
$S_1=R(\al)=\al^3=E_i\al^i$ and
$S_2=R(\al^2)=\al^7=E_i\al^{2i}$, $i$ the location in error,
$E_i$ the error value.
This gives, $\al^i=S_2/S_1=\al^4$, i.e., $i=4$. Also,
$E_4=(S_1)^2/S_2=\al^{14}$. Hence, symbol 4 has to be replaced by
$\al^8 -\al^{14}=\al^6$. If the information is carried in the
first 13 bytes, the output of the decoder is
 
\begin{eqnarray*}
U(x)&=&
\al^3+\al x+\al^6x^2+\al^5x^3+\al^6x^4+x^5
+\al^3x^6+\al^8 x^7+\al^6x^8+\al^6x^9+\al^3x^{10}\\
&&+\al^4x^{11}
+\al^{12}x^{12}
\end{eqnarray*}
 
\vspace{.8cm}
 
{\bf Problem~\ref{pr2.19'}}
 
\begin{enumerate}
 
\item Let $f(x)=\sum_{i=0}^ma_ix^i$ and
$g(x)=\sum_{i=0}^mb_ix^i$, so,
$(f+g)(x)
\sum_{i=0}^m\,(a_i+b_i)x^i$ and
$(f+g)'(x)=\sum_{i=0}^{m-1}(i+1)(a_{i+1}+b_{i+1})x^i=
(\sum_{i=0}^{m-1}(i+1)a_{i+1}x^i)+
(\sum_{i=0}^{m-1}(i+1)b_{i+1}x^i)=f'(x)+g'(x)$.
 
Given the linearity of the derivative with respect to the sum,
it is enough to prove the result for $f(x)=x^i$ and $g(x)=x^j$,
$0\leq i,j$. Notice that $(fg)'(x)=(x^{i+j})'=(i+j)x^{i+j-1}=
(ix^{i-1})x^j+x^i(jx^{j-1})=
(f(x))'g(x)+f(x)(g(x))'$.
 
\item Let $f(x)=a_0+a_1x+\cdots +a_mx^m$, where $a_i\in GF(2)$.
Then, $f'(x)=a_1+2a_2x+\cdots +ma_mx^{m-1}=
a_1+a_3x^2+a_5x^4+\cdots$, since, in a field of characteristic 2,
$2j=0$ and $2j+1=1$. Differentiating this first derivative,
we obtain $f''(x)=0$.
 
\end{enumerate}
 
\vspace{.8cm}

{\bf Problem~\ref{pr2.20}}
 
If $\si(x)=\si_1(x)\si_2(x)$, the key equation is the same, i.e.,
 
$$\si(x)S(x)\eq -\om(x)+\mu(x)x^{n-k}.$$
 
However, now $\si_2(x)$ is known and we have to find both
$\si_1(x)$ and $\om(x)$. Defining a generalized syndrome polynomial
of degree $n-k+|\E_2|$ as
 
$$\hS(x)=\si_2(x)S(x)=\left(\prod_{i\in\E_2}(x-\al^{-i})\right)S(x),$$
 
we have to solve now the modified key equation
 
\begin{equation}
\label{modkey}
\si_1(x)\hS(x)\eq -\om(x)+\mu(x)x^{n-k}.
\end{equation}
 
In this case, $\si_1(x)$ has degree $|\E_1|$ and $\om(x)$ has
degree $\leq |\E_1|+|\E_2|-1$.
 
\vspace{.8cm}
 
{\bf Problem~\ref{pr2.81}}
 
In order to find the set of locations in error $\E$
and the corresponding error values $\{E_i:i\in\E\}$,
we define again the error locator polynomial as given
by~(\ref{eq2.7}).
However, the error evaluator polynomial needs a slightly different
definition (in fact, it is a generalization of~(\ref{eq2.8})) as follows: 
 
\begin{equation}
\label{eq2.82}
\om(x)=\sum_{i\in\E}E_i\al^{(m-1)i}\prod_{\scriptstyle
l\in\E\atop l\neq i}(x-\al^{-l}).
\end{equation}
 
Since an $[n,k]$ RS code corrects at most $(n-k)/2$ errors,
we assume that $|\E|=\deg(\si)\leq (n-k)/2$. Notice also that
$\deg(\om)\leq |\E|-1$, since $\om$ is a sum of polynomials of degree
$|\E|-1$. Moreover,
 
\begin{equation}
\label{eq2.9'}
E_i={\om(\al^{-i})\al^{-(m-1)i}\over \si'(\al^{-i})}.
\end{equation}
 
Similarly to Lemma~\ref{lemma2.2}, we can prove that
the polynomials $\si(x)$ and $\om(x)$ are relatively prime, and
the error values $E_i$ are given by~(\ref{eq2.9'}).
Therefore, if we find $\si(x)$ and $\om(x)$,
we can determine the error locations and their values.

Now, similarly to~(\ref{eq2.3}),  
we define the syndrome polynomial as the polynomial of degree
$\leq n-k-1$ whose coefficients are the $n-k$ syndromes $S_i$,
$m\leq i\leq m+n-k-1$.
Explicitly,
 
\begin{equation}
\label{eq2.3'}
S(x)=S_m+S_{m+1}x+S_{m+2}x^2+\ldots +S_{m+n-k-1}x^{n-k-1}=
\sum_{j=0}^{n-k-1}S_{m+j}x^j.
\end{equation}
 
As before, $R(x)$ is in $\C$ if and only if $S(x)=0$
and $\E$ denotes the set of locations in error. Then,
 
\begin{eqnarray*}
S(x)&=&\sum_{j=0}^{n-k-1}S_{m+j}x^j\\
    &=&\sum_{j=0}^{n-k-1}\left(\sum_{i\in\E}E_i\al^{i(m+j)}\right)x^j\\
    &=&\sum_{i\in\E}E_i\al^{mi}\sum_{j=0}^{n-k-1}(\al^ix)^j\\
    &=&\sum_{i\in\E}E_i\al^{mi}{(\al^ix)^{n-k}-1\over \al^ix-1}\\
    &=&\sum_{i\in\E}E_i\al^{(m-1)i}{(\al^ix)^{n-k}-1\over x-\al^{-i}},
\end{eqnarray*}
 
since $\sum_{l=0}^ma^l=(a^{m+1}-1)/(a-1)$ for $a\neq 1$.
Multiplying both sides by $\si(x)$, we obtain

\begin{eqnarray*}
\si(x)S(x)&=&\sum_{i\in\E}E_i\al^{(m-1)i}((\al^ix)^{n-k}-1)
\prod_{l\in\E\atop l\neq i}(x-\al^{-l})\\
&=&-\sum_{i\in\E}E_i\al^{(m-1)i}\prod_{l\in\E\atop l\neq i}(x-\al^{-l})+
\mu(x)x^{n-k}\\
&=&-\om(x)+\mu(x)x^{n-k},
\end{eqnarray*}

therefore, the key equation, as given by
Theorem~\ref{theo2.1}, looks the same, but $\om(x)$ is now given
by~(\ref{eq2.82}) and the error values by~(\ref{eq2.9'}).

As far as erasures are concerned, the treatment is completely analogous
to the one in Problem~\ref{pr2.19'}, except that $\om(x)$ is given
by~(\ref{eq2.82}) and the error values by~(\ref{eq2.9'}).

\section{The Peterson-Gorenstein-Zierler decoder}
\label{sec2.6}

Consider the key equation~(\ref{eq2.13}).
Assume that $s$ errors have occurred, where $2s\leq n-k$. Hence,
the code can correct these $s$ errors.
Let $\si(x)=\si_0+\si_1x+\ldots +\si_{s-1}x^{s-1}+x^s$ and
$\om(x)=\om_0+\om_1x+\ldots +\om_{s-1}x^{s-1}$.
Let $s\leq j\leq n-k-1$. According to~(\ref{eq2.13}), the $j$th
coefficient of $\si(x)S(x)$ is 0. But this $j$th coefficient is
given by

\begin{equation}
\label{eq2.14}
\sum_{l=0}^s\si_lS_{j+1-l}=0\;,\;s\leq j\leq n-k-1.
\end{equation}

Since $\si_s=1$,
(\ref{eq2.14}), is equivalent to

\begin{equation}
\label{eq2.15}
\sum_{l=0}^{s-1}\si_lS_{j+1-l}=-S_{j-s+1}\;,\;s\leq j\leq n-k-1.
\end{equation}

In matrix form, (\ref{eq2.15}) gives

\begin{eqnarray}
\label{eq2.16}
\begin{array}{cc}
\left(
\begin{array}{llll}
S_2&S_3&\ldots &S_{s+1}\\
S_3&S_4&\ldots &S_{s+2}\\
\vdots & \vdots &\ddots &\vdots\\
S_{n-k-s+1}&S_{n-k-s+2}&\ldots &S_{n-k}
\end{array}
\right)&
\left(
\begin{array}{l}
\si_{s-1}\\
\si_{s-2}\\
\vdots\\
\si_0
\end{array}
\right)\end{array}&= &
\left(
\begin{array}{l}
-S_1\\
-S_2\\
\vdots\\
-S_{n-k-s}\\
\end{array}\right).
\end{eqnarray}

In order to solve~(\ref{eq2.16}), it is enough to take the first $s$ rows
in the matrix at the left (the remaining rows may be used for verification),
thus, we obtain

\begin{eqnarray}
\label{eq2.17}
\begin{array}{cc}
\left(
\begin{array}{llll}
S_2&S_3&\ldots &S_{s+1}\\
S_3&S_4&\ldots &S_{s+2}\\
\vdots & \vdots &\ddots &\vdots\\
S_{s+1}&S_{s+2}&\ldots &S_{2s}
\end{array}
\right)&
\left(
\begin{array}{l}
\si_{s-1}\\
\si_{s-2}\\
\vdots\\
\si_0
\end{array}
\right)\end{array}&= &
\left(
\begin{array}{l}
-S_1\\
-S_2\\
\vdots\\
-S_{s}\\
\end{array}\right).
\end{eqnarray}

Finding $\si(x)$ using~(\ref{eq2.17}) provides the basis for the
so called Peterson-Gorenstein-Zierler decoder.
Let

\begin{eqnarray}
\label{eq2.17'}
\bS_r&=&
\left(
\begin{array}{llll}
S_2&S_3&\ldots &S_{r+1}\\
S_3&S_4&\ldots &S_{r+2}\\
\vdots & \vdots &\ddots &\vdots\\
S_{r+1}&S_{r+2}&\ldots &S_{2r}
\end{array}
\right),
\end{eqnarray}

where $2r\leq n-k$. Since $s$ errors have occurred and this is within
the error-correcting capability of the code, $\bS_s$ is non-singular.
We will prove that $\bS_r$ is singular for $s<r\leq (n-k)/2$.
Hence, the decoder starts checking if $\bS_r$ is non-singular for
the largest possible $r$ (i.e., $r=\lf (n-k)/2\rf$).
When it finds an $r$ such that $\bS_r$
is non-singular, this $r$ gives the number
of errors $s$.
Then, (\ref{eq2.17}) can be solved simply by inverting
$\bS_s$, i.e.,

\begin{eqnarray}
\label{eq2.18}
\left(
\begin{array}{c}
\si_{s-1}\\
\si_{s-2}\\
\vdots\\
\si_0
\end{array}
\right)
&= &
\begin{array}{cc}
(\bS_s)^{-1}&
\left(
\begin{array}{c}
-S_1\\
-S_2\\
\vdots\\
-S_{s}\\
\end{array}\right).
\end{array}
\end{eqnarray}

Once we have obtained $\si(x)$, by~(\ref{eq2.13}),
we can compute $\om(x)$ by calculating
the coefficients $j$, $0\leq j\leq s-1$, of $\si(x)S(x)$ and
changing their sign.
We then find the error values using~(\ref{eq2.9}).

The roots of the polynomial $\si(x)$ are found using an exhaustive
search algorithm called {\em Chien search}. Once the roots are found,
we know the locations of errors. However, we must not forget that
if a root $\al^{-i}$ has been found, the error is in location $i$, not
in location $-i$. For instance, in $GF(256)$, if $\al^{85}$ is a root of
$\si(x)$, since $\al^{85}=\al^{-170}$, the error is in location 170.

Another possibility for finding the error values $E_l$, $1\leq l\leq s$,
once we have obtained
the $s$ error locations, is the following (the decoding process has
been transformed into a problem of correcting erasures only):
since the syndromes are given by

\begin{eqnarray}
\label{eq2.18'}
S_j&=& \sum_{l\in\E}\al^{jl}E_l\;,\;1\leq j\leq s,
\end{eqnarray}

this is a system of $s$ linear equations with $s$ unknowns, which
can be solved by inverting the matrix of coefficients
$(\al^{jl})\;,\;l\in\E\;,\;1\leq j\leq s$.

The next lemma proves that $\bS_r$ is singular for $s<r\leq (n-k)/2$.

\begin{lemma}
\label{sing}
{\em
Matrix $\bS_r$ given by~(\ref{eq2.17'}) is singular for $s<r\leq (n-k)/2$.
}\end{lemma}

\pf Let $s<r\leq (n-k)/2$.
Let the error set be $\E=\{i_1,i_2,\ldots,i_s\}$, and consider the errors
$E_{i_1},E_{i_2},\ldots,E_{i_s}$.
Consider the $r\times r$ matrices

\begin{eqnarray}
\label{mA}
A&=&\left(\begin{array}{llllllll}
1&1&\ldots &1&1&1&\ldots &1\\
\al^{i_1}&\al^{i_2}&\ldots &\al^{i_s}&0&0&\ldots &0\\
\al^{2i_1}&\al^{2i_2}&\ldots &\al^{2i_s}&0&0&\ldots &0\\
\vdots &\vdots &\vdots &\vdots &\vdots &\vdots &\vdots &\vdots \\
\al^{(r-1)i_1}&\al^{(r-1)i_2}&\ldots &\al^{(r-1)i_s}&0&0&\ldots &0
\end{array}\right)
\end{eqnarray}

and

\begin{eqnarray}
\label{mB}
B&=&\left(\begin{array}{llllllll}
E_{i_1}\al^{2i_1}&0                 &\ldots &0                 &0&0&\ldots &0\\
0                 &E_{i_2}\al^{2i_2}&\ldots &0                 &0&0&\ldots &0\\
\vdots            &\vdots&\vdots &\vdots &\vdots &\vdots &\vdots &\vdots \\
0                 &0                 &\ldots &E_{i_s}\al^{2i_s}&0&0&\ldots &0\\
0                 &0                 &\ldots &0                 &0&0&\ldots &0\\
\vdots            &\vdots&\vdots &\vdots &\vdots &\vdots &\vdots &\vdots \\
0                 &0                 &\ldots &0                 &0&0&\ldots &0
\end{array}\right).
\end{eqnarray}

Notice that

\begin{eqnarray}
\label{BAT}
BA^T&=&
\left(\begin{array}{llll}
E_{i_1}\al^{2i_1}&E_{i_1}\al^{3i_1}&\ldots &E_{i_1}\al^{(r+1)i_1}\\
E_{i_2}\al^{2i_2}&E_{i_2}\al^{3i_2}&\ldots &E_{i_2}\al^{(r+1)i_2}\\
\vdots &\vdots &\ddots &\vdots \\
E_{i_s}\al^{2i_s}&E_{i_s}\al^{3i_s}&\ldots &E_{i_s}\al^{(r+1)i_s}\\
0&0&\ldots &0\\
\vdots &\vdots &\ddots &\vdots \\
0&0&\ldots &0
\end{array}\right),
\end{eqnarray}

From~(\ref{mA}) and~(\ref{BAT}), $ABA^T$ is given by

\begin{eqnarray*}
ABA^T&=&
\left(\begin{array}{llll}
\sum_{l=1}^sE_{i_l}\al^{2i_l}&\sum_{l=1}^sE_{i_l}\al^{3i_l}&\ldots&
\sum_{l=1}^sE_{i_l}\al^{(r+1)i_l}\\
\sum_{l=1}^sE_{i_l}\al^{3i_l}&\sum_{l=1}^sE_{i_l}\al^{4i_l}&\ldots&
\sum_{l=1}^sE_{i_l}\al^{(r+2)i_l}\\
\vdots &\vdots &\ddots &\vdots \\
\sum_{l=1}^sE_{i_l}\al^{(r+1)i_l}&\sum_{l=1}^sE_{i_l}\al^{(r+2)i_l}&\ldots&
\sum_{l=1}^sE_{i_l}\al^{2ri_l}\\
\end{array}\right)\\
&=&
\left(\begin{array}{llll}
S_2&S_3&\ldots &S_{r+1}\\
S_3&S_4&\ldots &S_{r+2}\\
\vdots &\vdots &\ddots &\vdots \\
S_{r+1}&S_{r+2}&\ldots &S_{2r}
\end{array}\right)\\
&=& \bS_r,
\end{eqnarray*}

the last two equalities by~(\ref{eq2.18'}) and~(\ref{eq2.17'}) respectively.
Since $\bS_r\eq ABA^T$,

$$\det(\bS_r)\eq \det(A)^2\det(B),$$

and since $\det(B)=0$ by~(\ref{mB}),
then $\det(\bS_r)\eq 0$ and $\bS_r$ is singular, as claimed.\qed

Next, we apply the Peterson-Gorenstein-Zierler algorithm to the
particular cases of 1-error correcting and 2-error correcting RS codes.

Consider an $[n,n-2]$ 1-byte correcting RS code. Assume that one
error occurred.
In this case, $s=1$ and $n-k=2$, so, if the syndrome polynomial is given
by $S(x)=S_1+S_2x$,
(\ref{eq2.16}) gives $S_2\si_0=-S_1$, i.e.,
$\si(x)=(-S_1/S_2)+x$. The polynomial $\om(x)$ has degree 0,
and is given by minus the 0 coefficient of $\si(x)S(x)$, i.e.,
$\om(x)=(S_1)^2/S_2$. This gives the error value $E_1$.
Notice that the root of $\si(x)$ is $S_1/S_2$. The error location
is given by the value $i$ such that $\al^{-i}=S_1/S_2$, hence,
$\al^i=S_2/S_1$. These results were derived in the previous section
by direct syndrome calculation.

Consider next an $[n,n-4]$ 2-byte correcting RS code and
assume that two errors have occurred.
In this case, $s=2$ and $n-k=4$, so, if the syndrome polynomial is given
by $S(x)=S_1+S_2x+S_3x^2+S_4x^3$,
(\ref{eq2.16}) gives

\begin{eqnarray}
\label{eq2.19}
\begin{array}{cc}
\left(
\begin{array}{ll}
S_2&S_3\\
S_3&S_4
\end{array}
\right)&
\left(
\begin{array}{c}
\si_{1}\\
\si_{0}
\end{array}
\right)\end{array}&= &
\left(
\begin{array}{l}
-S_1\\
-S_2
\end{array}\right),
\end{eqnarray}

and, since two errors have occurred,

$$
\det\left(
\begin{array}{ll}
S_2&S_3\\
S_3&S_4
\end{array}
\right)=S_2S_4-(S_3)^2\neq 0.$$

Solving for $\si_1$ and $\si_0$ in~(\ref{eq2.19}), say by Cramer's rule,
we obtain

\begin{eqnarray}
\label{eq2.20}
\si_0&=&{S_1S_3-(S_2)^2\over
S_2S_4-(S_3)^2} \\
\label{eq2.21}
\si_1&=&{S_2S_3-S_1S_4\over
S_2S_4-(S_3)^2}
\end{eqnarray}

If $\om(x)=\om_0+\om_1x$, then the coefficients of $\om$ are the
coefficients 0 and 1 of $\si(x)S(x)$ with the sign changed, i.e.,

\begin{eqnarray}
\label{eq2.22}
\om_0&=&-\si_0S_1 \\
\label{eq2.23}
\om_1&=&-\si_0S_2-\si_1S_1,
\end{eqnarray}

where $\si_0$ and $\si_1$ are given by ~(\ref{eq2.20})
and~(\ref{eq2.21}).

\begin{ex}
\label{ex6.1}
{\em
Consider the $[7,3,5]$ RS code over $GF(8)$.

Assume that we want to decode the received vector

$$\ur=(0 1 1\;1 0 1\;1 1 1\;1 1 1\;1 1 1\;1 0 1\;0 1 0),$$

which in polynomial form is

$$R(x)=\al^4+\al^6x+\al^5 x^2+\al^5x^3+\al^5x^4+\al^6 x^5+\al x^6.$$

Evaluating the syndromes, we obtain $S_1=R(\al)=\al^5$,
$S_2=R(\al^2)=\al$,
$S_3=R(\al^3)=0$ and
$S_4=R(\al^4)=\al^3$.
By (\ref{eq2.20}) and~(\ref{eq2.21}),
we obtain
$\si_0=\al^5$ and
$\si_1=\al^4$, i.e.,
$\si(x)=\al^5+\al^4x+x^2$.
Searching the roots of $\si(x)$, we verify that these roots are
$\al^0=1$ and $\al^5$; hence, the errors are in locations 0 and 2.
Using (\ref{eq2.22}) and~(\ref{eq2.23}), we obtain
$\om_0=\si_0S_1=\al^3$ and
$\om_1=\si_0S_2+\si_1S_1=1$; hence,
$\om(x)=\al^3+x$.
The derivative of $\si(x)$ is
$\si'(x)=\al^4$. By~(\ref{eq2.9}), we obtain
$E_0=\om(1)/\si'(1)=\al^4$ and
$E_2=\om(\al^5)/\si'(\al^5)=\al^5$.
Adding $E_0$ and $E_2$ to the received locations 0 and 2, the
decoder concludes that the transmitted polynomial was

$$F(x)=\al^6x+\al^5x^3+\al^5x^4+\al^6 x^5+\al x^6,$$

which in vector form is

$$\uc=(0 0 0\;1 0 1\;0 0 0\;1 1 1\;1 1 1\;1 0 1\;0 1 0).$$

If the information is carried in the first 3 bytes, then the output
of the decoder is

$$\uu=(0 0 0\;1 0 1\;0 0 0).$$\qed
}
\end{ex}

Below we state explicitly the
Peterson-Gorenstein-Zierler algorithm.

\begin{alg}[Peterson-Gorenstein-Zierler Decoder]
\label{alg6.1}
{\em
Consider an $[n,k]$ RS code. Assume that we want to correct up to $s$
errors, where $2s\leq n-k$. Let $R(x)$
be a received vector (in polynomial form). Then:

\begin{tabbing}
Compute the syndromes $S_i=R(\al^i)$, $1\leq i\leq n-k$.\\
If $S_i\eq 0$ for $1\leq i\leq n-k$, then output $R(x)$. \\
Else, set $r\la \lf (n-k)/2\rf$.\\
{\bf START:} \=
Let $\bS_r$ be given by~(\ref{eq2.17'}).\\
\> If $\det(\bS_r)\neq 0$, then go to {\bf NEXT}.\\
\> Else, \= set $r\la r-1$.\\
\>\> If $r=0$, then declare an uncorrectable error and stop.\\
\>\> Else, go to {\bf START}.\\
{\bf NEXT:} \> Compute $(\si_{r-1},\si_{r-2},\ldots,\si_0)=
(-S_1,-S_2,\ldots,-S_r)(\bS_r)^{-1}$.\\
\> Let $\si(x)=\si_0+\si_1x+\cdots+\si_{r-1}x^{r-1}+x^r$.\\
\> Compute $\om_j$ as the $j$th coefficient of
$-\si(x)S(x)$, $0\leq j\leq r-1$.\\
\> Let $\om(x)=\om_0+\om_1x+\cdots+\om_{r-1}x^{r-1}$.\\
\> Find $\E=\{l:\si(\al^{-l})=0\}$ by searching the roots of $\si(x)$.\\
\> Compute the error values $E_l=\om(\al^{-l})/\si'(\al^{-l})$ for
$l\in\E$.\\
\> Define $E(x)$ as the polynomial with coefficients
$E_l$ when $l\in\E$, 0 elsewhere.\\
\>\> If $E(\al^i)\neq S_i=R(\al^i)$ \= for some $i$, $1\leq i\leq n-k$,
then declare \\
\>\>\> an uncorrectable error and stop.\\
\>\> Else, output $R(x)-E(x)$ as the estimate of the transmitted polynomial.
\end{tabbing}
}
\end{alg}

By looking at Algorithm~\ref{alg6.1}, we can see that we have added a
step: before releasing the output, we check if the syndromes of the
error polynomial coincide with the original syndromes, therefore, the
output of the decoder is in the code. This step is important to avoid
a miscorrection for cases in which
the number of errors that the code can handle has
been exceeded. It assures that the decoder will not output anything that
is not a codeword.

The Peterson-Gorenstein-Zierler algorithm is important both historically
and conceptually. It is also efficient to handle a small number of
errors. However, when the number of errors is relatively large, it
becomes too complex. One of the reasons is that we have to check
repeatedly if the matrix $\bS_r$ is non-singular, until we find the
correct number of errors. This process may involve too many
multiplications in a finite field. A more efficient decoding algorithm,
and a one widely used in practice, is the so called Berlekamp-Massey
algorithm~\cite{be}. This algorithm exploits the particular structure of the
matrix $\bS_r$. Another efficient decoding algorithm is obtained by
using Euclid's algorithm for division of polynomials. We present
Euclid's algorithm in the next section.

\pr

\begin{prob}
\label{pr2.21'}
{\em
Consider the $[10,4]$ (shortened) RS code over the
finite field $GF(16)$ generated by $1+x+x^4$.
Decode the received vector
 
$$\ur=
(1 1 1 0\;1 1 1 0\;0 0 1 0\;0 1 1 0\;1 1 1 0\;0 1 0 1\;0 1 1 0\;
0 0 0 1\;0 0 0 1\;0 0 1 1),$$
 
Notice that the code is shortened, therefore, the first four bytes
correspond to information while the last six correspond to the redundancy.
In polynomial form, the first 4 bytes are followed by 5 0-bytes, therefore,
the polynomial form of $\ur$ with coefficients as powers of $\al$
is given by
 
$$R(x)=\al^{10}+\al^{10}x+
\al^{2 }x^2+
\al^{5 }x^3+
\al^{10}x^9+
\al^{9 }x^{10}+
\al^{5 }x^{11}+
\al^{3 }x^{12}+
\al^{3 }x^{13}+
\al^{6 }x^{14}.$$
 
}\end{prob}
 
\begin{prob}
\label{pr2.21}
{\em
Consider the $[8,4]$ RS code over the
finite field $GF(9)$ generated by $2+x+x^2$.
Decode the received vector
 
$$\ur=(1 1\;0 1\;2 2\;2 0\;1 1\;2 1\;2 1\;1 2).$$
 
}\end{prob}
 
\begin{prob}
\label{pr2.22}
{\em
Consider the $[8,2]$ RS code over the
finite field $GF(9)$ generated by $2+x+x^2$.
Decode the received vector
 
$$\ur=(2 1\;0 0\;2 2\;1 1\;2 0\;0 0\;0 2\;0 1).$$
 
}\end{prob}
 
\begin{prob}
\label{pr2.23}
{\em
Consider the $[10,2]$ RS code over the
finite field $GF(11)$ generated by
$g(x)\eq (x-2)(x-2^2)\ldots (x-2^8)$ (notice that 2 is
primitive in $GF(11)$).
Decode the received vector
 
$$\ur=(7\;1\;3\;3\;4\;7\;10\;5\;6\;8).$$
}\end{prob}
 
\begin{prob}
\label{pr2.24}
{\em
Using the key equation for errors and erasures obtained in
Problem~\ref{pr2.20}, obtain a version of the
Peterson-Gorenstein-Zierler decoder for errors and erasures.
Use it to decode
 
$$\ur=(0011\;1100\;1111\;0110\;????\;1101\;1010\;????\;0001\;1110),$$
 
over the $[10,4]$ (shortened) RS code of Problem~\ref{pr2.21'}
(the symbol $?$ denotes an erased bit).
 
}\end{prob}
 
\begin{prob}
\label{pr2.241}
{\em
As in Problems~\ref{pr2.16'} and~\ref{pr2.81}, consider
an $[n,k]$ RS code as the set
of polynomials of degree $\leq n-1$ having as roots the
consecutive powers $\al^m,\al^{m+1},\ldots,\al^{m+n-k-1}$.
Give an equivalent form of the Peterson-Gorenstein-Zierler decoder 
for this case. 

Consider the $[15,9]$ RS code over $GF(16)$, $GF(16)$ generated by $1+x+x^4$,
whose roots are $1,\al,\ldots,\al^5$ (i.e., $m=0$ in the description above). 
Use the modified 
Peterson-Gorenstein-Zierler decoder to decode the received polynomial
 
$$R(x)=1+\al^{12}x+\al^{10}x^2+\al x^4+\al^{8}x^5+\al^{10}x^6+\al^{6}x^7+
\al^{8}x^8+\al^{5}x^{10}+\al^{14}x^{11}+\al^{10}x^{12}+x^{13}+\al^{12}x^{14}.$$

}\end{prob}

\begin{prob}
\label{pr2.242}
{\em
As in Problem~\ref{pr2.241}, consider
an $[n,k]$ RS code as the set
of polynomials of degree $\leq n-1$ having as roots the
consecutive powers $\al^m,\al^{m+1},\ldots,\al^{m+n-k-1}$.
Give an equivalent form of the Peterson-Gorenstein-Zierler decoder 
for errors and erasures, as in Problem~\ref{pr2.24}.
As in Problem~\ref{pr2.241}, consider the $[15,9]$ RS code over $GF(16)$
generated by $1+x+x^4$
whose roots are $1,\al,\ldots,\al^5$. Use the error-erasure version 
of the Peterson-Gorenstein-Zierler decoder to decode 

$$\begin{array}{ccl}
R(x)&=&1+?x+\al^{10}x^2+?x^3+\al x^4
+\al^{8}x^5+\al^{10}x^6+\al^{6}x^7+\al^{8}x^8+
\al^{6}x^9+\al^{5}x^{10}\\
&&+\al^{6}x^{11}+\al^{7}x^{12}+x^{13}+\al^{6}x^{14}.
\end{array}$$

}\end{prob}

\begin{prob}
\label{pr2.25}
{\em
Write a computer program implementing the Peterson-Gorenstein-Zierler
decoder.
}\end{prob}
 
\sol

\vspace{.8cm} 

{\bf Problem~\ref{pr2.21'}}
 
We apply Algorithm~\ref{alg6.1}.
The 6 syndromes of the received vector $R(x)$ are
 
\begin{eqnarray*}
S_1=R(\al^{\phantom{2}})&=&\al^3\\
S_2=R(\al^2)&=&\al^2\\
S_3=R(\al^3)&=&\al^{12}\\
S_4=R(\al^4)&=&0       \\
S_5=R(\al^5)&=&\al^{12}\\
S_6=R(\al^6)&=&\al
\end{eqnarray*}
 
Next, we verify that
 
$$\det
\left(\begin{array}{ccc}
S_2&S_3&S_4\\
S_3&S_4&S_5\\
S_4&S_5&S_6\end{array}\right)\eq
\det
\left(\begin{array}{ccc}
\al^2&\al^{12}&0\\
\al^{12}&0&\al^{12}\\
0&\al^{12}&\al\end{array}\right)\eq\al^{14}\neq 0.$$
 
This means, the decoder will assume that 3 errors have occurred.
Solving for
 
\begin{eqnarray*}
\left(\begin{array}{ccc}
\al^2&\al^{12}&0\\
\al^{12}&0&\al^{12}\\
0&\al^{12}&\al\end{array}\right)
\left(\begin{array}{c}
\si_2\\ \si_1\\ \si_0
\end{array}\right)
&=&
\left(\begin{array}{c}
\al^3\\ \al^2\\ \al^{12}
\end{array}\right),
\end{eqnarray*}
 
we obtain $\si_2=\al^2$, $\si_1=\al^{10}$ and $\si_0=\al$.
Therefore, $\si(x)\eq \al+\al^{10}x+\al^2x^2+x^3$.
The roots of $\si(x)$ are $1=\al^0$, $\al^{13}=\al^{-2}$ and
$\al^{3}=\al^{-12}$, so,
the set $\E$ of locations in error is
$\E\eq\{0,2,12\}$.
 
The coefficients 0, 1 and 2 of the product $-\si(x)S(x)$ are:
 
$$\begin{array}{lllll}
\om_0&=&-\si_0S_1&=&\al^4       \\
\om_1&=&-(\si_0S_2+\si_1S_1)&=&\al^8\\
\om_2&=&-(\si_0S_3+\si_1S_2+\si_2S_1)&=&\al^2
\end{array}$$
 
Therefore, the error evaluator polynomial is $\om(x)\eq\al^4+\al^8x+\al^2x^2$.
The derivative of $\si(x)$ is $\si'(x)=\al^{10}+x^2$.
 
The error values are:
 
$$\begin{array}{lllll}
E_0&=&\om(1)/\si'(1)&=&\al^{11}       \\
E_1&=&\om(\al^{-2})/\si'(\al^{-2})&=&\al^{2}       \\
E_2&=&\om(\al^{-12})/\si'(\al^{-12})&=&\al^{11}
\end{array}$$
 
Finally, substracting the values $E_0$, $E_1$ and $E_2$ from
$R_0$, $R_2$ and $R_{12}$, we obtain the estimate for $R(x)$
 
$$C(x)=\al^{14}+\al^{10}x+
\al^{5 }x^3+
\al^{10}x^9+
\al^{9 }x^{10}+
\al^{5 }x^{11}+
\al^{5 }x^{12}+
\al^{3 }x^{13}+
\al^{6 }x^{14}.$$
 
Taking only the information part in vector form, i.e., the first four
bytes, the output of the decoder is
 
$$\uu=
(1 0 0 1\;1 1 1 0\;0 0 0 0\;0 1 1 0).$$
 
\vspace{.8cm}
 
{\bf Problem~\ref{pr2.21}}
 
The finite field is described in Problem~\ref{pr2.9}.
If we write the received vector in polynomial form, we obtain
 
$$R(x)=\al^{7}+\al x+
\al^{3 }x^2+
\al^{4 }x^3+
\al^{7 }x^4+
\al^{6 }x^5+
\al^{6 }x^6+
\al^{2 }x^7
.$$
 
The syndromes are:
 
\begin{eqnarray*}
S_1=R(\al^{\phantom{1}})&=&\al^5\\
S_2=R(\al^2)&=&0\\
S_3=R(\al^3)&=&1\\
S_4=R(\al^4)&=&\al^7
\end{eqnarray*}
 
We can see that
 
$$\det
\left(\begin{array}{cc}
S_2&S_3\\
S_3&S_4
\end{array}\right)\eq
\det
\left(\begin{array}{cc}
0&1\\
1&\al^{7}
\end{array}\right)\eq 2\neq 0.$$
 
Thus, the decoder assumes that two errors have occurred.
Solving for
 
\begin{eqnarray*}
\left(\begin{array}{cc}
0&1\\
1&\al^{7}
\end{array}\right)
\left(\begin{array}{c}
\si_1\\ \si_0
\end{array}\right)
&=&
\left(\begin{array}{c}
\al\\ 0
\end{array}\right),
\end{eqnarray*}
 
we obtain $\si_1=\al^4$ and $\si_0=\al$.
Therefore, $\si(x)\eq \al+\al^{4}x+x^2$.
The roots of $\si(x)$ are $\al^6=\al^{-2}$ and
$\al^{3}=\al^{-5}$, so,
the set $\E$ of locations in error is
$\E\eq\{2,5\}$.
 
The coefficients 0 and 1 of the product $-\si(x)S(x)$ are:
 
$$\begin{array}{lllll}
\om_0&=&-\si_0S_1&=&\al^2       \\
\om_1&=&-(\si_0S_2+\si_1S_1)&=&\al^5
\end{array}$$
 
Therefore, the error evaluator polynomial is $\om(x)\eq\al^2+\al^5x$.
The derivative of $\si(x)$ is $\si'(x)=\al^{10}+\al^4x^2$.
 
The error values are:
 
$$\begin{array}{lllll}
E_0&=&\om(\al^{-2})/\si'(\al^{-2})&=&\al^{4}       \\
E_1&=&\om(\al^{-5})/\si'(\al^{-5})&=&\al^{2}
 
\end{array}$$
 
Substracting the values $E_0$ and $E_1$ from
$R_2$ and $R_{5}$, we obtain the estimate for $R(x)$
 
$$C(x)=\al^{7}+\al x+
\al^{5 }x^2+
\al^{4 }x^3+
\al^{7 }x^4+
\al^{2 }x^5+
\al^{6 }x^6+
\al^{2 }x^7
.$$
 
In vector form, this gives
 
$$\uc=( 1 1\;0 1\;0 2\;2 0\;1 1\;1 2\;2 1\;1 2).$$
 
If we are interested only in the information part, the final output of
the decoder is
 
$$\uu=( 1 1\;0 1\;0 2\;2 0).$$

\vspace{.8cm}
 
{\bf Problem~\ref{pr2.22}}
 
If we write the received vector in polynomial form, we obtain
 
$$R(x)=\al^{6}+
\al^{3 }x^2+
\al^{7 }x^3+
\al^{4 }x^4+
\al^{5 }x^6+
\al x^7
.$$
 
The syndromes are given by
 
\begin{eqnarray*}
S_1=R(\al^{\phantom{1}})&=&\al^7\\
S_2=R(\al^2)&=&0\\
S_3=R(\al^3)&=&\al^7\\
S_4=R(\al^4)&=&\al^4\\
S_5=R(\al^5)&=&0\\
S_6=R(\al^6)&=&1
\end{eqnarray*}
 
Next, we verify that
 
$$\det
\left(\begin{array}{ccc}
S_2&S_3&S_4\\
S_3&S_4&S_5\\
S_4&S_5&S_6\end{array}\right)\eq
\det
\left(\begin{array}{ccc}
0&\al^{7}&\al^4\\
\al^{7}&\al^4&0\\
\al^4&0&1\end{array}\right)
\neq 0.$$
 
This means, the decoder will assume that 3 errors have occurred.
Solving for
 
\begin{eqnarray*}
\left(\begin{array}{ccc}
0&\al^{7}&\al^4\\
\al^{7}&\al^4&0\\
\al^4&0&1\end{array}\right)
\left(\begin{array}{c}
\si_2\\ \si_1\\ \si_0
\end{array}\right)
&=&
\left(\begin{array}{c}
\al^3\\ 0\\ \al^{3}
\end{array}\right),
\end{eqnarray*}
 
we obtain $\si_2=1$, $\si_1=\al^{7}$ and $\si_0=\al^5$.
Therefore, $\si(x)\eq \al^5+\al^{7}x+x^2+x^3$.
The roots of $\si(x)$ are $1=\al^0$, $\al^{6}=\al^{-2}$ and
$\al^{3}=\al^{-5}$, so,
the set $\E$ of locations in error is
$\E\eq\{0,2,5\}$.
 
The coefficients 0, 1 and 2 of the product $-\si(x)S(x)$ are:
 
$$\begin{array}{lllll}
\om_0&=&-\si_0S_1&=&1       \\
\om_1&=&-(\si_0S_2+\si_1S_1)&=&\al^2\\
\om_2&=&-(\si_0S_3+\si_1S_2+\si_2S_1)&=&\al^5
\end{array}$$
 
Therefore, the error evaluator polynomial is $\om(x)\eq 1+\al^2x+\al^5x^2$.
The derivative of $\si(x)$ is $\si'(x)=\al^{7}+\al^4x$.
 
The error values are:
 
$$\begin{array}{lllll}
E_0&=&\om(1)/\si'(1)&=&\al^{5}       \\
E_1&=&\om(\al^{-2})/\si'(\al^{-2})&=&\al^{2}       \\
E_2&=&\om(\al^{-5})/\si'(\al^{-5})&=&\al^{6}
\end{array}$$
 
Finally, substracting the values $E_0$, $E_1$ and $E_2$ from
$R_0$, $R_2$ and $R_{5}$, we obtain the estimate for $R(x)$
 
$$C(x)=\al^{3}+
x^2+
\al^7x^3+
\al^{4 }x^4+
\al^{2 }x^5+
\al^{5 }x^6+
\al x^7
.$$
 
In vector form, this gives
 
$$\uc=( 2 2\;0 0\;1 0\;1 1\;2 0\;1 2\;0 2\;0 1).$$
 
If we are interested only in the information part, the output of the
decoder is
 
$$\uu=( 2 2\;0 0).$$
 
\vspace{.8cm}
 
{\bf Problem~\ref{pr2.23}}
 
In polynomial form, the received vector can be written as
 
$$R(x)=7+x+3x^2+3x^3+4x^4+7x^5+10x^6+5x^7+6x^8+8x^9.$$
 
Since 2 is a primitive element in $GF(11)$, we define
$\al=2$. In effect, a table for the non-zero elements of
$GF(11)$ is given by
 
$$\begin{array}{|c|c|c|c|c|c|c|c|c|c|}
\hline
2^0&2^1&2^2&2^3&2^4&2^5&2^6&2^7&2^8&2^9\\
\hline
1&2&4&8&5&10&9&7&3&6\\
\hline
\end{array}$$
 
The 8 syndromes corresponding to $R(x)$
are given by
 
\begin{eqnarray*}
S_1=R(2^{\phantom{1}})&=&7\\
S_2=R(2^2)&=&6\\
S_3=R(2^3)&=&8\\
S_4=R(2^4)&=&6\\
S_5=R(2^5)&=&6\\
S_6=R(2^6)&=&1\\
S_7=R(2^7)&=&8\\
S_8=R(2^8)&=&3
\end{eqnarray*}
 
Next, we verify that
 
$$\det
\left(\begin{array}{cccc}
S_2&S_3&S_4&S_5\\
S_3&S_4&S_5&S_6\\
S_4&S_5&S_6&S_7\\
S_5&S_6&S_7&S_8
\end{array}\right)\eq
\det
\left(\begin{array}{cccc}
6&8&6&6\\
8&6&6&1\\
6&6&1&8\\
6&1&8&3
\end{array}\right)
\neq 0.$$
 
This means, the decoder will assume that 4 errors have occurred.
Solving for
 
\begin{eqnarray*}
\left(\begin{array}{cccc}
6&8&6&6\\
8&6&6&1\\
6&6&1&8\\
6&1&8&3
\end{array}\right)
\left(\begin{array}{c}
\si_3\\ \si_2\\ \si_1\\ \si_0
\end{array}\right)
&=&
\left(\begin{array}{c}
4\\5\\3\\5
\end{array}\right),
\end{eqnarray*}
 
we obtain $\si_3=10$, $\si_2=5$,
$\si_1=2$ and $\si_0=4$.
Therefore, $\si(x)\eq 4+2x+5x^2+10x^3+x^4$.
The roots of $\si(x)$ are $1=2^0$, $6=2^{9}=2^{-1}$, $7=2^{7}=2^{-3}$,
and $9=2^{6}=2^{-4}$, so, the set $\E$ of locations in error is
$\E\eq\{0,1,3,4\}$.
 
The coefficients 0, 1, 2 and 3 of the product $-\si(x)S(x)$ are:
 
$$\begin{array}{lllll}
\om_0&=&-\si_0S_1&=&5       \\
\om_1&=&-(\si_0S_2+\si_1S_1)&=&6\\
\om_2&=&-(\si_0S_3+\si_1S_2+\si_2S_1)&=&9\\
\om_3&=&-(\si_0S_4+\si_1S_3+\si_2S_2+\si_3S_1)&=&3
\end{array}$$
 
Therefore, the error evaluator polynomial is $\om(x)\eq 5+6x+9x^2+3x^3$.
The derivative of $\si(x)$ is $\si'(x)=2+10x+8x^2+4x^3$.
 
The error values are:
 
$$\begin{array}{lllll}
 
E_0&=&\om(1)/\si'(1)&=&6       \\
E_1&=&\om(2^{-1})/\si'(2^{-1})&=&3      \\
E_2&=&\om(2^{-3})/\si'(2^{-3})&=&1\\
E_3&=&\om(2^{-4})/\si'(2^{-4})&=&4
\end{array}$$
 
Finally, substracting the values $E_0$, $E_1$, $E_2$ and $E_3$ from
$R_0$, $R_1$, $R_3$ and $R_{4}$, we obtain the estimate for $\ur$
 
$$\uc=(1\;9\;3\;2\;0\;7\;10\;5\;6\;8).$$
 
If we are interested only in
the information part, the output of the decoder is
 
$$\uu=(1\;9).$$
 
\vspace{.8cm}
 
{\bf Problem~\ref{pr2.24}}
 
In this case, we use the modified key equation~(\ref{modkey}) obtained
in Problem~\ref{pr2.20}. We refer to the notation in that problem.
 
Assume that $s$ errors and $t$ erasures have occurred such that
$2s+t\leq n-k-1$, i.e., we are within the error correcting capability
of the code.
Note that $\si_1(x)$ has degree $s$, $\hS(x)$ has degree $n-k-1+t$
and $\om(x)$ has degree $\leq s+t-1$.
Moreover, let
 
$$\si_1(x)=\si_0^{(1)}+\si_1^{(1)}x+\ldots +\si_{s-1}^{(1)}x^{s-1}+x^s.$$
 
Also, let
 
$$\hS(x)=\si_2(x)S(x)=\hS_1+\hS_2x+\cdots +\hS_{n-k+t}x^{n-k+t-1}.$$
 
From the modified key equation~(\ref{modkey}), we notice that the
coefficients $s+t+i$, $0\leq i\leq n-k-s-t-1$, of $\si_1(x)\hS(x)$
are 0. Hence, writing explicitly the coefficient $s+t+i$ of this
polynomial product, we obtain
 
$$\si_0^{(1)}\hS_{s+t+i+1}+
  \si_1^{(1)}\hS_{s+t+i}+\cdots +\si_{s-1}^{(1)}\hS_{t+i}+
  \hS_{t+i+1}=0\;,\;0\leq i\leq n-k-s-t-1.$$
 
If we just consider the first $s$ equations above, and we keep the
rest for verification, since
$\si_s^{(1)}=1$, we can express them as the matrix multiplication
 
\begin{eqnarray}
\label{meq2.17}
\begin{array}{cc}
\left(
\begin{array}{llll}
\hS_{t+2}&\hS_{t+3}&\ldots &\hS_{s+t+1}\\
\hS_{t+3}&\hS_{t+4}&\ldots &\hS_{s+t+2}\\
\vdots & \vdots &\ddots &\vdots\\
\hS_{s+t+1}&\hS_{s+t+2}&\ldots &\hS_{2s+t}
\end{array}
\right)&
\left(
\begin{array}{l}
\si_{s-1}^{(1)}\\
\si_{s-2}^{(1)}\\
\vdots\\
\si_0^{(1)}
\end{array}
\right)\end{array}&= &
\left(
\begin{array}{l}
-\hS_{t+1}\\
-\hS_{t+2}\\
\vdots\\
-\hS_{s+t}\\
\end{array}\right).
\end{eqnarray}
 
Now we can
find $\si_1(x)$ using~(\ref{meq2.17}). This gives an error-erasure
Peterson-Gorenstein-Zierler decoder.
 
In effect, let
 
\begin{eqnarray}
\label{meq2.17'}
\hat{\bS}_r&=&
\left(
\begin{array}{llll}
\hS_{t+2}&\hS_{t+3}&\ldots &\hS_{t+r+1}\\
\hS_{t+3}&\hS_{t+4}&\ldots &\hS_{t+r+2}\\
\vdots & \vdots &\ddots &\vdots\\
\hS_{t+r+1}&\hS_{t+r+2}&\ldots &\hS_{t+2r}
\end{array}
\right),
\end{eqnarray}
 
where $2r\leq n-k-t$. Since $s$ errors and $t$ erasures
have occurred and this is within
the error-correcting capability of the code, $\hat{\bS}_s$ is non-singular.
Similarly to Lemma~\ref{sing}, we can prove that $\hat{\bS}_r$ is singular for
$s<r\leq (n-k-t)/2$.
Hence, the decoder starts checking if $\hat{\bS}_r$ is non-singular for
the largest possible $r$ (i.e., $r=\lf (n-k-t)/2\rf$).
The moment it finds an $r$ such that $\hat{\bS}_r$
is non-singular, this $r$ gives the number
of errors $s$.
Then, (\ref{meq2.17}) can be solved simply by inverting
$\hat{\bS}_s$, i.e.,
 
\begin{eqnarray}
\label{meq2.18}
\left(
\begin{array}{l}
\si^{(1)}_{s-1}\\
\si^{(1)}_{s-2}\\
\vdots\\
\si^{(1)}_0
\end{array}
\right)
&= &
\begin{array}{cc}
(\hat{\bS}_s)^{-1}&
\left(
\begin{array}{l}
-\hS_{t+1}\\
-\hS_{t+2}\\
\vdots\\
-\hS_{s+t}\\
\end{array}\right).
\end{array}
\end{eqnarray}
 
Once we have obtained $\si_1(x)$,
we can compute $\om(x)$ by calculating
the coefficients $j$, $0\leq j\leq t+s-1$, of $\si_1(x)\hS(x)$ and
changing their sign.
We then find the error values using~(\ref{eq2.9}).
 
Consider now $\ur$ as given in the problem. Since this is
a shortened code, we add 0's in appropriate information bytes
as in Problem~\ref{pr2.21'}. Thus, in polynomial form $\ur$ becomes
 
$$R(x)=\al^{6}+\al^{4}x+
\al^{12}x^2+
\al^{5 }x^3+
?x^9+
\al^{7}x^{10}+
\al^{8 }x^{11}+
?x^{12}+
\al^{3 }x^{13}+
\al^{10}x^{14},$$
 
where $?$ denotes an erased byte.
Evaluating the syndromes $S_1,S_2,\ldots,S_6$ assuming that the erased
bytes are equal to 0, we obtain the syndrome polynomial
 
$$S(x)=
\al^{14}+\al^{11}x+
\al^{10}x^2+
\al^{12}x^3+
\al^{12}x^4+
\al^{14}x^5.$$
 
The erasure locator polynomial is
$\si^{(2)}(x)=(x+\al^{-9})(x+\al^{-12})=
(x+\al^{3})(x+\al^{6})$, so,
 
$$\hS(x)=\si^{(2)}(x)S(x)=
\al^{8}+\al^{2}x+
\al^{10}x^2+
\al^{13}x^3+
\al x^4+
\al^{4}x^5+
\al^{13}x^6+
\al^{14}x^7
.$$
 
Observe that
 
$$\det\left(\begin{array}{ll}
\hS_4&\hS_5\\
\hS_5&\hS_6\end{array}\right)=
\det\left(\begin{array}{cc}
\al^{13}&\al \\
\al     &\al^4
\end{array}\right)=0.$$
 
Since $\hS_4\neq 0$, one error has occurred and the error locator
polynomial $\si^{(1)}(x)$ has degree 1. Applying the algorithm,
we obtain
$\hS_4\si_0^{(1)}=\hS_3$, i.e.,
$\al^{13}\si_0^{(1)}=\al^{10}$ and
$\si_0^{(1)}=\al^{12}$. So, the error locator polynomial
is
 
$$\si^{(1)}(x)=\al^{12}+x=\al^{-3}+x.$$
 
This means, the error is in location 3 (the erasures were in
locations 9 and 12).
The error-erasure locator polynomial is now
 
$$
\si(x)=\si^{(1)}(x)\si^{(2)}(x)=
\al^{6}+\al^{4}x+
\al^{7 }x^2+
x^3.$$
 
The derivative of $\si(x)$ is $\si'(x)=\al^4+x^2$.
The error evaluator polynomial
$\om(x)=\om_0+\om_1x+\om_2x^2$ is obtained as the coefficients
0, 1 and 2 of the product of the polynomials $\si(x)$ and $S(x)$.
Evaluating these coefficients, we obtain
 
$$\om(x)=\al^5+\al^6x+\al^{12}x^2.$$
 
The error values are given by:
 
$$\begin{array}{lllll}
E_3&=&\om(\al^{-3})/\si'(\al^{-3})&=&\al^2\\
E_9&=&\om(\al^{-9})/\si'(\al^{-9})&=&\al^{13}\\
E_{12}&=&\om(\al^{-12})/\si'(\al^{-12})&=&\al^{5}
\end{array}.$$
 
Finally, substracting the error values from the corresponding locations,
we decode $R(x)$ as
 
$$C(x)=
\al^{6}+\al^{4}x+
\al^{12}x^2+
\al x^3+
\al^{13}x^9+
\al^{7}x^{10}+
\al^{8 }x^{11}+
\al^5x^{12}+
\al^{3 }x^{13}+
\al^{10}x^{14}.$$
 
Considering only the four information bytes, the output of the decoder is
 
$$0011\;1100\;1111\;0100.$$
 
\vspace{.8cm}
 
{\bf Problem~\ref{pr2.241}}

As in Problem~\ref{pr2.81}, we have the modified key equation
 
\begin{equation}
\label{eq2.131}
\si(x)S(x)=-\om(x)+ \mu(x) x^{n-k},
\end{equation}
 
in which $S(x)$ is given by~(\ref{eq2.3'}) and $\om(x)$ by~(\ref{eq2.82}).

Assume that $s$ errors have occurred, where $2s\leq n-k$. Hence,
the code can correct these $s$ errors.
Let $\si(x)=\si_0+\si_1x+\ldots +\si_{s-1}x^{s-1}+x^s$ and
$\om(x)=\om_0+\om_1x+\ldots +\om_{s-1}x^{s-1}$.
Let $s\leq j\leq n-k-1$. According to~(\ref{eq2.131}), the $j$th
coefficient of $\si(x)S(x)$ is 0. But this $j$th coefficient is
given by
 
\begin{equation}
\label{eq2.141}
\sum_{l=0}^s\si_lS_{j+m-l}=0\;,\;s\leq j\leq n-k-1.
\end{equation}
 
Since $\si_s=1$,
(\ref{eq2.141}), is equivalent to
 
\begin{equation}
\label{eq2.151}
\sum_{l=0}^{s-1}\si_lS_{j+m-l}=-S_{j+m-s}\;,\;s\leq j\leq n-k-1.
\end{equation}
 
In matrix form, (\ref{eq2.151}) gives
 
\begin{eqnarray*}
\begin{array}{cc}
\left(
\begin{array}{llll}
S_{m+1}&S_{m+2}&\ldots &S_{m+s}\\
S_{m+2}&S_{m+3}&\ldots &S_{m+s+1}\\
\vdots & \vdots &\ddots &\vdots\\
S_{m+n-k-s}&S_{m+n-k-s+1}&\ldots &S_{m+n-k-1}
\end{array}
\right)&
\left(
\begin{array}{l}
\si_{s-1}\\
\si_{s-2}\\
\vdots\\
\si_0
\end{array}
\right)\end{array}&= &
\left(
\begin{array}{l}
-S_m\\
-S_{m+1}\\
\vdots\\
-S_{m+n-k-s-1}\\
\end{array}\right).
\end{eqnarray*}
 
In order to solve this system, it is enough to take the first $s$ rows
in the matrix at the left (the remaining rows may be used for verification),
thus, we obtain
 
\begin{eqnarray}
\label{eq2.171}
\begin{array}{cc}
\left(
\begin{array}{llll}
S_{m+1}&S_{m+2}&\ldots &S_{m+s}\\
S_{m+2}&S_{m+3}&\ldots &S_{m+s+1}\\
\vdots & \vdots &\ddots &\vdots\\
S_{m+s}&S_{m+s+1}&\ldots &S_{m+2s-1}
\end{array}
\right)&
\left(
\begin{array}{l}
\si_{s-1}\\
\si_{s-2}\\
\vdots\\
\si_0
\end{array}
\right)\end{array}&= &
\left(
\begin{array}{l}
-S_{m}\\
-S_{m+1}\\
\vdots\\
-S_{m+s-1}\\
\end{array}\right).
\end{eqnarray}
 
Let
 
\begin{eqnarray}
\label{eq2.17'1}
\bS_{m,r}&=&
\left(
\begin{array}{llll}
S_{m+1}&S_{m+2}&\ldots &S_{m+r}\\
S_{m+2}&S_{m+3}&\ldots &S_{m+r+1}\\
\vdots & \vdots &\ddots &\vdots\\
S_{m+r}&S_{m+r+1}&\ldots &S_{m+2r-1}
\end{array}
\right),
\end{eqnarray}
 
where $2r\leq n-k$. Since $s$ errors have occurred and this is within
the error-correcting capability of the code, $\bS_{m,s}$ is non-singular.
We can prove that $\bS_{m,r}$ is singular for $s<r\leq (n-k)/2$ as
in Lemma~\ref{sing}. 
Hence, the decoder starts checking if $\bS_{m,r}$ is non-singular for
the largest possible $r$ (i.e., $r=\lf (n-k)/2\rf$).
When it finds an $r$ such that $\bS_{m,r}$
is non-singular, this $r$ gives the number
of errors $s$.
Then, (\ref{eq2.171}) can be solved simply by inverting
$\bS_{m,s}$, i.e.,
 
\begin{eqnarray}
\label{eq2.181}
\left(
\begin{array}{l}
\si_{s-1}\\
\si_{s-2}\\
\vdots\\
\si_0
\end{array}
\right)
&= &
\begin{array}{cc}
(\bS_{m,s})^{-1}&
\left(
\begin{array}{l}
-S_m\\
-S_{m+1}\\
\vdots\\
-S_{m+s-1}\\
\end{array}\right).
\end{array}
\end{eqnarray}
 
Once we have obtained $\si(x)$, by~(\ref{eq2.131}),
we can compute $\om(x)$ by calculating
the coefficients $j$, $0\leq j\leq s-1$, of $\si(x)S(x)$ and 
changing their sign.
We then find the error values using~(\ref{eq2.9'}).

Consider now the polynomial $R(x)$ described in the problem.
We do the computations using the table of the field
described in Problem~\ref{pr2.8}.
The syndromes are:

$$
\begin{array}{lclcl}
S_0&=&R(1)    &=&\al^5\\
S_1&=&R(\al)  &=&\al^7\\
S_2&=&R(\al^2)&=&\al^8\\
S_3&=&R(\al^3)&=&\al^6\\
S_4&=&R(\al^4)&=&\al^9\\
S_5&=&R(\al^5)&=&1,
\end{array}
$$

therefore, 

\begin{eqnarray*}
S(x)&=&\al^5+\al^7x+\al^8x^2+\al^6x^3+\al^9x^4+x^5.
\end{eqnarray*}

Next we compute the determinant

\begin{eqnarray*}
\det(\bS_{0,3})&=&\det\left(\begin{array}{lll}
S_1&S_2&S_3\\
S_2&S_3&S_4\\
S_3&S_4&S_5
\end{array}\right).
\end{eqnarray*}

We can verify that $\det(\bS_{0,3})\eq 0$, thus $\bS_{0,3}$ is singular. 
Next, we can see that

\begin{eqnarray*}
\det(\bS_{0,2})&=&\det\left(\begin{array}{ll}
S_1&S_2\\
S_2&S_3
\end{array}\right)\neq 0.
\end{eqnarray*}

Solving the system 

\begin{eqnarray*}
\left(\begin{array}{ll}
S_1&S_2\\
S_2&S_3
\end{array}\right)
\left(\begin{array}{l}
\si_1\\
\si_0
\end{array}\right)
&=&
\left(\begin{array}{l}
S_0\\
S_1
\end{array}\right), 
\end{eqnarray*}

we obtain $\si_0=\al^5$ and $\si_1=1$, therefore, 

\begin{eqnarray*}
\si(x)&=&\al^5+x+x^2.
\end{eqnarray*}

The roots of this polynomial are $\al=\al^{-14}$ and $\al^4=\al^{-11}$, therefore,
the errors are in locations 11 and 14. In order to find $\om(x)$, 
we need to estimate the coefficients 0 and 1 of $\si(x)S(x)$. This gives
$\om_0=\al^{10}$ and $\om_1=\al^{14}$, thus, 

\begin{eqnarray*}
\om(x)&=&\al^{10}+\al^{14}x.
\end{eqnarray*}

Also, we obtain

\begin{eqnarray*}
\si'(x)&=&1.
\end{eqnarray*}

Now, using~(\ref{eq2.9'}) to estimate the errors, we obtain

\begin{eqnarray*}
E_{11}&=&{\om(\al^{4})\al^{11}\over \si'(\al^{4})}\eq \al^8\\
E_{14}&=&{\om(\al)\al^{14}\over \si'(\al)}\eq \al^4\\
\end{eqnarray*}

Finally, substracting the errors from $R(x)$ at locations 11 and 14, 
we obtain the decoded polynomial

$$C(x)=1+\al^{12}x+\al^{10}x^2+\al x^4+\al^{8}x^5+\al^{10}x^6+\al^{6}x^7+
\al^{8}x^8+\al^{5}x^{10}+\al^{6}x^{11}+\al^{10}x^{12}+x^{13}+\al^{6}x^{14}.$$

\vspace{.8cm}

{\bf Problem~\ref{pr2.242}}

We use the modified key equation~(\ref{modkey}) obtained
in Problem~\ref{pr2.20}, but 
$S(x)$ is given by~(\ref{eq2.3'},
$\om(x)$ is given by~(\ref{eq2.82}) and the error values by~(\ref{eq2.9'}).

Assume that $s$ errors and $t$ erasures have occurred such that
$2s+t\leq n-k-1$, i.e., we are within the error correcting capability
of the code.
Note that $\si_1(x)$ has degree $s$, $\hS(x)$ has degree $n-k-1+t$
and $\om(x)$ has degree $\leq s+t-1$.
Moreover, let
 
$$\si_1(x)=\si_0^{(1)}+\si_1^{(1)}x+\ldots +\si_{s-1}^{(1)}x^{s-1}+x^s.$$
 
Also, let
 
$$\hS(x)=\si_2(x)S(x)=\hS_m+\hS_{m+1}x+\cdots +\hS_{m+n-k+t-1}x^{n-k+t-1}.$$
 
From the modified key equation~(\ref{modkey}), we notice that the
coefficients $s+t+i$, $0\leq i\leq n-k-s-t-1$, of $\si_1(x)\hS(x)$
are 0. Hence, writing explicitly the coefficient $s+t+i$ of this
polynomial product, we obtain
 
$$\si_0^{(1)}\hS_{m+s+t+i}+
  \si_1^{(1)}\hS_{m+s+t+i-1}+\cdots +\si_{s-1}^{(1)}\hS_{m+t+i-1}+
  \hS_{m+t+i}=0\;,\;0\leq i\leq n-k-s-t-1.$$
 
If we just consider the first $s$ equations above, and we keep the
rest for verification, since
$\si_s^{(1)}=1$, we can express them as the matrix multiplication
 
\begin{eqnarray}
\label{meq2.171}
\begin{array}{cc}
\left(
\begin{array}{llll}
\hS_{m+t+1}&\hS_{m+t+2}&\ldots &\hS_{m+s+t}\\
\hS_{m+t+2}&\hS_{m+t+3}&\ldots &\hS_{m+s+t+1}\\
\vdots & \vdots &\ddots &\vdots\\
\hS_{m+s+t}&\hS_{m+s+t+1}&\ldots &\hS_{m+2s+t-1}
\end{array}
\right)&
\left(
\begin{array}{l}
\si_{s-1}^{(1)}\\
\si_{s-2}^{(1)}\\
\vdots\\
\si_0^{(1)}
\end{array}
\right)\end{array}&= &
\left(
\begin{array}{l}
-\hS_{m+t}\\
-\hS_{m+t+1}\\
\vdots\\
-\hS_{m+s+t-1}\\
\end{array}\right).
\end{eqnarray}
 
Now we can
find $\si_1(x)$ using~(\ref{meq2.171}). This gives an error-erasure
Peterson-Gorenstein-Zierler decoder.
 
In effect, let
 
\begin{eqnarray}
\label{meq2.17'1}
\hat{\bS}_{m,r}&=&
\left(
\begin{array}{llll}
\hS_{m+t+1}&\hS_{m+t+2}&\ldots &\hS_{m+t+r}\\
\hS_{m+t+2}&\hS_{m+t+3}&\ldots &\hS_{m+t+r+1}\\
\vdots & \vdots &\ddots &\vdots\\
\hS_{m+t+r}&\hS_{m+t+r+1}&\ldots &\hS_{m+t+2r-1}
\end{array}
\right),
\end{eqnarray}
 
where $2r\leq n-k-t$. Since $s$ errors and $t$ erasures
have occurred and this is within
the error-correcting capability of the code, $\hat{\bS}_{m,s}$ is non-singular.
Similarly to Lemma~\ref{sing}, we can prove that $\hat{\bS}_{m,r}$ is singular for
$s<r\leq (n-k-t)/2$.
Hence, the decoder starts checking if $\hat{\bS}_{m,r}$ is non-singular for
the largest possible $r$ (i.e., $r=\lf (n-k-t)/2\rf$).
The moment it finds an $r$ such that $\hat{\bS}_{m,r}$
is non-singular, this $r$ gives the number
of errors $s$.
Then, (\ref{meq2.171}) can be solved simply by inverting
$\hat{\bS}_{m,s}$, i.e.,
 
\begin{eqnarray*}
\left(
\begin{array}{l}
\si^{(1)}_{s-1}\\
\si^{(1)}_{s-2}\\
\vdots\\
\si^{(1)}_0
\end{array}
\right)
&= &
\begin{array}{cc}
(\hat{\bS}_{m,s})^{-1}&
\left(
\begin{array}{l}
-\hS_{t+1}\\
-\hS_{t+2}\\
\vdots\\
-\hS_{s+t}\\
\end{array}\right).
\end{array}
\end{eqnarray*}
 
Once we have obtained $\si_1(x)$,
we can compute $\om(x)$ by calculating
the coefficients $j$, $0\leq j\leq t+s-1$, of $\si_1(x)\hS(x)$ and
changing their sign.
We then find the error values using~(\ref{eq2.9'}).

Consider now the polynomial $R(x)$ given in the problem. It has erasures
in locations 1 and 3, therefore, the erasure-locator polynomial is 

$$\si_2(x)\eq (x-\al^{-1})(x-\al^{-3})\eq \al^{11}+\al^5x+x^2.$$

Taking as 0 the erased locations, the syndromes are

$$
\begin{array}{lclcl}
S_0&=&R(1)    &=&\al^{12}\\
S_1&=&R(\al)  &=&\al^2\\
S_2&=&R(\al^2)&=&\al\\
S_3&=&R(\al^3)&=&\al^5\\
S_4&=&R(\al^4)&=&\al^{10}\\
S_5&=&R(\al^5)&=&\al^2,
\end{array}
$$

therefore, 

\begin{eqnarray*}
S(x)&=&\al^{12}+\al^2x+\al x^2+\al^5x^3+\al^{10}x^4+\al^2x^5.
\end{eqnarray*}

The generalized syndrome polynomial is

$$\hS(x)\eq \si_2(x)S(x)\eq
\al^{8}+\al^{14}x+\al^7 x^2+\al^9x^3+\al^{14}x^4+\al^9x^5+\al^6x^6+\al^2x^7.$$

Since $t$, the number of erasures, is equal to 2 and $m=0$, 
according to~(\ref{meq2.17'1}), we have to estimate first

$$
\hat{\bS}_{0,2}\eq \left(
\begin{array}{ll}
\hS_{3}&\hS_{4}\\
\hS_{4}&\hS_{5}
\end{array}
\right)\eq
\left(
\begin{array}{ll}
\al^9&\al^{14}\\
\al^{14}&\al^9
\end{array}
\right)\neq 0.
$$

Therefore, we have to solve the system given by~(\ref{meq2.171}), which in this
particular case is

\begin{eqnarray*}
\left(
\begin{array}{ll}
\al^9&\al^{14}\\
\al^{14}&\al^9
\end{array}
\right)
\left(
\begin{array}{l}
\si_1^{(1)}\\
\si_0^{(1)}
\end{array}
\right)&=&
\left(
\begin{array}{ll}
\al^7\\
\al^9
\end{array}
\right).
\end{eqnarray*}

Solving this system, we obtain the following error-locator polynomial:

$$\si_1(x)\eq \al^{9}+\al^2x+x^2.$$

The roots of this polynomial are $\al^3$ and $\al^6$, therefore, the errors
are in locations 9 and 12. The error-erasure locator polynomial is given by 

$$\si(x)\eq \si_1(x)\si_2(x)\eq \al^{5}+\al^2x+\al^{12}x^2+\al x^3+x^4.$$

The error evaluator polynomial is given by the coefficients 0 to 3 of the
product $\si(x)S(x)$. This gives 

$$\om(x)\eq \al^{2}+\al x+\al^{8}x^2+\al^7 x^3.$$

The derivative of $\si(x)$ is

$$\si'(x)\eq \al^2+\al x^2.$$

Using~(\ref{eq2.9'}), the error values are:

$$\begin{array}{lclcl}
E_{1}&=&{\om(\al^{14})\al \over \si'(\al^{14})}&=& \al^{12}\\
E_{3}&=&{\om(\al^{12})\al^{3}\over \si'(\al^{12})}&=& 0\\
E_{9}&=&{\om(\al^{6})\al^{9}\over \si'(\al^{6})}&=& \al^{6}\\
E_{12}&=&{\om(\al^{3})\al^{12}\over \si'(\al^{3})}&=& \al^{6}
\end{array}$$

Substracting these error values from $R(x)$ at locations 1, 3, 9 and 12, we
obtain the decoded vector 

$$C(x)=1+\al^{12}x+\al^{10}x^2+
\al x^4+\al^{8}x^5+\al^{10}x^6+\al^{6}x^7+\al^{8}x^8+
\al^{5}x^{10}+\al^{6}x^{11}+\al^{10}x^{12}+x^{13}+\al^{6}x^{14}.$$

\section{Decoding RS Codes with Euclid's Algorithm}
\label{sec2.7}

Given two polynomials or integers $A$ and $B$, Euclid's algorithm
provides a recursive procedure to find the greatest common divisor $C$
between $A$ and $B$, denoted $C=\gcd(A,B)$. Moreover, the algorithm
also finds two polynomials or integers $S$ and $T$ such that
$C=SA+TB$.

Recall that we want to solve the key equation

$$\mu(x)x^{n-k}+\si(x)S(x)=-\om(x).$$

In the recursion, $x^{n-k}$ will have the role of $A$ and
$S(x)$ the role of $B$; $\si(x)$ and $\om(x)$ will be obtained at
a certain step of the recursion.

Let us describe Euclid's algorithm for integers or polynomials.
Consider $A$ and $B$ such that $A\geq B$ if they are integers
and $\deg(A)\geq\deg(B)$ if they are polynomials. We start from
the initial conditions $r_{-1}=A$ and $r_0=B$.

We perform a recursion in steps
$1, 2,\ldots,i,\ldots$. At step $i$ of the recursion,
we obtain $r_i$ as the
residue of dividing $r_{i-2}$ by $r_{i-1}$, i.e.,
$r_{i-2}=q_ir_{i-1}+r_i$,
where $r_i<r_{i-1}$ for integers and
$\deg(r_i)<\deg(r_{i-1})$ for polynomials.
The recursion is then given by

\begin{equation}
\label{eq2.23'}
r_{i}=r_{i-2}-q_ir_{i-1}.
\end{equation}

We also obtain values $s_i$ and $t_i$ such that
$r_i=s_iA+t_iB$. Hence, the same recursion is valid
for $s_i$ and $t_i$ as well:

\begin{eqnarray}
\label{eq2.24}
s_i&=&s_{i-2}-q_is_{i-1}\\
\label{eq2.25}
t_i&=&t_{i-2}-q_it_{i-1}
\end{eqnarray}

Since $r_{-1}=A=(1)A+(0)B$ and
$r_{ 0}=B=(0)A+(1)B$, we set the initial conditions
$s_{-1}=1$, $t_{-1}=0$, $s_0=0$ and $t_0=1$.

Let us illustrate the process with $A=124$ and $B=46$.
We will find $\gcd(124,46)$. The idea is to
divide recursively by the residues of the division until
obtaining a last residue 0. Then, the last divisor is the $\gcd$.
The procedure works as follows:

$$\begin{array}{rcrcr}
124&=&( 1)124&+& (0)46\\
46 &=&( 0)124&+& (1)46\\
32 &=&( 1)124&+&(-2)46\\
14 &=&(-1)124&+&(3)46\\
 4 &=&( 3)124&+&(-8)46\\
 2 &=&(-10)124&+&(27)46
\end{array}$$

Since 2 divides 4, 2 is the greatest common divisor between
124 and 46.

The best way to develop the process above,
is to construct
a table for $r_i$, $q_i$, $s_i$ and $t_i$ using the initial
conditions and recursions~(\ref{eq2.23'}),
(\ref{eq2.24}) and~(\ref{eq2.25}).

Let us do it again for 124 and 46.

$$\begin{array}{|r|r|r|crc|crc|}
\hline
i & r_i & q_i & s_i&=&s_{i-2}-q_is_{i-1} & t_i&=&t_{i-2}-q_it_{i-1}\\
\hline
-1 & 124 & && 1& && 0&\\
0  & 46  & && 0& & &1&\\
1  & 32  &2& &1& &&-2&\\
2  & 14  &1&&-1& && 3&\\
3  &  4  &2&& 3& &&-8&\\
4  &  2  &3&&-10&&&27&\\
5  &  0  &2&& 23&&&-62&\\
\hline
\end{array}$$

From now on, let us concentrate on Euclid's algorithm for polynomials.
If we want to solve the key equation

$$\mu(x)x^{n-k}+\si(x)S(x)=-\om(x),$$

and the error correcting capability of the code has not been exceeded,
then applying Euclid's algorithm to $x^{n-k}$ and to $S(x)$,
at a certain point of the recursion we obtain

$$r_i(x)=s_i(x)x^{n-k}+t_i(x)S(x),$$

where $\deg (r_i)\leq \lf (n-k)/2\rf-1$, and $i$ is the first with
this property. Then, $\om(x)=-\lambda
r_i(x)$ and $\si(x)=\lambda t_i(x)$,
where $\lambda$ is a constant that makes $\si(x)$ monic.
For a proof that Euclid's algorithm gives the right solution,
see~\cite{bl} or~\cite{mc}.

We illustrate the decoding of RS codes using
Euclid's algorithm with an example. Notice that
we are interested in $r_i(x)$ and $t_i(x)$ only.

\begin{ex}
\label{ex7.1}
{\em
Consider the $[7,3,5]$ RS code over $GF(8)$ of Example~\ref{ex6.1},
and assume that we want to decode the received vector

$$\ur=(0 1 1\;1 0 1\;1 1 1\;1 1 1\;1 1 1\;1 0 1\;0 1 0),$$

which in polynomial form is

$$R(x)=\al^4+\al^6x+\al^5 x^2+\al^5x^3+\al^5x^4+\al^6 x^5+\al x^6.$$

This vector was decoded in Example~\ref{ex6.1} using the
Peterson-Gorenstein-Zierler decoder.
We will decode it next using Euclid's algorithm.
Evaluating the syndromes, we obtain

\begin{eqnarray*}
S_1=R(\al^{\phantom{2}})&=&\al^5\\
S_2=R(\al^2)&=&\al\\
S_3=R(\al^3)&=&0\\
S_4=R(\al^4)&=&\al^3
\end{eqnarray*}

Therefore, the syndrome polynomial is $S(x)=\al^5+\al x+\al^3x^3.$

Next, we apply Euclid's algorithm with respect to $x^4$ and
to $S(x)$. When we find the first $i$ for which $r_i(x)$ has
degree $\leq 1$, we stop the algorithm and we obtain $\om(x)$
and $\si(x)$. The process is tabulated below.

$$\begin{array}{|r|l|l|l|}
\hline
i & r_i=r_{i-2}-q_ir_{i-1} & q_i & t_i=t_{i-2}-q_it_{i-1}\\
\hline
-1 & x^4                 & &           0     \\
0  &\al^5+\al x+\al^3x^3 &             & 1     \\
1  &\al^2x+\al^5x^2      &\al^4x       & \al^4x\\
2  &\al^5+\al^2x          &\al^2+\al^5x & 1+\al^6x+\al^2x^2\\
\hline
\end{array}$$

So, for $i=2$, we obtain a polynomial
$r_2(x)=\al^5+\al^2x$ of degree 1. Now, multiplying
both $r_2(x)$ and $t_2(x)$ by $\lambda=\al^5$, we obtain
$\om(x)=\al^3+x$ and $\si(x)=\al^5+\al^4x+x^2$.
This result coincides with the one of Example~\ref{ex6.1}, so
the rest of the solution proceeds the same way.\qed
}
\end{ex}

We end this section by stating the Euclid's Algorithm Decoder explicitly.

\begin{alg}[Euclid's Algorithm Decoder]
\label{algeu}
{\em
Consider an $[n,k]$ RS code. Assume that we want to correct up to $s$
errors, where $2s\leq n-k$. Let $R(x)$
be a received vector (in polynomial form). Then:

\begin{tabbing}
Compute the syndromes $S_j=R(\al^j)$, $1\leq j\leq n-k$,
and let $S(x)\eq \sum_{j=1}^{n-k}S_jx^{j-1}$.\\
If $S_j\eq 0$ for $1\leq j\leq n-k$, then output $R(x)$ and stop. \\
Else, \= set $r_{-1}(x)\la x^{n-k}$, $r_0(x)\la S(x)$,
             $t_{-1}(x)\la 0$, $t_0(x)\la 1$ and $i\la 1$.\\
\> {\bf LOOP:} \= Using Euclid's algorithm, find $r_i(x)$
such that $r_{i-2}(x)\eq r_{i-1}(x)q_i(x)+r_i(x)$\\
\>\> with \= $\deg (r_i)<\deg (r_{i-1})$
and set $t_{i}(x)\la t_{i-2}(x)-t_{i-1}(x)q_i(x)$.\\
\>\>\> If $\deg (r_i)\geq s$, then set $i\la i+1$ and go to {\bf LOOP}.\\
\>\>\> Else,\= find $\lambda$ such that $\lambda t_i(x)$ is monic, and let
$\sigma(x)\eq \lambda t_i(x)$ and \\
\>\>\>\> $\omega(x)\eq -\lambda r_i(x)$. \\
\>\>\> Find $\E=\{l:\si(\al^{-l})=0\}$ by searching the roots of $\si(x)$.\\
\>\>\> Compute the error values $E_l=\om(\al^{-l})/\si'(\al^{-l})$ for
$l\in\E$.\\
\>\>\> Define \= $E(x)$ \= as the polynomial with coefficients
$E_l$ when $l\in\E$, \\
\>\>\>\>\> 0 elsewhere.\\
\>\>\>\> If \= $E(\al^i)\neq S_i=R(\al^i)$ \= for some $i$, $1\leq i\leq n-k$,
then declare \\
\>\>\>\>\> an uncorrectable error and stop.\\
\>\>\>\> Else, output $R(x)-E(x)$ as the
estimate of the \\
\>\>\>\>\> transmitted polynomial.
\end{tabbing}
}
\end{alg}

\pr

\begin{prob}
\label{pr2.26}
{\em
Solve problems \ref{pr2.21'}-\ref{pr2.23} using Euclid's algorithm.
}
\end{prob}
 
\begin{prob}
\label{pr2.27}
{\em
Using the key equation for errors and erasures obtained in
Problem~\ref{pr2.20}, obtain a version of
Euclid's algorithm for decoding errors and erasures.
Use it to decode $\ur$, where $\ur$ is the same as in
Problem~\ref{pr2.24}.
}\end{prob}
 
\begin{prob}
\label{pr2.271}
{\em
As in Problem~\ref{pr2.242}, consider the $[15,9]$ RS code over $GF(16)$
whose roots are $1,\al,\ldots,\al^5$. Use the error-erasure version 
Euclid's algorithm to decode $R(x)$, $R(x)$ being the same polynomial
as in Problem~\ref{pr2.242}.
}\end{prob}

\begin{prob}
\label{pr2.28}
{\em
Write a computer program implementing Euclid's algorithm for decoding
both errors and erasures.
}\end{prob}

\sol
\vspace{.8cm}

{\bf Problem~\ref{pr2.26}}
 
Consider Problem~\ref{pr2.21'}.
Using the syndromes found in this problem,
the syndrome polynomial is given by
 
$$S(x)\,=\,\al^3+\al^2x+\al^{12}x^2+\al^{12}x^4+\al x^5.$$
 
We apply now Euclid's algorithm with respect to $x^6$ and $S(x)$.
Proceeding as in Example~\ref{ex7.1}, we obtain the following
table:
 
$$\begin{array}{|r|l|l|l|}
\hline
i & r_i=r_{i-2}-q_ir_{i-1} & q_i & t_i=t_{i-2}-q_it_{i-1}\\
\hline
-1 & x^6                 & &           0     \\
0  & \al^3+\al^2x+\al^{12}x^2+\al^{12}x^4+\al x^5&             & 1     \\
1  &\al^{13}+\al^7x+\al^{14}x^2\al^{11}x^3+\al^7x^4 &\al^{10}+\al^{14}x
&\al^{10}+\al^{14}x \\
2  &\al^{11}+\al^5x+x^2\al^{13}x^3
&\al^7+\al^9x &\al^8+\al^{12}x+\al^8x^2      \\
3  &\al^6+\al^{10}x+\al^4x^2 &\al^4+\al^9x
&\al^3+\al^{12}x+\al^4x^2+\al^2x^3  \\
\hline
\end{array}$$
 
Multiplying $t_3(x)$ and $r_3(x)$ by $\al^{13}$,
we obtain $\si(x)=\al+\al^{10}x+\al^2x^2+x^3$ and
$\om(x)=\al^4+\al^8x+\al^2 x^2$. These are the same values of $\si(x)$
and of $\om(x)$ found in Problem~\ref{pr2.21'}, so the rest of the
decoding proceeds the same way.
 
Consider Problem~\ref{pr2.21}.
Using the syndromes found in this problem,
the syndrome polynomial is given by
 
$$S(x)\,=\,\al^5+x^2+\al^7x^3.$$
 
We apply now Euclid's algorithm with respect to $x^4$ and $S(x)$.
Proceeding as in Example~\ref{ex7.1}, we obtain the following
table:
 
$$\begin{array}{|r|l|l|l|}
\hline
i & r_i=r_{i-2}-q_ir_{i-1} & q_i & t_i=t_{i-2}-q_it_{i-1}\\
\hline
-1 & x^4                 & &           0     \\
0  &\al^5+x^2+\al^7x^3 &             & 1     \\
1  & \al^7+\al^2x+\al^2x^2 &\al^6+\al x          &\al^2+\al^5x \\
2  &\al^4+\al^7x &\al^3+\al^5x &\al^7+\al^2x+\al^6x^2      \\
\hline
\end{array}$$
 
Multiplying $t_2(x)$ by $\al^2$ and $r_2(x)$ by $-\al^2=\al^6$,
we obtain $\si(x)=\al+\al^4x+x^2$ and
$\om(x)=\al^2+\al^5 x$. These are the same values of $\si(x)$
and of $\om(x)$ found in Problem~\ref{pr2.21}, so the rest of the
decoding proceeds the same way.
 
Consider Problem~\ref{pr2.22}.
Using the syndromes found in this problem,
the syndrome polynomial is given by
 
$$S(x)\,=\,\al^7+\al^7x^2+\al^4x^3+x^5.$$
 
We apply now Euclid's algorithm with respect to $x^6$ and $S(x)$.
Proceeding as in Example~\ref{ex7.1}, we obtain the following
table:
 
$$\begin{array}{|r|l|l|l|}
\hline
i & r_i=r_{i-2}-q_ir_{i-1} & q_i & t_i=t_{i-2}-q_it_{i-1}\\
\hline
-1 & x^6                 & &           0     \\
0  & \al^7+\al^7x^2+\al^4x^3+x^5 &             & 1     \\
1  &\al^3x+\al^3x^3+x^4 & x           &\al^4x \\
2  &\al^7+\al^6x+\al^3x^2+\al^7x^3 &\al^7+x &1+\al^7x+x^2      \\
3  &\al+\al^3x+\al^6x^2 &\al^6+x &\al^2+\al^4x+\al^5x^2+\al^5x^3  \\
\hline
\end{array}$$
 
Multiplying $t_3(x)$ by $\al^3$ and $r_3(x)$ by $-\al^3=\al^7$,
we obtain $\si(x)=\al^5+\al^7x+x^2+x^3$ and
$\om(x)=1+\al^2x+\al^5 x^2$. These are the same values of $\si(x)$
and of $\om(x)$ found in Problem~\ref{pr2.22}, so the rest of the
decoding proceeds the same way.
 
Consider Problem~\ref{pr2.23}.
$S(x)$ is obtained using the syndromes calculated there.
Applying Euclid's algorithm with respect to $x^8$ and $S(x)$,
we obtain the following table:
 
$$\begin{array}{|r|l|l|l|}
\hline
i & r_i=r_{i-2}-q_ir_{i-1} & q_i & t_i=t_{i-2}-q_it_{i-1}\\
\hline
-1 & x^8                 & &           0     \\
0  &7+6x+8x^2+6x^3++6x^4+x^5+8x^6+3x^7 &             & 1     \\
1  &5+3x+10x^2+10x^3+7x^4+5x^5+8x^6 &4+4x         &7+7x \\
2  &3+2x+3x^2+8x^3+6x^4+4x^5 &3+10x &2+8x+7x^2      \\
3  &2+6x+3x^2+7x^3+7x^4&1+2x &5+6x+10x^2+8x^3  \\
4  &4+7x+5x^2+9x^3     &5+10x&10+5x+7x^2+3x^3+8x^4  \\
\hline
\end{array}$$
 
We then obtain
$\si(x)=7t_4(x)=4+2x+5x^2+10x^3+x^4$ and
$\om(x)=(-7)r_4(x)=5+6x+9x^2+3x^3$.
These values of $\si(x)$
and $\om(x)$ are the same as those found in
Problem~\ref{pr2.23}, so the rest of the
decoding proceeds the same way.
 
\vspace{.8cm}
 
{\bf Problem~\ref{pr2.27}}
 
Again we refer to the modified key equation~(\ref{modkey}) obtained
in Problem~\ref{pr2.20} and to the notation in that problem
and in Problem~\ref{pr2.24}.
Assume that $t$ erasures have occurred, with $t\geq 1$.
Therefore, the code can correct up to $s=\lf (n-k-t)/2\rf$ errors,
where $n-k$ is the redundancy of the code.
 
Writing the modified key equation~(\ref{modkey}) as an equality, we have
to solve
 
\begin{eqnarray}
\label{modkey'}
\si_1(x)\hS(x)+\mu(x)x^{n-k}&=&-\om(x).
\end{eqnarray}
 
Next we apply the Euclid's algorithm process with respect to
$r_{-1}(x)=\hS(x)$ and to $r_0(x)=x^{n-k}$. Notice that, since $t\geq 1$,
$\deg(\hS(x))\geq n-k$. At step $n$ of the algorithm,
we find $s_n(x)$ and $t_n(x)$ such that
$r_n(x)\eq s_n(x)\hS(x)+t_n(x)x^{n-k}$, where $r_n(x)$ is the residue
of dividing $r_{n-2}(x)$ by $r_{n-1}(x)$. The algorithm
stops when $\deg(r_n(x))\leq s+t-1$ (recall that in the case of no
erasures, i.e., $t=0$, the algorithm stopped when
$\deg(r_n(x))\leq s-1$).
Therefore, for that $n$, $\om(x)=-\lambda r_n(x)$ and
$\si_1(x)=\lambda s_n(x)$, where $\lambda$ is a constant making $\si_1(x)$
monic. Finally, the error-erasure locator polynomial is
$\si(x)=\si_1(x)\si_2(x)$, and the rest of the algorithm proceeds like	
in Problem~\ref{pr2.24}.
 
Next we apply Euclid's algorithm to decode the received vector given
in Problem~\ref{pr2.24}.
We had found in that problem that
$r_{-1}(x)=\hS(x)=
\al^{8}+\al^{2}x+\al^{10}x^2+\al^{13}x^3+\al x^4+
\al^{4}x^5+\al^{13}x^6+\al^{14}x^7$.
Also, $r_0(x)=x^6$.
Notice that $s=\lf (n-k-t)/2\rf=\lf (6-2)/2\rf=2$, thus,
the algorithm will stop when $\deg(r_n(x))\leq s+t-1=3$.
 
Applying Euclid's algorithm, we obtain
 
$$\begin{array}{|r|l|l|l|}
\hline
i & r_i=r_{i-2}-q_ir_{i-1} & q_i & s_i=s_{i-2}-q_is_{i-1}\\
\hline
-1 & \al^{8}+\al^{2}x+\al^{10}x^2+\al^{13}x^3+\al x^4+& & 1     \\
&\al^{4}x^5+\al^{13}x^6+\al^{14}x^7 & &      \\
0  &x^6 &             & 0  \\
1  &\al^{8}+\al^{2}x+\al^{10}x^2+\al^{13}x^3+\al x^4+\al^{4}x^5  &
\al^{13}+\al^{14}x &1 \\
2  &\al +\al^{2}x+\al^{8}x^2  &
\al^{8}+\al^{11}x &\al^{8}+\al^{11}x  \\
\hline
\end{array}$$
 
Therefore, $-\om(x)=\al^4 r_2(x)=\al^{5}+\al^{6}x+\al^{12}x^2$
and $\si_1(x)=\al^4 s_2(x)=\al^{12}+x$. These values coincide with those
obtained in Problem~\ref{pr2.24}, so the rest of the solution proceeds
the same way.

\vspace{.8cm}
 
{\bf Problem~\ref{pr2.271}}
 
Next we apply Euclid's algorithm to decode the received vector given
in Problem~\ref{pr2.242}.
We had found in that problem that

$$r_{-1}(x)=\hS(x)=
\al^{8}+\al^{14}x+\al^7 x^2+\al^9x^3+\al^{14}x^4+\al^9x^5+\al^6x^6+\al^2x^7.$$

Also, $r_0(x)=x^6$.
Notice that $s=\lf (n-k-t)/2\rf=\lf (6-2)/2\rf=2$, thus,
the algorithm will stop when $\deg(r_n(x))\leq s+t-1=3$.
 
Applying Euclid's algorithm, we obtain
 
$$\begin{array}{|r|l|l|l|}
\hline
i & r_i=r_{i-2}-q_ir_{i-1} & q_i & s_i=s_{i-2}-q_is_{i-1}\\
\hline
-1 & \al^{8}+\al^{14}x+\al^7 x^2+\al^9x^3+\al^{14}x^4+& & 1     \\
&\al^9x^5+\al^6x^6+\al^2x^7& &      \\
0  &x^6 &             & 0  \\
1  &\al^{8}+\al^{14}x+\al^{7}x^2+\al^{9}x^3+\al^{14} x^4+\al^{9} x^5  &
\al^{6}+\al^{2}x &1 \\
2  &\al^{4}+\al^{11}x+\al^{11}x^2+\al^{7}x^3+\al^{5}x^4  &
\al^{11}+\al^{6}x &\al^{11}+\al^{6}x  \\
3  &\al^{12}+\al^{11}x+\al^{3}x^2+\al^{2}x^3 &
\al^{5}+\al^{4}x &\al^{4}+\al^{12}x+\al^{10}x^2  \\
\hline
\end{array}$$
 
Therefore, $-\om(x)=\al^5 r_3(x)=\al^{2}+\al x+\al^{8}x^2+\al^{7}x^3$
and $\si_1(x)=\al^5 s_3(x)=\al^{9}+\al^{2}x+x^2$. 
These values coincide with those
obtained in Problem~\ref{pr2.242}, so the rest of the solution proceeds
the same way.

\section{BCH Codes}
\label{sec2.8}

Bose-Chaudhuri-Hocquenghem codes, or more briefly, BCH codes,
are important historically because they are the first class of
codes correcting any number of errors, extending the results of Hamming
on codes correcting only one error.

Given a field $F$, we say that
a set $F'$ is a {\em subfield} of $F$ if $F'\subseteq F$ and
$F'$ is also a field. In particular, since we are dealing only with
finite fields, $F\eq GF(q)$ and $F'\eq GF(q')$, with
$q=p^b$, $q'=p^{b'}$, $b'\leq b$, $p$ a prime. Also,
$F'-\{0\}$ is a multiplicative subgroup of
$F-\{0\}$. Since the order (i.e., number of elements)
of a subgroup divides the order of the group,
$|F'-\{0\}|\eq p^{b'}-1$ divides
$|F-\{0\}|\eq p^b-1$, and thus $b'$ divides $b$ (see Problem~\ref{pr2.29}).
For instance, $GF(2^5)$ has no subfields except the trivial ones
(i.e., $GF(2)$ and $GF(2^5)$).

The converse can also be proven: given a finite field $GF(q)$,
$q\eq p^b$, if $b'$ divides $b$ and $q'\eq p^{b'}$, then $GF(q')$
is a subfield of $GF(q)$. In effect, if $\al$ is a primitive element
in $GF(q)$, and $l\eq (q-1)/(q'-1)$, it can be shown that $\be\eq \al^l$
is a primitive element in $GF(q')$. In effect, notice that
$\be$ has order $q'-1$, since
$\be^{q'-1}\eq \al^{l(q'-1)}\eq \al^{q-1}\eq 1$.
Also, if $1\leq t\leq q'-2$, then
$\be^t\eq \al^{lt}\neq 1$ since $1\leq lt\leq q-2$ and $\al$ is
primitive in $GF(q)$.

In order to prove that 0 together with the
$q'-1$ powers of $\be$ form a field, we need the concept of minimal
polynomial, to be defined below. But now we are ready to define
BCH codes.

Given a linear code $\C$, we say that $\C'$ is a subcode of $\C$ if
$\C'\subseteq \C$ and $\C'$ is also linear.

\begin{defin}
\label{defBCH}
{\em Let $GF(q)$ be a field and
$GF(q')$ a subfield of $GF(q)$.
Consider an $[n,k]$ RS code $\C$ over
$GF(q)$, where $n=q-1$, and the generator polynomial has the form
$g(x)=\prod_{i=1}^{n-k}(x-\al^i)$, $\al$ a primitive element in
$GF(q)$. A BCH code $\C'$ over $GF(q')$ corresponding to $\C$
is the subcode of $\C$ consisting of those codewords
whose entries are in $GF(q')$, i.e.,
$C(x)\eq \sum_{i=0}^{n-1}c_ix^i\in\C'$ if and only if
$C(x)\eq \sum_{i=0}^{n-1}c_ix^i\in\C$ and $c_i\in GF(q')$, $0\leq i\leq n-1$.
}
\end{defin}

Notice that the BCH code $\C'$ as given by Definition~\ref{defBCH} is
still cyclic, but it is not
clear yet what its dimension and minimum distance are. We can only say
at this point that
$d'\geq n-k+1$, where
$d'$ denotes the minimum distance of $\C'$.
Apparently, if $d'> n-k+1$, it looks like we are breaking the
Singleton bound, but the dimension of the BCH code, let's call it $k'$,
goes down, i.e., $k'<k$, so there will be no violation. In a while, we
will show how to find $k'$.

Often, the minimum distance $n-k+1$ of the underlying RS code
is called the {\em designed distance} of the BCH code. Notice also that,
in particular, if
$GF(q')=GF(q)$, then $\C\eq \C'$, so RS codes may be
considered as special cases of BCH codes.

An important case is when $q\eq 2^b$ and we take as a subfield $GF(2)$,
so we obtain binary BCH codes. Notice also that the consecutive powers
of $\al$ do not need to start at $\al\eq 1$, but at any power $\al^m$.
The same considerations given for RS codes apply here.

In order to determine the dimension of a BCH code, we need
to obtain its generator polynomial, since the degree of the generator
polynomial is equal to the redundancy $n-k'$.

We need a couple of definitions. Given an element $\be$ in $GF(q)$,
consider the smallest degree polynomial with coefficients in $GF(q')$
having $\be$ as a root. We call such a polynomial the {\em minimal
polynomial} of $\be$ with respect to $GF(q')$, and we denote it
$f_{\beta}(x)$.
In other words, $f_{\beta}(x)$ is a polynomial
with coefficients in $GF(q')$ such that
$f_{\beta}(\be)\eq 0$ and, if $g(x)$
is a polynomial with coefficients in $GF(q')$ such that $g(\be)\eq 0$,
then $\deg(f_{\beta})\leq\deg(g)$.
When we refer to a minimal polynomial of $\be$, we will omit the
``with respect to $GF(q')$'' when the context is clear.

So, consider $f_{\beta}(x)$. An important observation is that $f_{\beta}(x)$
is irreducible over $GF(q')$. In effect,
assume that $f_{\beta}(x)\eq h(x)q(x)$,
where both $h(x)$ and $q(x)$ have degree
smaller than the degree of $f_{\beta}(x)$ and
their coefficients are in $GF(q')$.
In particular, $f_{\beta}(\be)\eq h(\be)q(\be)\eq 0$, so, either
$h(\be)\eq 0$ or $q(\be)\eq 0$. This contradicts the minimality of
the degree of $f_{\beta}(x)$.

Also, if $g(x)$
is a polynomial with coefficients in $GF(q')$ such that $g(\be)\eq 0$,
then $f_{\beta}(x)$ divides $g(x)$. In effect, assume that it does not,
then, by Euclid's algorithm,
$g(x)\eq q(x)f_{\beta}(x)+r(x)$, where $\deg (r)<\deg (f_{\beta})$.
Thus,
$0\eq g(\be)\eq q(\be)f_{\beta}(\be)+r(\be)$, i.e.,
$r(\be)\eq 0$, contradicting the minimality of $\deg (f_{\beta})$.

Since $\be\in GF(q)$, in particular, $\be^{q-1}\eq 1$, i.e.,
$\be$ is a root of the polynomial $x^{q-1}-1$. Therefore,
$f_{\beta}(x)$ divides $x^{q-1}-1$ for each $\be\in GF(q)$.

Consider now $f_{\beta}(x)$ with respect to $GF(p)$.
If $q\eq p^b$, observe that
$b'\eq \deg(f_{\beta})\leq b$. In effect,
let $f_{\beta}(x)\eq \sum_{i=0}^{b'}a_ix^i$, $a_i\in GF(p)$ and $a_{b'}\neq 0$.
Since $f_{\beta}(\be)\eq 0$, the elements $1,\be,\be^2,\ldots ,\be^{b'}$
are linearly dependent. However, the elements $1,\be,\be^2,\ldots ,\be^{b'-1}$
are linearly independent, otherwise we would have a non-zero linear
combination of them equal to 0, contradicting the minimality of
$\deg (f_{\beta})\eq b'$. The total number of linear combinations of
$1,\be,\be^2,\ldots ,\be^{b'-1}$ over $GF(p)$
is $p^{b'}\leq p^b$, so $b'\leq b$.

An easy corollary of this observation is that, if $\be$ is primitive
in $GF(q)$, then $b'\eq \deg(f_{\beta})\eq b$, since each power of $\be$
can be expressed as a linear combination of
$1,\be,\be^2,\ldots ,\be^{b'-1}$ over $GF(p)$,
and the powers of $\be$ generate
all the non-zero elements in $GF(q)$.

Consider $GF(q)$, $q\eq p^b$,
and assume that $\al$ is primitive in $GF(q)$. Let
$q'\eq p^{b'}$, where $b'$ divides $b$. As before, let
$l\eq (q-1)/(q'-1)$ and $\be\eq\al^l$.
We have seen that the powers of $\be$ generate
$q'-1$ different elements in $GF(q)$. The minimal
polynomial of $\be$, $f_{\beta}(x)$, generates a
subfield $GF(q')\subseteq GF(q)$.
Moreover, $f_{\beta}(x)$ is primitive in $GF(q')$, thus, it has
degree $b'$.

Next we want to show how to find explicitly the minimal polynomial
$f_{\beta}(x)$.
Consider the set

\begin{eqnarray}
\label{eqcon}
S_{\beta}(q')&=&\{\be,\be^{q'},\be^{(q')^2},\ldots,\be^{(q')^i},\ldots,
\be^{(q')^{(b/b')-1}}\}
\end{eqnarray}

The set $S_{\beta}(q')$ is called the set of {\em conjugates}
of $\be$ with respect to $GF(q')$ (notice that
$\be^{(q')^{(b/b')}}\eq \be^{p^{b'(b/b')}}\eq \be^{p^b}\eq \be^q\eq \be$).
Any two elements in $S_{\beta}(q')$ are said to be conjugates with
respect to $GF(q')$ (we will omit the ``with respect to $GF(q')$'' when
the context is clear). If $\be$ and $\be'$ are conjugates, it can
be proven that $S_{\beta}(q')\eq S_{\beta'}(q')$ (Problem~\ref{pr2.30'}).
Problem~\ref{pr2.30'} also
shows that the different sets of conjugates give a
partition of the non-zero elements of $GF(q)$.

\begin{ex}
\label{excon}
{\em
Consider $GF(8)$ and $\al$ a primitive element. We have,

\begin{eqnarray*}
S_{1}(2)&=&\{1\}\\
S_{\alpha}(2)&=&\{\al,\al^2,\al^4\}\\
S_{\alpha^3}(2)&=&\{\al^3,\al^6,\al^5\}
\end{eqnarray*}

Similarly, considering $GF(16)$ and $\al$ a primitive element, we obtain

\begin{eqnarray*}
S_{1}(2)&=&\{1\}\\
S_{\alpha}(2)&=&\{\al,\al^2,\al^4,\al^8\}\\
S_{\alpha^3}(2)&=&\{\al^3,\al^6,\al^{12},\al^9\}\\
S_{\alpha^5}(2)&=&\{\al^5,\al^{10}\}\\
S_{\alpha^7}(2)&=&\{\al^7,\al^{14},\al^{13},\al^{11}\}
\end{eqnarray*}

Notice that $GF(4)$ as a subfield of $GF(16)$ consists of
the elements $\{0,1,\al^5,\al^{10}\}$ (Problem~\ref{pr2.30}).
The sets of conjugates with respect to
$GF(4)$ are given by

\begin{eqnarray*}
S_{1}(4)&=&\{1\}\\
S_{\alpha}(4)&=&\{\al,\al^4\}\\
S_{\alpha^2}(4)&=&\{\al^2,\al^8\}\\
S_{\alpha^3}(4)&=&\{\al^3,\al^{12}\}\\
S_{\alpha^5}(4)&=&\{\al^5\}\\
S_{\alpha^6}(4)&=&\{\al^6,\al^{9}\}\\
S_{\alpha^7}(4)&=&\{\al^7,\al^{13}\}\\
S_{\alpha^{10}}(4)&=&\{\al^{10}\}\\
S_{\alpha^{11}}(4)&=&\{\al^{11},\al^{14}\}
\end{eqnarray*}\qed
}\end{ex}

Assume that $C(x)$ is a polynomial whose coefficients are in $GF(q')$.
We can prove that if $\be\in GF(q)$
and $C(\be)\eq 0$, then $C(\be^{q'})\eq 0$ (Problem~\ref{pr2.31}).
In particular, if \\
$C(\al)\eq C(\al^2)\eq \ldots \eq C(\al^{n-k})\eq 0$
(i.e., $C(x)$ is in the BCH code), then the conjugates of the roots
$\al,\al^2,\ldots,\al^{n-k}$ are also roots of $C(x)$.

Let $\be\in GF(q)$.
We can show that the minimal polynomial $f_{\beta}(x)$ is given by

$$f_{\beta}(x)\eq \prod_{\ga\in S_{\beta}(q')}(x-\ga)$$

We have to prove that the coefficients of
$\prod_{\ga\in S_{\beta}(q')}(x-\ga)$
are in $GF(q')$ (Problem~\ref{pr2.31}), that it
is irreducible over $GF(q')$, and that it is the smallest degree
polynomial with coefficients in $GF(q')$ having $\be$ as a root
(Problem~\ref{pr2.321}).
Also, by Problem~\ref{pr2.30'}, if $\be$ and $\be'$ are conjugates,
then $f_{\beta}(x)\eq f_{\beta'}(x)$.

\begin{ex}
\label{excyc}
{\em
Consider $GF(8)$ as given by Table~\ref{table1}. Using
Example~\ref{excon}, we have

$$\begin{array}{lclcl}
f_{1}(x)&&&=&1+x\\
f_{\alpha}(x)&=&(x+\al)(x+\al^2)(x+\al^4)&=& 1+x+x^3\\
f_{\alpha^3}(x)&=&(x+\al^3)(x+\al^6)(x+\al^5)&=& 1+x^2+x^3
\end{array}$$

Using $GF(16)$ generated by the primitive polynomial $1+x+x^4$
(Problem~\ref{pr2.8}), again by
Example~\ref{excon}, we have

$$\begin{array}{lclcl}
f_{1}(x)&&&=&1+x\\
f_{\alpha}(x)&=&(x+\al)(x+\al^2)(x+\al^4)(x+\al^8)&=& 1+x+x^4\\
f_{\alpha^3}(x)&=&(x+\al^3)(x+\al^6)(x+\al^{12})(x+\al^9)&=& 1+x+x^2+x^3+x^4\\
f_{\alpha^5}(x)&=&(x+\al^5)(x+\al^{10})&=& 1+x+x^2\\
f_{\alpha^7}(x)&=&(x+\al^7)(x+\al^{14})(x+\al^{13})(x+\al^{11})&=& 1+x^3+x^4
\end{array}$$

Finally, following Example~\ref{excon}, the minimal polynomials
with respect to $GF(4)$ are given by

$$\begin{array}{lclcl}
f_{1}(x)&&&=&1+x\\
f_{\alpha}(x)&=&(x+\al)(x+\al^4)&=& \al^5+x+x^2\\
f_{\alpha^2}(x)&=&(x+\al^2)(x+\al^8)&=& \al^{10}+x+x^2\\
f_{\alpha^3}(x)&=&(x+\al^3)(x+\al^{12})&=& 1+\al^{10}x+x^2\\
f_{\alpha^5}(x)&&&=&\al^5+x\\
f_{\alpha^6}(x)&=&(x+\al^6)(x+\al^{9})&=& 1+\al^{5}x+x^2\\
f_{\alpha^7}(x)&=&(x+\al^7)(x+\al^{13})&=& \al^5+\al^{5}x+x^2\\
f_{\alpha^{10}}(x)&&&=&\al^{10}+x\\
f_{\alpha^{11}}(x)&=&(x+\al^{11})(x+\al^{14})&=& \al^{10}+\al^{10}x+x^2
\end{array}$$\qed
}\end{ex}

Assume that $C(x)$ is in the BCH code
$\C'$ as given by Definition~\ref{defBCH}, then
$C(\al)=C(\al^2)=\ldots =C(\al^{n-k})=0$, $\al$ primitive in $GF(q)$,
and the coefficients of $C(x)$ are in $GF(q')$.
Consider the minimal polynomials
$f_{\alpha}(x),f_{\alpha^2}(x),\ldots,f_{\alpha^{n-k}}(x)$. Each one of them
has its coefficients in $GF(q')$ and
divides $C(x)$, then, the least common multiple of these
minimal polynomials also divides $C(x)$. Since, by
Problem~\ref{pr2.31}, for two different powers of $\al$, their
minimal polynomials are either the same
or relatively prime, then the
least common multiple is the product of the distinct minimal
polynomials. The generator polynomial $g(x)$ of the BCH code,
in particular, is also a codeword, so
this product of minimal polynomials
divides $g(x)$. Therefore, it has to coincide with $g(x)$.

\begin{ex}
\label{exBCH1}
{\em
Consider $GF(8)$ as given by Table~\ref{table1}, and BCH codes
over $GF(2)$. Take a [7,5,3] RS code $\C$ over $GF(8)$.
The corresponding BCH code $\C'$ over $GF(2)$ is given by all the codewords
$C(x)\eq\sum_{i=0}^6c_ix^i$ such that
$C(\al)\eq C(\al^2)\eq 0$ and $c_i\in GF(2)$, $0\leq i\leq 6$.
Let us find its generator polynomial $g(x)$, and thus its dimension
$k\eq 7-\deg(g)$.

Notice that, by Example~\ref{excyc},
$f_{\alpha}(x)\eq f_{\alpha^2}(x)$. Thus, the generator polynomial
is given by

$$g(x)\eq f_{\alpha}(x)\eq (x+\al)(x+\al^2)(x+\al^4)\eq 1+x+x^3.$$

So, $\C'$ has dimension 4 and minimum distance at least 3.
Since there are codewords of weight 3, like $g(x)$,
then the minimum distance is exactly 3.
This code is a cyclic version of a Hamming code,
and we have seen it in Problem~\ref{pr2.13}.

Consider now a [7,3,5] RS code $\C$ over $GF(8)$.
The corresponding BCH code $\C'$ over $GF(2)$ is given by all the codewords
$C(x)\eq\sum_{i=0}^6c_ix^i$ such that
$C(\al)\eq C(\al^2)\eq C(\al^3)\eq C(\al^4)\eq 0$
and $c_i\in GF(2)$, $0\leq i\leq 6$.
Notice that $f_{\alpha}(x)\eq f_{\alpha^2}(x)\eq f_{\alpha^4}(x)$.
Again, let us find its generator polynomial $g(x)$, and thus its dimension
$k\eq 7-\deg(g)$.

Notice that, by Example~\ref{excyc},
$f_{\alpha^3}(x)\eq (x+\al^3)(x+\al^6)(x+\al^5)\eq 1+x^2+x^3$.
So, the generator polynomial is given by

$$g(x)\eq f_{\alpha}(x)f_{\alpha^3}(x)\eq (1+x+x^3)(1+x^2+x^3)\eq
1+x+x^2+x^3+x^4+x^5+x^6.$$

The resulting BCH code is $[7,1,d]$ with $d\geq 5$. In fact, the obtained
code in this case is the repetition code, that has $d\eq 7$.\qed

}\end{ex}

\begin{ex}
\label{exBCH2}
{\em
Consider $GF(16)$ generated by the primitive polynomial $1+x+x^4$
(Problem~\ref{pr2.8}).
Take a [15,13,3] RS code $\C$ over $GF(16)$.
The corresponding BCH code $\C'$ over $GF(2)$ is given by all the codewords
$C(x)\eq\sum_{i=0}^{14}c_ix^i$ such that
$C(\al)\eq C(\al^2)\eq 0$ and $c_i\in GF(2)$.
By Example~\ref{excyc},
$f_{\alpha}(x)\eq f_{\alpha^2}(x)\eq 1+x+x^4$. Thus, the generator polynomial
is given by $g(x)\eq 1+x+x^4$ and the code $\C'$ is a $[15,11,d]$ code
with $d\geq 3$. In fact, since $g(x)$ has weight 3, then $d\eq 3$, and we
obtain a cyclic version of the $[15,11,3]$ Hamming code.

Take a [15,11,5] RS code $\C$ over $GF(16)$.
The corresponding BCH code $\C'$ over $GF(2)$ is given by all the codewords
$C(x)$ with coefficients in $GF(2)$ such that
$C(\al^j)\eq 0$ for $1\leq j\leq 4$.
By Example~\ref{excyc},
$f_{\alpha}(x)\eq f_{\alpha^2}(x)\eq f_{\alpha^4}(x)\eq 1+x+x^4$ and
$f_{\alpha^3}(x)\eq 1+x+x^2+x^3+x^4$.
Thus, the generator polynomial
is given by
$g(x)\eq f_{\alpha}(x)f_{\alpha^3}(x)$, which has degree 8, so
the code $\C'$ is a $[15,7,d]$ BCH code with $d\geq 5$.

Take a [15,9,7] RS code $\C$ over $GF(16)$.
The corresponding BCH code $\C'$ over $GF(2)$ is given by all the codewords
$C(x)$ with coefficients in $GF(2)$ such that
$C(\al^j)\eq 0$ for $1\leq j\leq 6$ and $c_i\in GF(2)$, $0\leq i\leq 6$.
By Example~\ref{excyc},
$f_{\alpha^3}(x)\eq f_{\alpha^6}(x)$.
Thus, the generator polynomial
is given by
$g(x)\eq f_{\alpha}(x)f_{\alpha^3}(x)f_{\alpha^5}(x)$, which has degree 10, so
the code $\C'$ is a $[15,5,d]$ BCH code with $d\geq 7$. \qed

}\end{ex}

\begin{ex}
\label{exBCH3}
{\em
Let us consider now BCH codes over $GF(4)$ when $GF(4)$ is taken
as a subfield of $GF(16)$, $GF(16)$ being the same as in
Example~\ref{exBCH2}.

Take a [15,13,3] RS code $\C$ over $GF(16)$.
The corresponding BCH code $\C'$ over $GF(4)$ is given by all the codewords
$C(x)\eq\sum_{i=0}^{14}c_ix^i$ such that
$C(\al)\eq C(\al^2)\eq 0$ and $c_i\in GF(4)$.
By Example~\ref{excyc},
$g(x)\eq f_{\alpha}(x)f_{\alpha^2}(x)\eq (\al^5+x+x^2)(\al^{10}+x+x^2)$,
thus, since $g(x)$ has degree 4, the BCH code
$\C'$ is a $[15,11,d]$ code over $GF(4)$
with $d\geq 3$.

Take a [15,11,5] RS code $\C$ over $GF(16)$.
The corresponding BCH code $\C'$ over $GF(4)$ is given by all the codewords
$C(x)$ with coefficients in $GF(4)$ such that
$C(\al^j)\eq 0$ for $1\leq j\leq 4$.
By Example~\ref{excyc},
$f_{\alpha}(x)\eq f_{\alpha^4}(x)$, so,
$g(x)\eq f_{\alpha}(x)f_{\alpha^2}(x)f_{\alpha^3}(x)$, which has degree 6.
Thus, the code $\C'$ is a $[15,9,d]$ BCH code over $GF(4)$ with $d\geq 5$.

Take a [15,9,7] RS code $\C$ over $GF(16)$.
The corresponding BCH code $\C'$ over $GF(4)$ is given by all the codewords
$C(x)$ with coefficients in $GF(4)$ such that
$C(\al^j)\eq 0$ for $1\leq j\leq 6$.
By Example~\ref{excyc},
$g(x)\eq f_{\alpha}(x)f_{\alpha^2}(x)f_{\alpha^3}(x)
f_{\alpha^5}(x)f_{\alpha^6}(x)$,
which has degree 9.
Thus, the code $\C'$ is a $[15,6,d]$ BCH code over $GF(4)$ with $d\geq 7$.
\qed
}\end{ex}

BCH codes can be decoded using the decoding algorithms of RS codes. In
some cases, the decoding is going to be easier. For instance, if we
are correcting errors using a BCH code over $GF(2)$, it is enough to
find the error locator polynomial $\si(x)$: by finding the roots of
$\si(x)$, we know the error locations, and then we simply flip the
bits in those locations. We don't need to worry about finding the
error evaluator polynomial $\om(x)$.

Let us point out that even when the minimum distance of a BCH code exceeds
the designed distance $n-k+1$, the decoding algorithm decodes up to
the designed distance, since in fact it is correcting the underlying
RS code.

Let us end this section by indicating how to find the isomorphism
between two versions of $GF(q)$, say, $F_1$ and $F_2$.
Assume that $\al$ is in $F_1$ with minimal polynomial
$f_{\alpha}(x)$ over $GF(p)$, which divides $x^q-1$. The degree of
$f_{\alpha}(x)$ is equal to the size of the set of conjugates of $\al$, i.e.,
$\deg (f_{\alpha})\eq |S_{\alpha}(p)|$. Now, consider
$\al_1,\al_2,\ldots,\al_m\in F_1$ such that
$F_1-\{0\}\eq\cup_{i=1}^m S_{\alpha_i}(p)$ and
$S_{\alpha_i}(p)\cap S_{\alpha_j}(p)\eq \emptyset$ for $i\neq j$.
In particular,
$\sum_{i=1}^m |S_{\alpha_i}(p)|\eq \sum_{i=1}^m \deg (f_{\alpha_i})\eq q-1$.
Since each $f_{\alpha_i}(x)$ divides $x^q-1$, the product of the
$f_{\alpha_i}$'s also divides $x^q-1$, since they are relatively prime.
Since the sum of their degrees equals $q-1$, this means, the product
of the $f_{\alpha_i}$'s equals $x^q-1$, giving a unique prime factorization
of $x^q-1$. Explicitly,

\begin{eqnarray*}
x^q-1&=&\prod_{i=1}^mf_{\alpha_i}(x).
\end{eqnarray*}

If we consider $F_2$, we can repeat the process and find $m'$ elements
$\be_1,\be_2,\ldots,\be_{m'}\in F_2$ such that

\begin{eqnarray*}
x^q-1&=&\prod_{j=1}^{m'}f_{\beta_j}(x).
\end{eqnarray*}

But since the factorization of $x^q-1$ over $GF(p)$ in irreducible
factors must be unique, this means, $m\eq m'$, and for each $\al_i\in F_1$,
there is a $\be_j\in F_2$ such that $f_{\alpha_i}(x)\eq f_{\beta_j}(x)$.

Now, let $\al$ be primitive in $F_1$.
By the previous observation, we know that there is
an element $\be\in F_2$
such that $f_{\beta}(x)\eq f_{\alpha}(x)$. The isomorphism
$h:F_1\ra F_2$ is determined by $h(\al)\eq \be$. Since this is an
isomorphism, $h(\al^i)\eq (h(\al))^i\eq \be^i$. The fact that
$f_{\beta}(x)\eq f_{\alpha}(x)$ determines that
$h(\al^i  +\al^j)\eq h(\al^i)+h(\al^j)\eq \be^i+\be^j$.
Let us illustrate the isomorphism with an example.

\begin{ex}
\label{exiso}
{\em Let $F_1$ be $GF(9)$ generated by the primitive polynomial
$2+x+x^2$ (Problem~\ref{pr2.9}) and
$F_2$ be $GF(9)$ generated by the primitive polynomial
$2+2x+x^2$. Let $\al$ be a primitive element in $F_1$ and
$\be$ be a primitive element in $F_2$.
We give $F_1$ and $F_2$ below.

$$
\begin{array}{cc}
\begin{array}{|c|c|c|}
\hline
{\rm Vector}&{\rm Polynomial}&{\rm Power\;of}\;\al\\
\hline
00&0&0\\
10&1&1\\
01&\al &\al\\
12&1+2\al&\al^2\\
22&2+2\al&\al^3\\
20&2&\al^4\\
02&2\al&\al^5\\
21&2+\al&\al^6\\
11&1+\al&\al^7\\
\hline
\end{array}
&
\begin{array}{|c|c|c|}
\hline
{\rm Vector}&{\rm Polynomial}&{\rm Power\;of}\;\be\\
\hline
00&0&0\\
10&1&1\\
01&\be &\be\\
11&1+\be&\be^2\\
12&1+2\be&\be^3\\
20&2&\be^4\\
02&2\be&\be^5\\
22&2+2\be&\be^6\\
21&2+\be&\be^7\\
\hline
\end{array}
\end{array}
$$

Now we want to find an isomorphism $h:F_1\ra F_2$. This isomorphism
cannot be given by $h(\al)\eq \be$, since $\al$ and $\be$ have different
minimal polynomials. So, we need to find an element $\ga\in F_2$ such
that the minimal polynomial of $\ga$ is $2+x+x^2$, i.e.,
$2+\ga+\ga^2\eq 0$. Since $\ga\in F_2$ and $\be$ is primitive in $F_2$,
in particular, $\ga\eq \be^i$. Also, $\ga$ must be primitive,
so $\gcd(i,8)\eq 1$. So, consider $i\eq 3$. Replacing $\be^3$ in
$2+x+x^2$, we obtain $2+\be^3+\be^6\eq \be^7\neq 0$, using the table
of $F_2$. Consider next $i\eq 5$. Replacing $\be^5$ in
$2+x+x^2$, we obtain $2+\be^5+\be^{10}\eq 2+\be^2+\be^5\eq 0$.

Thus, the isomorphism $h:F_1\ra F_2$ is given by $h(\al)\eq \be^5$.
If we write it element to element, we obtain

\begin{eqnarray*}
h(0)&=&0\\
h(1)&=&1\\
h(\al)&=&\be^5\\
h(\al^2)&=&\be^2\\
h(\al^3)&=&\be^7\\
h(\al^4)&=&\be^4\\
h(\al^5)&=&\be\\
h(\al^6)&=&\be^6\\
h(\al^7)&=&\be^3
\end{eqnarray*}
}\end{ex}

\pr

\begin{prob}
\label{pr2.29}
{\em
Let $p$ be a prime. Prove that $p^{b'}-1$ divides $p^b-1$ if and only
if $b'$ divides $b$.
}\end{prob}

\begin{prob}
\label{pr2.30}
{\em
Find all the subfields of $GF(16)$ and $GF(64)$.
}\end{prob}

\begin{prob}
\label{pr2.30'}
{\em
Consider the field $GF(q)$, $q\eq p^b$, and let 
$GF(q')$ be a subfield of $GF(q)$, $q'\eq p^{b'}$ and $b'$ divides $b$.
Let $\be$ and $\be'$ be two elements in $GF(q)$.

\begin{enumerate}
\item
If $\be$ and $\be'$ are conjugates, 
prove that $S_{\beta}(q')\eq S_{\beta'}(q')$.

\item If $\be$ and $\be'$ are not conjugates, 
prove that $S_{\beta}(q')\cap S_{\beta'}(q')\eq\emptyset$.
\end{enumerate}

}\end{prob}

\begin{prob}
\label{pr2.31}
{\em
Consider the field $GF(q)$, $q\eq p^b$, and let 
$GF(q')$ be a subfield of $GF(q)$, $q'\eq p^{b'}$ and $b'$ divides $b$.

\begin{enumerate}
\item
Let $a\in GF(q)$.
Prove that $a\in GF(q')$ if and only if $a^{q'}\eq a$.

\item
Let $f(x)$ be a polynomial 
with coefficients in $GF(q)$. Prove that the coefficients of $f(x)$ 
are in $GF(q')$ if and only if $f(x^{q'})\eq (f(x))^{q'}$.

\item
Let $f(x)$ be a polynomial with coefficients in $GF(q')$
and let $\be\in GF(q)$. Prove that
if $f(\be)\eq 0$, then $f(\ga)\eq 0$, for any $\ga\in S_{\beta}(q')$.

\item 
Let $\be\in GF(q)$. Prove that the coefficients of the
polynomial

\begin{eqnarray}
\label{eqmin}
f_{\beta}(x)&=& \prod_{\ga\in S_{\beta}(q')}(x-\ga)
\end{eqnarray}

are in $GF(q')$.

\item
Let $\be,\ga\in GF(q)$. Prove that either
$f_{\beta}(x)\eq f_{\gamma}(x)$ or $\gcd (f_{\beta}(x),
f_{\gamma}(x))\eq 1$.
\end{enumerate}
}\end{prob}

\begin{prob}
\label{pr2.321}
{\em
Let $\beta\in GF(q)$.
Consider the polynomial $f_{\beta}(x)$ given by~(\ref{eqmin}).
Assume that $g(x)$ is a polynomial with coefficients in $GF(q')$
and $g(\be)\eq 0$. Prove that $f_{\beta}(x)$ divides $g(x)$.

}\end{prob}

\begin{prob}
\label{pr2.331}
{\em Find the dimensions of the binary BCH codes of length 31 with 
designed distances 3, 4, 5, 6, 7, 8 and 9.
}\end{prob}

\begin{prob}
\label{pr2.34}
{\em Consider $GF(16)$ generated by the primitive polynomial $1+y+y^4$
(Problem~\ref{pr2.8}), and the $[15,5]$ BCH code with designed distance 7
(Example~\ref{exBCH2}). Decode the received vector

$$\ur\eq (1\;0\;1\;0\;1\;0\;1\;0\;0\;1\;1\;1\;0\;0\;0),$$

which in polynomial form is

$$R(x)\eq 1+x^2+x^4+x^6+x^9+x^{10}+x^{11}.$$

}\end{prob}

\sol
\vspace{.8cm}

{\bf Problem~\ref{pr2.29}}

Dividing $b$ by $b'$ and finding the residue, we can write
$b\eq lb'+r$, where $0\leq r<b'$.

Now, notice that

\begin{eqnarray*}
p^b-1&=&p^{lb'+r}-1\\
&=&p^r((p^{b'})^l-1)+p^r-1\\
&=&p^r(p^{(l-1)b'}+p^{(l-2)b'}+\cdots +1)(p^{b'}-1)+p^r-1,
\end{eqnarray*}

this last equality by Problem~\ref{pr2.18}. Since $r<b'$, then
$p^r-1<p^{b'}-1$, therefore, by Euclid's algorithm, 
$p^r-1$, is the residue of dividing $p^b-1$ by $p^{b'}-1$. 
Thus, $p^{b'}-1$ divides $p^b-1$ if and only if
$p^r-1\eq 0$, if and only if $r\eq 0$, if and only if
$b'$ divides $b$.

\vspace{.8cm}
{\bf Problem~\ref{pr2.30}}

Let us start with $GF(16)\eq GF(2^4)$. The subfields of
$GF(2^4)$ are all those $GF(2^b)$ such that $b$ divides 4. 
The divisors of 4 are 1, 2 and 4 itself, so the subfields of
$GF(16)$ are $GF(2)$, $GF(4)$ and $GF(16)$ itself. 
Let us look at $GF(4)$. If $\al$ is a primitive element in $GF(16)$, 
then $\al^5$ is a primitive element in $GF(4)$ when taken as a subfield
of $GF(16)$. Therefore, $GF(4)\eq\{0,1,\al^5,\al^{10}\}$.

Similarly, $GF(64)\eq GF(2^6)$. The subfields of
$GF(2^6)$ are all those $GF(2^b)$ such that $b$ divides 6. 
The divisors of 6 are 1, 2, 3 and 6 itself, so the subfields of
$GF(64)$ are $GF(2)$, $GF(4)$, $GF(8)$ and $GF(64)$ itself. 
If $\al$ is a primitive element in $GF(64)$, 
then $\al^{21}$ is a primitive element in $GF(4)$ 
and $\al^{9}$ is a primitive element in $GF(8)$ 
when $GF(4)$ and $GF(8)$ are taken as subfields of $GF(64)$.
Therefore, 

\begin{eqnarray*}
GF(4)&=&\{0,1,\al^{21},\al^{42}\}\\
GF(8)&=&\{0,1,\al^{9},\al^{18},\al^{27},\al^{36},\al^{45},\al^{54}\}
\end{eqnarray*}

\vspace{.8cm}
{\bf Problem~\ref{pr2.30'}}

Consider the field $GF(q)$, $q\eq p^b$, and let 
$GF(q')$ be a subfield of $GF(q)$, $q'\eq p^{b'}$ and $b'$ divides $b$.
Let $\be$ and $\be'$ be two elements in $GF(q)$.

\begin{enumerate}
\item
If $\be$ and $\be'$ are conjugates, then, by~(\ref{eqcon}), there
is a $\ga\in GF(q)$ such that

\begin{eqnarray*}
S_{\gamma}(q')&=&\{\ga,\ga^{q'},\ga^{(q')^2},\ldots,\ga^{(q')^i},\ldots,
\ga^{(q')^j},\ldots,\ga^{(q')^{(b/b')-1}}\}
\end{eqnarray*}

where $\be\eq \ga^{(q')^i}$ and $\be'\eq \ga^{(q')^j}$  
for, say, $0\leq i<j\leq (b/b')-1$.
So, 

\begin{eqnarray*}
S_{\beta}(q')&=&\{\be,\be^{q'},\be^{(q')^2},\ldots,\be^{(q')^{(b/b')-1}}\}\\
&=&\{\ga^{(q')^i},\ga^{(q')^{i+1}},\ldots,
\ga^{(q')^{(b/b')-1}},\ga^{(q')^{(b/b')}},\ga^{(q')^{(b/b')+1}},\ldots,
\ga^{(q')^{(b/b')+i-1}}\}\\
&=&\{\ga^{(q')^i},\ga^{(q')^{i+1}},\ldots,
\ga^{(q')^{(b/b')-1}},\ga,\ga^{q'},\ldots,
\ga^{(q')^{i-1}}\},
\end{eqnarray*}

since $\ga^{(q')^{(b/b')}}\eq\ga$. Therefore, 
$S_{\beta}(q')\eq S_{\gamma}(q')$, and similarly,
$S_{\beta'}(q')\eq S_{\gamma}(q')$, proving the assertion.

\item If $\ga\in S_{\beta}(q')\cap S_{\beta'}(q')$,
by the previous part, 
$S_{\beta}(q')\eq S_{\gamma}(q')$ and
$S_{\beta'}(q')\eq S_{\gamma}(q')$.
Therefore, $S_{\beta}(q')\eq S_{\beta'}(q')$ and in particular, 
$\be$ and $\be'$ are conjugates, a contradiction. 

\end{enumerate}

\vspace{.8cm}
{\bf Problem~\ref{pr2.31}}

\begin{enumerate}
\item
Assume that $a\in GF(q')$. 
If $a\eq 0$, certainly $0^{q'}\eq 0$. If $a\neq 0$, then $GF(q')-\{0\}$
is a multiplicative group of order $q'-1$, and $a^{q'-1}\eq 1$,
therefore, $a^{q'}\eq a$.

Conversely, assume that $a^{q'}\eq a$, which is certainly satisfied for
$a\eq 0$. Now, consider the $q'-1$ non-zero elements in $GF(q')$. 
They constitute a (unique) subgroup of the cyclic multiplicative group
$GF(q)-\{0\}$. This subgroup $GF(q')-\{0\}$
has order $q'-1$ (remember, $q'-1$ divides $q-1$).
Now, if we take $a\neq 0$, then $a^{q'-1}\eq 1$, so $a\in GF(q')-\{0\}$.
Thus, the set of elements $a$ such that $a^{q'}\eq a$ coincides with
$GF(q')$.

\item
Let $f(x)\eq \sum_{i=0}^ma_ix^i$.
Since the field has characteristic $p$, taking powers of $p$ is a
distributive (or linear) operation on sums. Therefore, 

\begin{eqnarray*}
(f(x))^{q'}\eq (\sum_{i=0}^ma_ix^i)^{q'} \eq
\sum_{i=0}^ma_i^{q'}(x^{q'})^i 
\end{eqnarray*}

and

\begin{eqnarray*}
f(x^{q'})\eq \sum_{i=0}^ma_i(x^{q'})^i 
\end{eqnarray*}

So, $f(x^{q'})\eq (f(x))^{q'}$, if and only if $a_i^{q'}\eq a_i$ for
$0\leq i\leq m$, if and only if $a_i\in GF(q')$ by the first part.

\item
Let $f(x)\eq \sum_{i=0}^ma_ix^i$ and $a_i\in GF(q')$, $0\leq i\leq m$.
Let $f(\be)\eq 0$ and $\ga\in S_{\beta}(q')$, thus, 
$\ga\eq \be^{(q')^i}$. So, by the previous part,

\begin{eqnarray*}
f(\ga)\eq f(\be^{(q')^i})\eq (f(\be))^{(q')^i}\eq 0.
\end{eqnarray*}

\item 
Let us denote $S_{\beta}(q')$ by $S_{\beta}$, there cannot be confusion in
this case.

Notice that

\begin{eqnarray*}
(f_{\beta}(x))^{q'}&=&(\prod_{\ga\in S_{\beta}}(x-\ga))^{q'}\\
&=&\prod_{\ga\in S_{\beta}}(x^{q'}-\ga^{q'})\\
&=&\prod_{\ga\in S_{\beta^{q'}}}(x^{q'}-\ga)\\
&=&\prod_{\ga\in S_{\beta}}(x^{q'}-\ga)\;\;\;({\rm Problem~\ref{pr2.30'}})\\
&=&f_{\beta}(x^{q'}),
\end{eqnarray*} 

therefore, by part 2, the coefficients of $f_{\beta}(x)$ are in $GF(q')$. 

\item
By Problem~\ref{pr2.30'}, either 
$S_{\beta}(q')\eq S_{\gamma}(q')$, in which case 
$f_{\beta}(x)\eq f_{\gamma}(x)$, 
or $S_{\beta}(q')\cap S_{\gamma}(q')\eq \emptyset$, in which case
$\gcd (f_{\beta}(x),f_{\gamma}(x))\eq 1$, since
$f_{\beta}(x)$ and $f_{\gamma}(x))$ have no factors in common.
\end{enumerate}

\vspace{.8cm}
{\bf Problem~\ref{pr2.321}}

If $g(\be)\eq 0$, then $g(\ga)\eq 0$ for any $\ga\in S_{\beta}(q')$
by Problem~\ref{pr2.31}, part 3. So, $x-\ga$ divides $g(x)$ for
every $\ga\in S_{\beta}(q')$, therefore, $f_{\beta}(x)$ divides $g(x)$.
This shows that $f_{\beta}(x)$ as given by~(\ref{eqmin}) is the
minimal polynomial of $\be$ with respect to $GF(q')$.

\vspace{.8cm}
{\bf Problem~\ref{pr2.331}}


Let us write down the conjugacy sets of the non-zero elements in $GF(32)$.
Since $\al^{31}\eq 1$, we have

\begin{eqnarray*}
S_{1}(2)&=&\{1\}\\
S_{\alpha}(2)&=&\{\al,\al^2,\al^4,\al^8\,\al^{16}\}\\
S_{\alpha^3}(2)&=&\{\al^3,\al^6,\al^{12},\al^{24},\al^{17}\}\\
S_{\alpha^5}(2)&=&\{\al^5,\al^{10},\al^{20},\al^{9},\al^{18}\}\\
S_{\alpha^7}(2)&=&\{\al^7,\al^{14},\al^{28},\al^{25},\al^{19}\}\\
S_{\alpha^{11}}(2)&=&\{\al^{11},\al^{22},\al^{13},\al^{26},\al^{21}\}\\
S_{\alpha^{15}}(2)&=&\{\al^{15},\al^{30},\al^{29},\al^{27},\al^{23}\}
\end{eqnarray*}

This means, each minimal polynomial $f_{\alpha^i}(x)$ has degree 5.

For designed distance 3, since the roots are $\al$ and $\al^2$, 
$g(x)\eq f_{\alpha}(x)$, thus, the BCH code has dimension $31-5\eq 26$.

For designed distance 4, since the roots are $\al$, $\al^2$  and $\al^3$, 
$g(x)\eq f_{\alpha}(x)f_{\alpha^3}(x)$, thus, 
the BCH code has dimension $31-10\eq 21$. The same is valid for 
designed distance 5, since $f_{\alpha^4}(x)\eq f_{\alpha}(x)$.

For designed distance 6, 
$g(x)\eq f_{\alpha}(x)f_{\alpha^3}(x)f_{\alpha^5}(x)$, thus, 
the BCH code has dimension $31-15\eq 16$. The same is valid for 
designed distance 7, since $f_{\alpha^6}(x)\eq f_{\alpha^3}(x)$.

For designed distance 8, 
$g(x)\eq f_{\alpha}(x)f_{\alpha^3}(x)f_{\alpha^5}(x)f_{\alpha^7}(x)$, thus, 
the BCH code has dimension $31-20\eq 11$. The same is valid for 
designed distance 9, since $f_{\alpha^8}(x)\eq f_{\alpha}(x)$.

\vspace{.8cm}
{\bf Problem~\ref{pr2.34}}





The syndromes are 

$$\begin{array}{ccccl}
S_1&=& R(\al)&=&\al^{14}\\
S_2&=& R(\al^2)&=&\al^{13}\\
S_3&=& R(\al^3)&=&\al^{6}\\
S_4&=& R(\al^4)&=&\al^{11}\\
S_5&=& R(\al^5)&=&1\\
S_6&=& R(\al^6)&=&\al^{12}
\end{array}$$

Applying Euclid's algorithm, we obtain

$$\begin{array}{|r|l|l|l|}
\hline
i & r_i=r_{i-2}-q_ir_{i-1} & q_i & t_i=t_{i-2}-q_it_{i-1}\\
\hline
-1 & x^6                 & &           0     \\
0  & \al^{14}+\al^{13}x+\al^{6}x^2+\al^{11}x^3+x^4+\al^{12}x^5 & & 1     \\
1  &\al^5+\al^{10}x+\al^{13}x^2+\al^{11}x^3+\al^8x^4 &\al^6+\al^3 x &
\al^6+\al^3 x \\
2  &\al^{14}+\al^{10}x+\al^{8}x^2+\al^{9}x^3 &\al^4 x &
1+\al^{10}x+\al^{7}x^2\\
3  &\al^{7}+\al^{6}x^2 &\al^{14}+\al^{14} x &
\al^{8}+\al^{7}x+\al^{5}x^2+\al^{6}x^3\\
\hline
\end{array}$$

Multiplying $t_3(x)$ by $\al^9$, we obtain
$\si(x)\eq \al^{2}+\al x+\al^{14}x^2+x^3$. The roots of $\si(x)$
are $\al^{13}\eq \al^{-2}$, $\al^{12}\eq \al^{-3}$, and
$\al^{7}\eq \al^{-8}$, therefore, the errors occurred in locations 2, 3
and 8. Since this is a binary code, it is not necessary to find $\om(x)$,
we simply change locations 2, 3 and 8 of the received vector
$R(x)$, giving

$$C(x)\eq 1+x^3+x^4+x^6+x^8+x^9+x^{10}+x^{11}.$$

In binary, this is                                                     

$$\uc\eq (1\;0\;0\;1\;1\;0\;1\;0\;1\;1\;1\;1\;0\;0\;0).$$

\section{Techniques for Correction of Bursts}
\label{secint}

In the previous sections we have studied Reed-Solomon codes and their
decoding. In this section, we will look into methods for their use
in burst correction. The two main methods that we will investigate are
interleaving and product codes.

Let us start by the definition of a burst. A burst of length $l$ is a
vector whose non-zero entries are among $l$ consecutive (cyclically)
entries, the first and last of them being non-zero. Although the entries
of the vector can be in any field, let us concentrate on binary bursts.
We will use the elements of larger fields (bytes) to correct them.
Below are some examples of bursts of length 4 in vectors of length 15:

$$\begin{array}{ccccccccccccccc}
0&0&0&1&0&1&1&0&0&0&0&0&0&0&0\\
0&0&0&0&0&0&1&1&1&1&0&0&0&0&0\\
1&0&0&0&0&0&0&0&0&0&0&0&1&0&0
\end{array}$$

There is a relationship between the burst-correcting capability of
a code and its redundancy. This relationship is called the Reiger
bound and is presented next.

\begin{theo}[Reiger Bound]
\label{reiger}{\em
Let $\C$ be an $[n,k]$ linear code over a field $GF(q)$
that can correct all bursts of length
up to $l$. Then $2l\leq n-k$.
}\end{theo}

\pf
Recall that the total number of syndromes is $q^{n-k}$
Consider the $q^{2l}$ vectors
whose first $n-2l$ coordinates are zero. Those $q^{2l}$ vectors have
different syndromes. Otherwise, if two such vectors have the same
syndrome, their difference is in the code. This difference is a burst
of length $\leq 2l$, which can be viewed as the sum of two bursts of length
$\leq l$ each. These two bursts of length $\leq l$ have the same syndrome,
a contradiction. Thus, the number of syndromes corresponding
to the $q^{2l}$ vectors whose first $n-2l$ coordinates are zero
is exactly $q^{2l}$, and this number cannot exceed the total number
of syndromes, $q^{n-k}$.

The result follows.\qed

Cyclic binary codes that can correct bursts were obtained by computer
search. A well known family of burst-correcting codes are the so called
Fire codes. For a description of Fire codes and lists of good binary
cyclic burst-correcting codes, we refer the reader to~\cite{bl,lin}.
Here, we will concentrate on the use of RS codes for burst correction.
There are good reasons for this. One of them is that, although good
burst-correcting codes have been found by computer search, there are
no known general constructions giving cyclic codes that approach the
Reiger bound. Interleaving of RS codes on the other hand, to be described
below, provides a burst-correcting code whose redundancy, asymptotically,
approaches the Reiger bound. The longer the burst we want to correct, the
more efficient interleaving of RS codes is. The second reason for choosing
interleaving of RS codes, and probably the most important one, is that,
by increasing the error-correcting capability of the individual RS codes,
we can correct multiple bursts, as we will see. The known cyclic codes
are designed, in general, to correct only one burst.

Let us start with the use of regular RS codes for correction of bursts.
Let $\C$ be an $[n,k]$ RS code over $GF(2^b)$ (i.e., $b$-bit bytes).
If this code can correct $s$ bytes, in particular, it can correct
a burst of length up to $(s-1)b+1$ bits. In effect, a burst of length
$(s-1)b+2$ bits may affect $s+1$ consecutive bytes, exceeding the
byte-correcting capability of the code. This happens when the burst
$(s-1)b+2$ bits starts in the last bit of a byte. How good are then RS
codes as burst-correcting codes?
Given a binary $[n,k]$ that can correct bursts of length up to $l$,
we define a parameter, called
the {\em burst-correcting efficiency} of the code, as follows:

\begin{eqnarray}
\label{eff}
e_l&=&{2l\over n-k}
\end{eqnarray}

Notice that, by the Reiger bound, $e_l\leq 1$. The closer $e_l$ is to 1,
the more efficient the code is for correction of bursts.
Going back to our $[n,k]$ RS code over $GF(2^b)$, it can be regarded
as an $[nb,kb]$ binary code.
Assuming that the code can correct $s$ bytes and its redundancy is
$n-k=2s$, its burst-correcting efficiency is

\begin{eqnarray*}
e_{(s-1)b+1}&=&{(s-1)b+1\over bs}.
\end{eqnarray*}

Notice that, for $s\ra\infty$, $e_{(s-1)b+1}\ra 1$, justifying our
assertion that for long bursts, RS codes are efficient as burst-correcting
codes. However, when $s$ is large, there is a problem regarding
complexity. It may not be practical to implement a RS code with too
much redundancy. An alternative would be to implement a 1-byte correcting
RS code interleaved $s$ times. Given an $[n,k]$ code
interleaved $m$ times, the scheme looks as follows:

$$\begin{array}{|l|l|l|l|l|}
\hline c_{0,0}&c_{0,1}&c_{0,2}&\ldots &c_{0,m-1}\\
\hline c_{1,0}&c_{1,1}&c_{1,2}&\ldots &c_{1,m-1}\\
\hline c_{2,0}&c_{2,1}&c_{2,2}&\ldots &c_{2,m-1}\\
\hline \vdots &\vdots &\vdots &\ddots &\vdots \\
\hline c_{k-1,0}&c_{k-1,1}&c_{k-1,2}&\ldots &c_{k-1,m-1}\\
\hline
\hline c_{k,0}&c_{k,1}&c_{k,2}&\ldots &c_{k,m-1}\\
\hline c_{k+1,0}&c_{k+1,1}&c_{k+1,2}&\ldots &c_{k+1,m-1}\\
\hline \vdots &\vdots &\vdots &\ddots &\vdots \\
\hline c_{n-1,0}&c_{n-1,1}&c_{n-1,2}&\ldots &c_{n-1,m-1}\\
\hline
\end{array}$$

Each column $c_{0,j},c_{1,j},\ldots,c_{n-1,j}$ is a codeword in an
$[n,k]$ code. In general, each symbol $c_{i,j}$ is a byte and the code
is a RS code. The first $k$ bytes carry information bytes and the
last $n-k$ bytes are redundant bytes.
The bytes are read in row order, and the parameter $m$ is
called the depth of interleaving. If each of the individual codes
can correct up to $s$ errors, then the interleaved scheme can correct
up to $s$ bursts of length up to $m$ bytes each, or $(m-1)b+1$ bits each.
This occurs because a burst of length up to $m$ bytes is distributed
among $m$ different codewords. Intuitively, interleaving
``randomizes'' a burst.

The drawback of interleaving is delay: notice that we need to read most
of the information bytes before we are able to calculate and write the
redundant bytes. Thus, we need enough buffer space to accomplish this.

A natural generalization of the interleaved scheme described above is
product codes. In effect, we may consider that both rows and columns are
encoded into error-correcting codes. The product of an $[n_1,k_1]$ code
$\C_1$ with an $[n_2,k_2]$ code $\C_2$ is as follows:

$$\begin{array}{|l|l|l|l|l||l|l|l|l|}
\hline c_{0,0}&c_{0,1}&c_{0,2}&\ldots &c_{0,k_2-1}&
c_{0,k_2}&c_{0,k_2+1}&\ldots &c_{0,n_2-1}\\
\hline c_{1,0}&c_{1,1}&c_{1,2}&\ldots &c_{1,k_2-1}&
c_{1,k_2}&c_{1,k_2+1}&\ldots &c_{1,n_2-1}\\
\hline c_{2,0}&c_{2,1}&c_{2,2}&\ldots &c_{2,k_2-1}&
c_{2,k_2}&c_{2,k_2+1}&\ldots &c_{2,n_2-1}\\
\hline \vdots &\vdots &\vdots &\ddots &\vdots &\vdots &\vdots &\ddots
&\vdots \\
\hline c_{k_1-1,0}&c_{k_1-1,1}&c_{k_1-1,2}&\ldots &c_{k_1-1,k_2-1}&
c_{k_1-1,k_2}&c_{k_1-1,k_2+1}&\ldots &c_{k_1-1,n_2-1}\\
\hline
\hline c_{k_1,0}&c_{k_1,1}&c_{k_1,2}&\ldots &c_{k_1,k_2-1}&
c_{k_1,k_2}&c_{k_1,k_2+1}&\ldots &c_{k_1,n_2-1}\\
\hline c_{k_1+1,0}&c_{k_1+1,1}&c_{k_1+1,2}&\ldots &c_{k_1+1,k_2-1}&
c_{k_1+1,k_2}&c_{k_1+1,k_2+1}&\ldots &c_{k_1+1,n_2-1}\\
\hline \vdots &\vdots &\vdots &\ddots &\vdots &\vdots &\vdots &\ddots
&\vdots \\
\hline c_{n_1-1,0}&c_{n_1-1,1}&c_{n_1-1,2}&\ldots &c_{n_1-1,k_2-1}&
c_{n_1-1,k_2}&c_{n_1-1,k_2+1}&\ldots &c_{n_1-1,n_2-1}\\
\hline
\end{array}$$

If $\C_1$ has minimum distance $d_1$ and $\C_2$ has minimum distance $d_2$,
it is easy to see that the product code, that we denote $\C_1\times \C_2$,
has minimum distance $d_1d_2$ (Problem~\ref{pr2.32}).

In general, the symbols are read out in row order (although other readouts,
like diagonal readouts, are also possible). For encoding, first the
column redundant symbols are obtained, and then the row redundant symbols.
For obtaining the checks on checks $c_{i,j}$, $k_1\leq i\leq n_1-1$,
$k_2\leq j\leq n_2-1$, it is easy to see that it is irrelevant if we
encode on columns or on rows first.
If the symbols are read in row order,
normally $\C_1$ is called the outer code and $\C_2$ the inner code.
For decoding, there are many possible procedures. The idea is to correct
long bursts together with random errors. The inner code $\C_2$ corrects
first. In that case, two events may happen when its error-correcting
capability is exceeded: either the code will detect the error event or it
will miscorrect. If the code detects an error event (that may well have
been caused by a long burst), one alternative is to declare an erasure
in the whole row, which will be communicated to the outer code $\C_1$.
The other event is a miscorrection, that cannot be detected. In this
case, we expect that the errors will be corrected by the error-erasure
decoder of the outer code.

Another alternative is to use part of the power of the inner code to
correct and the rest to detect. If the channel is dominated by long
bursts, we might want to increase the detection capability of the inner
code and its ability to declare erasures to the outer code. In that case,
the task of the outer code is facilitated, since it can correct roughly
double the number of erasures as errors. Then the outer code will correct
errors together with erasures. Finally, we can use the inner code once
more, this time with full correction capability, to wipe out any remaining
errors left out by the outer code.

The methods described above concentrate on correcting bursts.
Let us point out that there are methods to decode product codes up to
the full minimum distance~\cite{bl}.

Product codes are important in practical applications. For instance, the
code used in the DVD (Digital Video Disk)
is a product code where $\C_1$ is a $[208,192,17]$ RS code and
$\C_2$ is a $[182,172,11]$ RS code. Both RS codes are defined over
$GF(256)$, where $GF(256)$ is generated by the primitive polynomial
$1+x^2+x^3+x^4+x^8$.

\pr

\begin{prob}
\label{pr2.32}
{\em
Let $\C_1$ and $\C_2$ be linear codes with minimum distance
$d_1$ and $d_2$ respectively.
Prove that the minimum distance of $\C_1\times \C_2$ is $d_1d_2.$
}\end{prob}


\sol
\vspace{.8cm} 

{\bf Problem~\ref{pr2.32}}

Take a non-zero codeword in $\C_1\times \C_2$ and consider 
a row that is non-zero, say row $i$. Let the non-zero coordinates
in row $i$ be $j_1,j_2,\ldots,j_l$. Since, in particular, row $i$
is a codeword in $\C_2$, then $l\geq d_2$. Also, each of the columns
$j_1,j_2,\ldots,j_l$ corresponds to a non-zero codeword in $\C_1$,
thus, each of those columns has weight at least $d_1$. So, adding the
weights of columns $j_1,j_2,\ldots,j_l$, we have at least
$d_1l\geq d_1d_2$ 1's.

\end{document}